\documentclass{article}
\usepackage{color, xcolor, colortbl}
\usepackage{graphicx,epstopdf}
\usepackage{geometry}
\usepackage{enumerate}
\usepackage{amsmath,amssymb,amsthm}
\usepackage{algorithm}
\usepackage{algorithmic}
\usepackage{caption}
\usepackage{array}
\usepackage{bm}
\usepackage{appendix}
\usepackage{multirow}
\usepackage{braket}
\usepackage[english]{babel}
\usepackage{aliascnt}
\usepackage{hyperref}
\hypersetup{hidelinks}
\usepackage[capitalize]{cleveref}
\usepackage[sort&compress,square,numbers]{natbib}
\usepackage{adjustbox}
\usepackage{xspace}
\usepackage[roman]{complexity}
\usepackage{soul}
\usepackage{tikz}
\usetikzlibrary{quantikz}
\usepackage{rotating}
\usepackage{setspace}
\usepackage{fancyhdr}
\usepackage{float}
\usepackage{placeins}
\usepackage{subfig}
\newcommand{\la}{\langle}
\newcommand{\ra}{\rangle}

\usepackage{authblk}

\newcommand{\norm}[1]{\left\lVert#1\right\rVert}

\newcommand{\op}{{\rm op}}
\newcommand{\dd}{ \mathrm{d}}
\newcommand{\Or}{\mathcal{O}}

\newcommand{\Sc}{\mathcal{S}}

\renewcommand{\Re}{\operatorname{Re}}

\newcommand{\I}{\mathrm{i}}

\theoremstyle{plain}

\newtheorem{thm}{\protect\theoremname}

\newaliascnt{lemma}{thm}
\newtheorem{lemma}[lemma]{\protect\lemmaname}
\aliascntresetthe{lemma}

\newaliascnt{prop}{thm}
\newtheorem{prop}[prop]{\protect\propositionname}
\aliascntresetthe{prop}

\newaliascnt{cor}{thm}
\newtheorem{cor}[cor]{\protect\corollaryname}
\aliascntresetthe{cor}

\newaliascnt{rem}{thm}

\aliascntresetthe{rem}

\newaliascnt{assumption}{thm}

\aliascntresetthe{assumption}

\newaliascnt{defn}{thm}

\aliascntresetthe{defn}

\newaliascnt{notation}{thm}

\aliascntresetthe{notation}

\newaliascnt{fact}{thm}

\aliascntresetthe{fact}

\providecommand{\definitionname}{Definition}
\providecommand{\assumptionname}{Assumption}
\providecommand{\corollaryname}{Corollary}
\providecommand{\lemmaname}{Lemma}
\providecommand{\propositionname}{Proposition}
\providecommand{\remarkname}{Remark}
\providecommand{\theoremname}{Theorem}

\newcommand{\bbe}{\mathbb E}

\newcommand{\bbr}{\mathbb R}

\newcommand{\cO}{\mathcal O}

\title{Tackling Many-Body Coulomb Singularity by Randomization}

\author[1,2]{Di Fang}
\author[1,3]{Wook Yoon}

\affil[1]{Department of Mathematics, Duke University}
\affil[2]{Duke Quantum Center, Duke University}
\affil[3]{Research Institute of Mathematics, Seoul National University}

\date{}

\begin{document}

\maketitle

\begin{abstract}

Many-body Coulomb dynamics is central to quantum chemistry and molecular simulation, yet the Coulomb singularity undermines both the standard error analysis and performance of product formulas. Even for the simple hydrogen ground state, the celebrated Trotter formulas converge more slowly than their nominal order, and going to higher order does not help. Randomization by itself does not resolve this. We show that conventional qDRIFT suffers the same loss of accuracy, and that for general initial states, the errors of both Trotter formulas and qDRIFT can decrease more slowly than any power of the number of time steps (logarithmically).
To overcome these limitations, we introduce supersmoothing qDRIFT, a randomized product formula that turns 
a physical property of free particle propagation
into an algorithmic resource. 
Our algorithm achieves an error bound that decreases linearly with the time step, uniformly over \textit{all} initial states and without any smoothness or energy assumptions.
Combined with spatial discretization analysis, the resulting end-to-end gate count improves on those of qubitization and high-order Trotter formulas applied directly to the discretized Hamiltonian, despite first-order convergence in time. This work demonstrates new ways of utilizing continuum dynamics to achieve algorithmic advantage.

\end{abstract}

\section{Introduction}

Many-body Coulomb dynamics underlies electronic structure, molecular processes, and the coupled motion of electrons and nuclei. Its simulation is a central application of quantum computing~\cite{KassalWhitfieldPerdomoOrtizYungAspuruGuzik2011,
CaoRomeroOlsonDegrooteJohnsonEtAl2019,
McArdleEndoAspuruGuzikBenjaminYuan2020,
BauerBravyiMottaChan2020,
BidartVaishKabengelePangLiuRubenstein2026}. Substantial advances in first- and second-quantized circuit constructions~\cite{KassalJordanLoveMohseniAspuru-Guzik2008,BabbushWiebeMcCleanMcClainNevenChan2018,BabbushBerryMcCleanNeven2019,SuBerryWiebeRubinBabbush2021,BerryWanBaczewskiEklundTikkuEtAl2026,LowKingBerryHanDePrinceEtAl2025,
DutkiewiczWhiteLowDePrinceHarriganEtAl2026} have been accompanied by detailed resource estimates for various applications~\cite{McCleanBabbushLoveAspuru-Guzik2014,ReiherWiebeSvoreWeckerTroyer2017,KivlichanGidneyBerryWiebeEtAl2020,VonBurgLowHaenerSteigerReiherEtAl2021,RubinBerryKononovEtAlLeeNevenBabbushBaczewski2024,BabbushHugginsBerryUngZhaoEtAlBaczewskiLee2023,DutkiewiczWhiteLowDePrinceHarriganEtAl2026,PathakKononovBaczewski2026}. 
Underlying all such constructions is the continuum problem, which presents difficulties that are not captured by a finite-dimensional Hamiltonian alone.
The kinetic energy is unbounded, while the interactions are both singular at particle collisions and long-ranged. Understanding how these features affect a quantum algorithm is essential for connecting its complexity to the underlying physical problem. 
To assess many-body scalability, the dependence of algorithmic bounds on the system size $N$ must be explicit:
convergence at each fixed $N$ does not establish polynomial scaling, as the prefactors may grow superpolynomially with $N$.

A common approach first discretizes space and then simulates the resulting finite-dimensional Hamiltonian using bounded-operator methods, including qubitization and interaction-picture algorithms~\cite{BabbushBerryMcCleanNeven2019,
SuBerryWiebeRubinBabbush2021,
LowChuang2017,GilyenSuLowEtAl2019,
LowChuang2019,LowWiebe2019,
KieferovaSchererBerry2019}.
This separates spatial approximation from quantum simulation, but the operator norms entering the simulation bounds grow with spatial resolution. Consequently, an efficient algorithm for a fixed discretized Hamiltonian does not by itself determine the end-to-end cost of the continuum problem~\cite{KivlichanWiebeBabbushEtAl2017,AnFangLin2021,ChildsLengEtAl2022,BurgarthGalkeHahnvanLuijk2023}. That cost also depends on a spatial error analysis fixing the required resolution, as recently carried out for Coulomb systems~\cite{FangZhang2026,BerryTikkuSimonBabbush2026}.
An alternative is to analyze the time approximation directly in the continuum and discretize space afterward. This approach retains the structure of the differential operators and allows the dynamics itself, rather than a spatial cutoff, to control the Coulomb singularities. This can yield basis-independent convergence results~\cite{FangWuSoffer2025,FangWu2026,BeckerGalkeSalzmannLuijk2024,FangZhang2026yukawa}. However, working directly with the continuum Hamiltonian 
exposes a challenge beyond increasing matrix dimension: the analysis must confront the Coulomb singularity directly.

Product formulas provide a natural starting point: kinetic evolution and individual Coulomb interactions have simple physical interpretations and can be implemented separately. Their familiar convergence orders, however, need not survive in the continuum. Coulomb singularities invalidate the bounded-commutator estimates behind the usual analysis~\cite{ChildsSuTranEtAl2020} and can cause severe order reduction even for initial states in the Hamiltonian domain, $H^2(\mathbb{R}^{3N})$. This phenomenon has been observed both numerically and theoretically for the ground state of hydrogen atom~[\citenum{BurgarthFacchiHahnJohnssonYuasa2024}, \citenum{FangWu2026}]. In particular, both Lie and Strang splitting have a one-step error proportional to $h^{5/4}$ on the hydrogen ground state, rather than their nominal local orders $h^2$ and $h^3$~\cite{FangWu2026}. Thus increasing the formal order of a product formula does not necessarily remove the singularity-induced loss of accuracy. 
For general $L^2(\mathbb{R}^{3N})$ initial states, the issue is more basic: strong convergence (in the large-step limit) alone gives no quantitative rate, while estimates requiring domain regularity do not apply to states outside the Hamiltonian domain.

The fundamental question motivating this work is: \emph{How can we efficiently simulate continuum many-body Coulomb dynamics on a quantum computer?} Its unbounded Hamiltonian and singular, long-range interactions challenge conventional simulation methods. A central issue is whether the structure of the dynamics itself can guide algorithm design and overcome these difficulties. We ask three questions:
\begin{enumerate}
\item Can we design a product formula that overcomes Coulomb-induced \textit{order reduction} in Trotterization? Can any algorithm achieve a guaranteed convergence rate for continuum Coulomb dynamics with general initial quantum states, not necessarily in the Hamiltonian domain?
\item Can the smoothing structure of continuum dynamics be turned into an \textit{algorithmic resource}, replacing initial-state regularity in the control of Coulomb singularities?
\item Can this continuum advantage survive spatial discretization and circuit implementation, retaining its favorable precision dependence in the \textit{end-to-end gate complexity}?
\end{enumerate}
We answer these questions affirmatively by treating a property of continuum motion as an algorithmic resource. We develop supersmoothing qDRIFT, a randomized product formula specifically designed to exploit the continuum smoothing structure of many-body Coulomb dynamics. The algorithm preserves the exact free evolution and randomizes the interaction times, so that the averaged dynamics exploits Kato supersmoothing, which limits how long a freely evolving wave function can remain concentrated near a particle collision.
This yields global first-order convergence in the unconstrained diamond norm, without \textit{any} regularity or energy assumption on the initial state (!). The initial state can be any quantum state (i.e., those in $L^2(\mathbb{R}^{3N})$) that does not even need to be in the domain of the Hamiltonian.
We also prove that conventional qDRIFT~\cite{Campbell2019} shares the Trotter order barrier in diamond norm even after averaging over its randomness, and that neither Trotterization nor conventional qDRIFT admits a positive algebraic (power-law) convergence rate valid for general $L^2(\mathbb{R}^{3N})$ initial states. 
Together with spatial discretization analysis, the algorithm enjoys an improved end-to-end gate complexity, improving on those for direct qubitization and every fixed-order Trotter formula under the same input assumptions. 
These results show, perhaps surprisingly, that exploiting the structure of continuum dynamics before discretizing space can outperform the standard paradigm of discretizing first and then applying generic simulation algorithms.

\section{Algorithm Overview and Main Results}\label{sec:algorithm-main-results}

We consider the many-body Coulomb Hamiltonian with $N$ particles
\begin{equation}\label{eq:hamiltonian}\vspace{-0.1in}
H=A+B,\quad A=-\sum_{j=1}^N\Delta_{x_j},\quad
B=\sum_{j<k}\frac{c_{jk}}{|x_j-x_k|}  =: \sum_{j<k}B_{jk},
\end{equation}
where $x_j\in\mathbb R^3$ and $c_{jk}\in\mathbb R$ with $c_0=\max_{j<k}|c_{jk}|<\infty$. Both attractive and repulsive interactions are allowed, which covers both electron-electron and nucleus interactions. We note that the mass of each particle can be different and will only affect the constants in the following result. The Hamiltonian acts on $L^2(\mathbb R^{3N})$ and is self-adjoint on its domain $D(H) = H^2(\mathbb R^{3N})$. 

We approximate its evolution over a target final time $T>0$ using $L$ steps of size $t=T/L$.
At each step, independently sample a pair $(j,k)$ whose $c_{jk}\ne0$ with probability
\begin{equation}\label{eq:overview-probability}
\pi_{jk}=\frac{|c_{jk}|}{\sum_{a<b}|c_{ab}|},
\end{equation}
and sample $\theta$ uniformly from $[0,1]$. We then apply
\begin{equation}\label{eq:overview-step}
S_{jk,\theta}(t)
=e^{-i\theta tA}
 e^{-itB_{jk}/\pi_{jk}}
 e^{-i(1-\theta)tA}.
\end{equation}
If all couplings vanish, the evolution is exactly free. The factor $1/\pi_{jk}$ ensures that the sampled potential has mean $B$. Each step applies only one pair interaction and retains the full kinetic evolution for a total duration $t$. The additional random variable $\theta$ determines when this interaction is applied within the step. We call this algorithm \emph{supersmoothing qDRIFT}.
Like Trotter splitting, this product formula gives rise to both classical and quantum algorithms, depending on how the individual unitaries are implemented.

Let $S=S_L\cdots S_1$, where the steps are independent. The algorithm implements exactly $L$ sampled interactions. Combining adjacent free-evolution segments leaves at most $L+1$ kinetic exponentials. Averaging over the classical randomness gives the output channel
\begin{equation}\label{eq:overview-channels}
\Phi(\rho)=\mathbb E[S\rho S^\dagger],
\quad
\mathcal U_T(\rho)=e^{-iTH}\rho e^{iTH},
\end{equation}
where $\mathcal U_T$ is the exact evolution channel.
Our main theorem establishes a global first-order error bound for supersmoothing qDRIFT, defined in~\cref{eq:overview-step}, with explicit polynomial dependence on the particle number. The bound holds uniformly over all normalized initial states in $L^2(\mathbb R^{3N})$, including those outside the Hamiltonian domain, without any regularity or energy assumption.

\begin{thm}[convergence for arbitrary initial states]
\label{thm:overview-channel}
For every $T>0$ and every positive integer $L$, supersmoothing qDRIFT, defined in~\cref{eq:overview-step}, 
with time step $t=T/L$ satisfies
\begin{equation}\label{eq:overview-diamond}
\norm{\Phi-\mathcal U_T}_\diamond
\le 
\frac{C}{L}\left(c_0^2N^4T+c_0^4N^8T^2\right),
\end{equation}
where $C$ is an absolute constant. 
\end{thm}

In particular, the same first-order bound holds in trace norm for every initial density operator, including pure states with wave functions outside $D(H)$, without a regularity or energy restriction. 
To achieve $\norm{\Phi-\mathcal U_T}_\diamond\le\epsilon$ for $0<\epsilon\le1$, it suffices to use
\begin{equation}\label{eq:overview-step-count}
L=\Or \left(\frac{c_0^2N^4T+c_0^4N^8T^2}{\epsilon}\right)
\end{equation}
time steps. Each sampled circuit uses exactly $L$ single-interaction potential exponentials
and at most $L+1$ kinetic exponentials. The total number of exponentials is therefore at most $2L+1$.

We also show that Trotter product formulas and conventional qDRIFT both have convergence barriers. Neither admits a positive algebraic convergence rate valid for general $L^2(\mathbb{R}^{3N})$ initial states. Even for initial states in the Hamiltonian domain $H^2(\mathbb{R}^{3N})$, conventional qDRIFT cannot achieve a convergence order greater than $3/4$ in general. These qDRIFT barriers persist after averaging over its randomness. The first-order diamond-norm guarantee of supersmoothing qDRIFT therefore goes beyond randomization and channel averaging alone. It relies on the specific randomization scheme that preserves continuum structure and exploits Kato supersmoothing.

To state these barriers precisely, we consider Trotter formula and a continuum all-term qDRIFT scheme that randomly selects kinetic or potential evolution. For the two-body case of~\cref{eq:hamiltonian}, their one-step channels are
\begin{equation}\label{eq:overview-comparison-channels}
\begin{aligned}
\mathcal T_t(\rho)
&=e^{-itB}e^{-itA}\rho e^{itA}e^{itB},\\
\mathcal Q_{t,p}(\rho)
&=p\,e^{-itA/p}\rho e^{itA/p}
 +(1-p)e^{-itB/(1-p)}\rho e^{itB/(1-p)}.
\end{aligned}
\end{equation}
Here $p\in(0,1)$ can be any fixed probability. This explicitly specifies the continuum sampling rule, since norm-proportional sampling is undefined for the unbounded operators $A$ and $B$. 

\begin{thm}[No algebraic rate for $L^2$ initial states]
\label{thm:barriers-L2}
Consider \cref{eq:hamiltonian} with $N=2$ and $c_{12}\ne0$, and fix $T=1$. There exist normalized states $\Psi_1,\Psi_2\in L^2(\mathbb R^6)$
such that, with $\rho_j=|\Psi_j\rangle\langle\Psi_j|$,
\begin{equation}\label{eq:barriers-L2}
\begin{aligned}
\norm{\mathcal T_{1/L}^{\,L}(\rho_1)-\mathcal U_1(\rho_1)}_1
&\ge \frac{c}{\log L},\\
\norm{\mathcal Q_{1/L,p}^{\,L}(\rho_2)-\mathcal U_1(\rho_2)}_1
&\ge \frac{c_p}{\log L},
\end{aligned}
\end{equation}
for every fixed $p\in(0,1)$ and sufficiently large $L$ and $c,c_p>0$ are some constants.
\end{thm}
We also show that no Trotter formula of fixed order $k\ge 1$, however high, guarantees an algebraic convergence rate for all $L^2(\mathbb R^6)$ initial states. Even on the hydrogen ground state, its one-step trace-norm error scales as $t^{5/4}$, decaying much more slowly than the expected $t^{k+1}$ rate (see \cref{sec:trotter-all-orders}).

\begin{thm}[Convergence barrier for conventional qDRIFT for $H^2$ initial states]
\label{thm:barrier-H2-qdrift}
Under the same conditions of \cref{thm:barriers-L2}, there exist normalized states $\Psi_3\in H^2(\mathbb R^6)$ such that, with $\rho_j=|\Psi_j\rangle\langle\Psi_j|$,
\begin{equation}\label{eq:barrier-H2-qdrift}
\norm{\mathcal Q_{1/L,p}^{\,L}(\rho_3)-\mathcal U_1(\rho_3)}_1
\ge c_p L^{-3/4},
\end{equation}
for every fixed $p\in(0,1)$ and sufficiently large $L$ and $c_p$ is some constant.
\end{thm}

\begin{table}[t]
\centering
\begin{tabular}{lcc}
\hline\hline
Method & $H^2(\mathbb{R}^{3N})$ states & $L^2(\mathbb{R}^{3N})$ states \\
\hline
$p$-th order Trotter formulas & local order barrier & none \\
Conventional qDRIFT & $\le 3/4$ & none \\
\textbf{Supersmoothing qDRIFT} & $\mathbf{1}$ & $\mathbf{1}$ \\
\hline\hline
\end{tabular}
\caption{Convergence orders in time for many-body Coulomb dynamics.
Orders are global unless marked local; 
``none'' means convergence
without any algebraic rate in general.}
\label{tab:orders}
\end{table}

\begin{figure*}[hbtp]
\centering
\begin{minipage}[t]{0.45\textwidth}
    \centering
    \includegraphics[width=\textwidth]{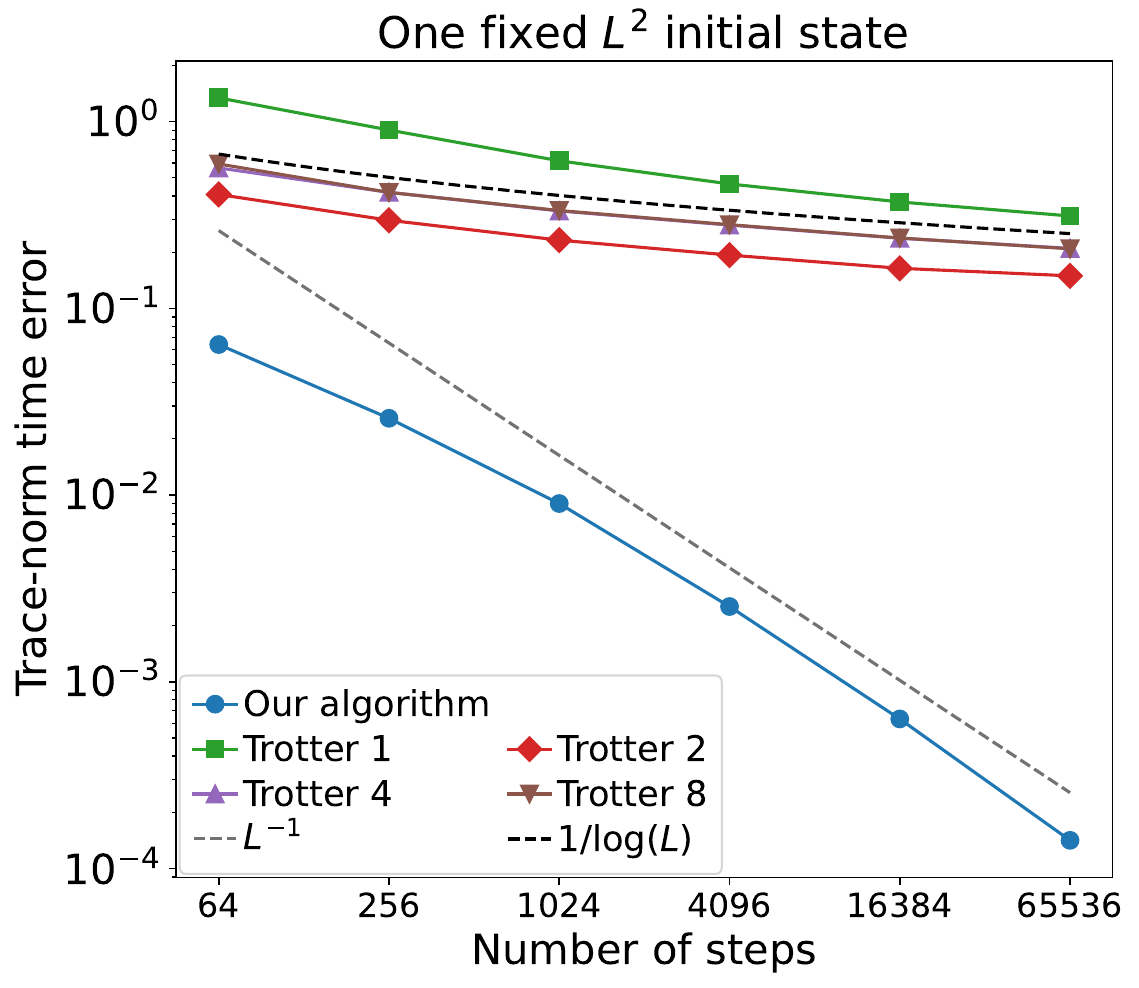}
    \par\smallskip
    (a)
\end{minipage}
\begin{minipage}[t]{0.45\textwidth}
    \centering
    \includegraphics[width=\textwidth]{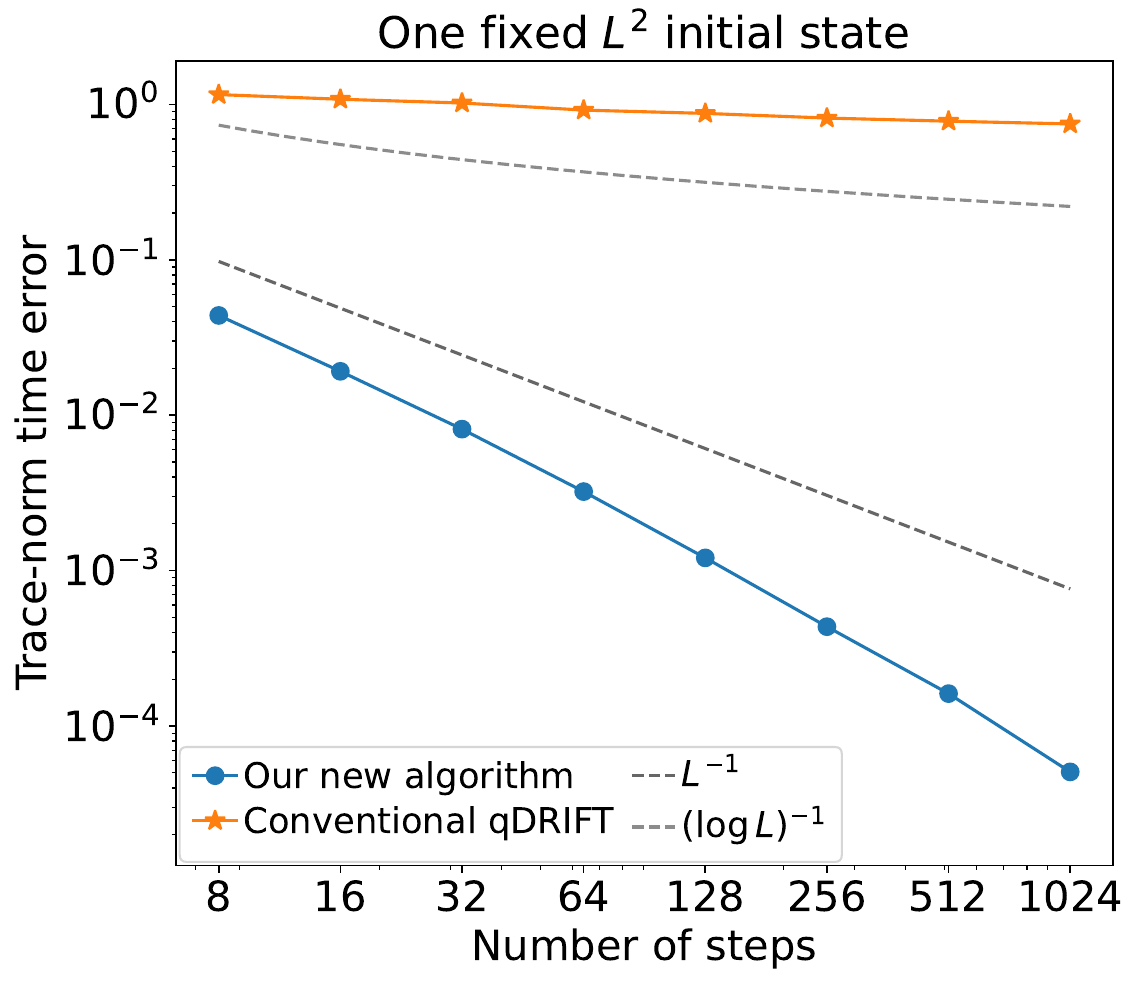}
    \par\smallskip
    (b)
\end{minipage}
\caption{Time-convergence comparison for two-body Coulomb dynamics at $T=1$. (a) Supersmoothing qDRIFT and Trotter formulas of orders $1$, $2$, $4$, and $8$. (b) Supersmoothing qDRIFT and conventional qDRIFT. Algorithms within each panel use the same initial state.
}
\label{fig:numerical-comparison}
\end{figure*}

\begin{figure}[h]
\centering
\includegraphics[width=.58\textwidth]{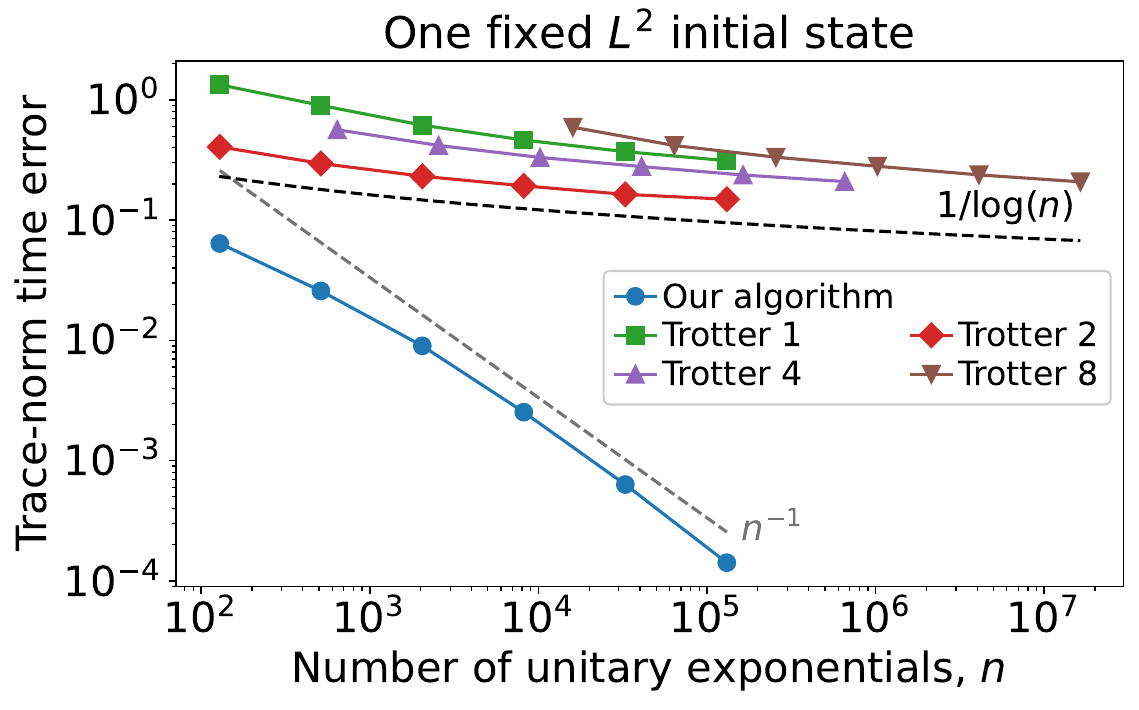}
\caption{Trace-norm time error versus the number $n$ of unitary exponentials for our algorithm and Trotter formulas of orders $1$, $2$, $4$, and $8$, using the same fixed initial state at $T=1$. Counts are per circuit realization after merging adjacent kinetic factors. The counts per circuit realization are $2L+1$ for supersmoothing qDRIFT and $2L$, $2L+1$, $10L+1$, and $250L+1$ for Trotter formulas of orders $1$, $2$, $4$, and $8$, respectively. For our algorithm, the error is estimated for the averaged channel. }
\label{fig:unitary-comparison}
\end{figure}

The convergence guarantees and barriers are summarized in \cref{tab:orders}. In the two-body examples of \cref{fig:numerical-comparison}, supersmoothing qDRIFT exhibits substantially faster error decay than the conventional methods over the tested range. Within each comparison, the initial state is held fixed as the number of time steps increases. The algorithmic efficiency is more pronounced in \cref{fig:unitary-comparison}, where the horizontal axis counts applications of the kinetic and potential exponentials, $e^{-isA}$ and $e^{-isB}$, per circuit realization after merging adjacent kinetic factors.

Note that for a general many-body system, the two methods use different potential exponentials. Our algorithm applies $e^{-isB_{jk}/\pi_{jk}}$, involving only one randomly selected interaction with sampling probability $\pi_{jk}$, whereas the $A+B$ Trotter formulas apply $e^{-isB}$, with $B=\sum_{j<k}B_{jk}$ containing all interactions. Counting each potential exponential as one application therefore does not reflect this difference in implementation cost. This distinction is absent in the two-body experiment above, where there is only one interaction term.

The $3/4$ barrier in \cref{thm:barrier-H2-qdrift} also has a numerical counterpart for a localized hydrogen atom in its internal ground state with a spreading center-of-mass wave packet. On this state, the trace-norm error decays approximately as $L^{-3/4}$ for conventional qDRIFT and substantially faster for supersmoothing qDRIFT (see \cref{fig:hydrogen-convergence}).

\section{Physical Intuition and Smoothing Mechanism}
\label{sec:physical-intuition}

The mechanism rests on a balance between Coulomb singularities and quantum dispersion. Bringing two particles close together makes their interaction stronger. However, localizing their relative wave function also increases its momentum uncertainty and accelerates its free spreading. For a freely evolving packet of width $\ell$, the squared Coulomb weight is of order $\ell^{-2}$, while the characteristic spreading time is of order $\ell^2$. The product is independent of $\ell$, suggesting that the time-integrated effect of the Coulomb singularity remains bounded even for arbitrarily narrow initial wave packets.

\begin{figure*}[h]
\centering
\includegraphics[width=0.4\textwidth]{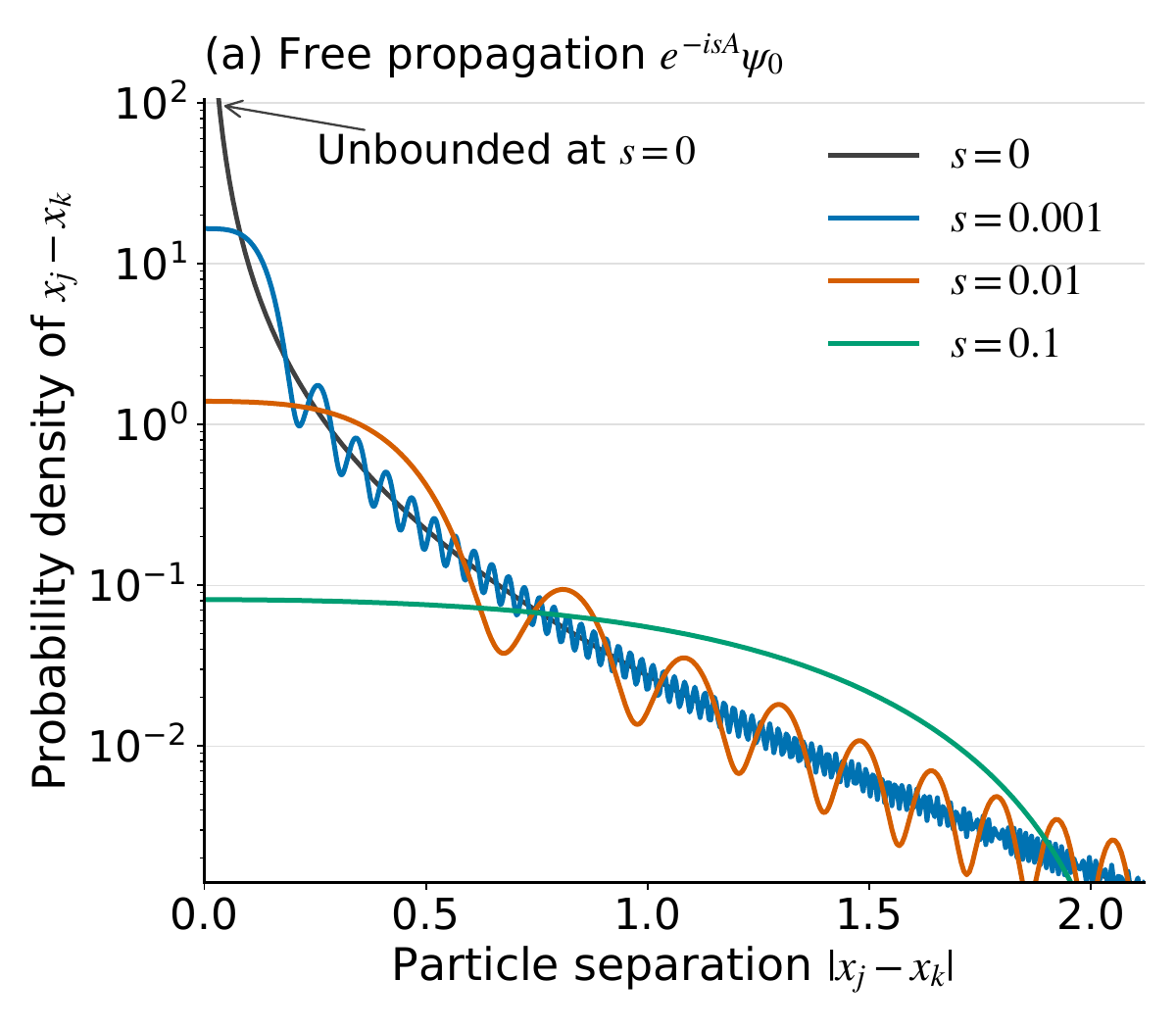}%
\includegraphics[width=0.4\textwidth]{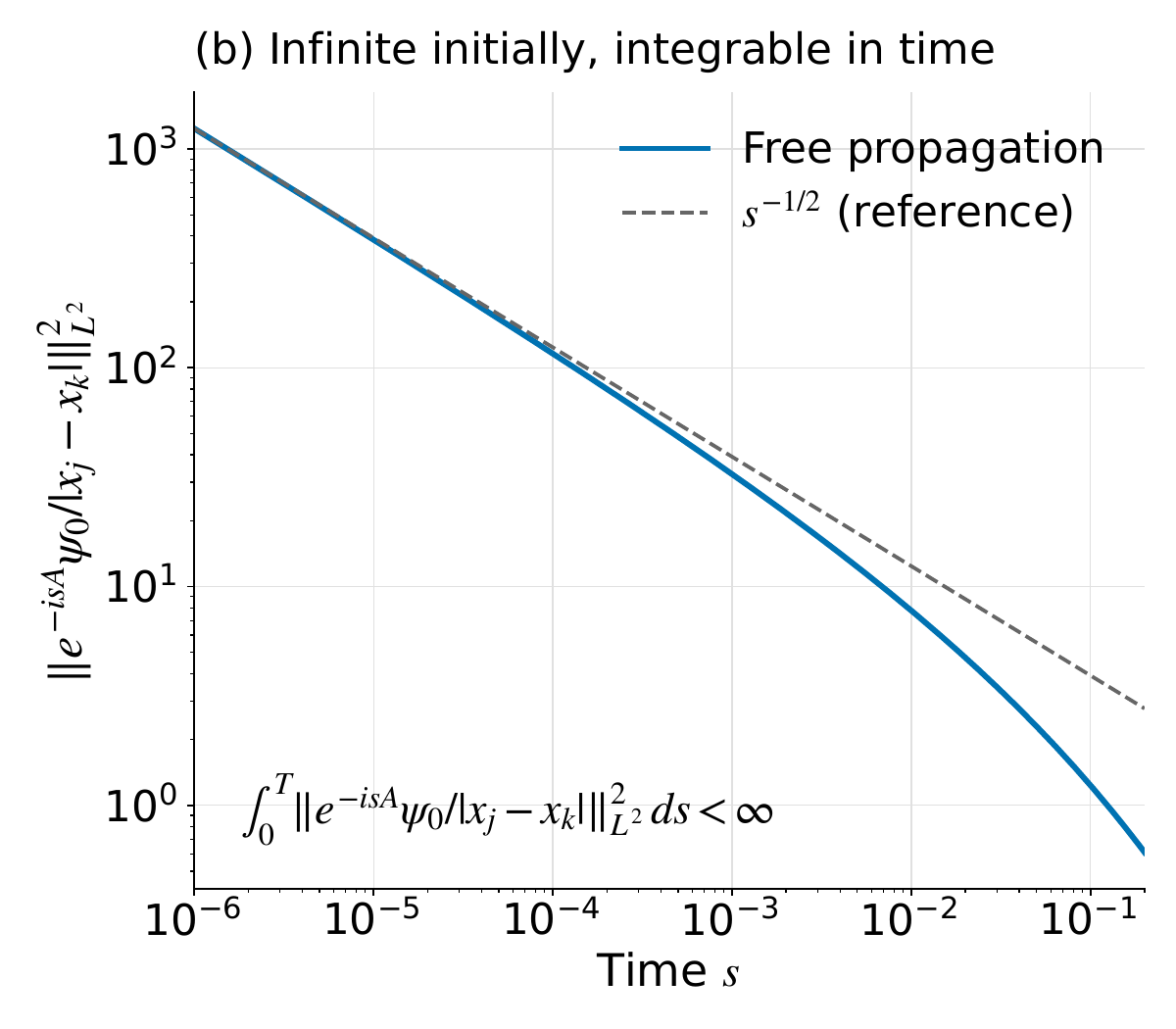}
\caption{
Kato supersmoothing effect illustration (free smoothing near a two-particle collision).
The normalized initial state has a relative-coordinate factor
proportional to $e^{-|x_j-x_k|/\sqrt{2}}/|x_j-x_k|$.
(a) The probability density of $x_j-x_k$, with all other
coordinates integrated out. The initially unbounded collision
peak becomes finite for every $s>0$ and broadens under free
propagation.
(b) The squared Coulomb-weighted norm
$\norm{e^{-isA}\psi_0/|x_j-x_k|}_{L^2}^2$.
For this initial state, it behaves like $s^{-1/2}$ near $s=0$
and is integrable in time.
}
\label{fig:free-smoothing}
\end{figure*}

Kato smoothing makes this intuition precise~\cite{KatoYajima1989,ReedSimonIV1978}. A consequence of the free supersmoothing property, as termed by Kato and Yajima~\cite{KatoYajima1989}, of the Coulomb weight is that, for every $\psi_0\in L^2(\mathbb R^{3N})$,
\begin{equation}\label{eq:mechanism-free-smoothing}
\int_{\mathbb R}
\left\|\frac{e^{-isA}\psi_0}{|x_j-x_k|}\right\|_{L^2}^2\,ds
\le C\|\psi_0\|_{L^2}^2,
\end{equation}
where $C$ is an absolute constant independent of $N$.
The left-hand side is the probability accumulated near a collision, weighted by the squared Coulomb singularity. Its time integral is finite even when the integrand is infinite at $s=0$. As illustrated in \cref{fig:free-smoothing}, an initially unbounded collision peak spreads under free propagation, and the squared Coulomb-weighted norm has only an integrable $s^{-1/2}$ singularity at $s=0$. In other words, this is a time-integrated smoothing effect of free particle propagation, and it does not require the wave function to become instantaneously smoother.

Our algorithm is designed to retain this physical mechanism. Each step applies the full kinetic evolution and inserts one sampled pair interaction at a uniformly random time. Averaging over the insertion time \textit{turns the step into an integral} along free propagation, precisely the structure required by smoothing, and the remaining error involves the squared Coulomb singularity, precisely the weight in~\cref{eq:mechanism-free-smoothing}. This error is therefore controlled by the $L^2$ norm alone, with no smoothness or energy constraints required for the initial state. Note that the time integral comes from classical randomness: each circuit realization still applies a single interaction between two free-evolution segments. Accumulating these step errors yields the first-order bound of \cref{thm:overview-channel}. By design, the smoothing comes from free propagation within each step, so it persists even when the full interacting dynamics stays localized, as for the bound states of atoms and molecules.

This is why \emph{the particular randomization} matters. A fixed insertion time, as in Trotter formulas including the Strang midpoint, generates no time average. Randomizing over all terms, including the kinetic one, as in conventional qDRIFT~\cite{Campbell2019}, makes the accumulated kinetic time fluctuate between realizations and dephases coherences between different kinetic energies. 
Random-time sampling~\cite{PoulinQarrySommaEtAl2011,BerryChildsSuEtAl2020,RajputRoggeroWiebe2021} and other forms of randomization~\cite{ChildsOstranderSu2019,NakajiBagherimehrabAspuruGuzik2024,GranetDreyer2024,SommaKingKothariEtAl2025,ChakrabortySambasivamShiraliEtAl2026} have been studied for bounded Hamiltonians.
For continuum Coulomb interactions, estimates based on the operator norms of the individual interaction terms give no finite bound.
Supersmoothing qDRIFT instead fixes the kinetic duration and randomizes only the interaction and its insertion time, preserving free propagation exactly and converting its smoothing into control of the simulation error.

\section{End-to-end complexity}
\label{sec:end-to-end}

\begin{table}[htb]
\centering
\begin{tabular}{lc}
\hline\hline
Method & Gate complexity \\
\hline
Spatial discretization $\to$ qubitization
& $\widetilde{\mathcal{O}}(T^{2}N^{8}\epsilon^{-3/2})$ \\
Spatial discretization $\to$ first-order Trotter
& $\widetilde{\mathcal{O}}(T^{7/2}N^{12}\epsilon^{-13/4})$ \\
Spatial discretization $\to$ $p$-th order Trotter
& $\widetilde{\mathcal{O}}_p(T^{2+\frac{3}{2p}}N^{7+\frac{5}{p}}\epsilon^{-\frac{3}{2}-\frac{7}{4p}})$ \\
\textbf{Supersmoothing qDRIFT $\to$ spatial discretization}
& {\boldmath $\widetilde{\mathcal{O}}(T^{2}N^{7}\epsilon^{-1})$} \\
\hline\hline
\end{tabular}
\caption{
End-to-end gate complexity for many-body Coulomb dynamics with error tolerance $\epsilon$ and final time $T\geq1$. Here $N$ denotes the number of electrons when the nuclei are fixed and the total number of electrons and nuclei for coupled electron-nuclear dynamics.
For all algorithms, errors are measured in the trace norm of the density operator. For randomized algorithms, the output is averaged over the circuit randomness.
Here, $\widetilde{\Or}(g)=\Or(g\,\polylog(g))$.}
\label{tab:complexity}
\end{table}

We combine the time-discretization estimate with the spatial error analysis to obtain the end-to-end algorithm complexity. We consider neutral molecules with $N$ electrons and fixed nuclei in a bounded region. The same gate bound holds for dynamical nuclei, 
with $N$ now counting both electrons and nuclei. For normalized initial states satisfying $\norm{\psi_0}_{H^2}=\Or(1)$ and $\norm{|x|\psi_0}_2=\Or(N^{3/2})$, combining our time-error estimate with the spatial discretization analysis of the wavefunction dynamics~\cite{FangZhang2026} gives 
averaged-output 
trace-norm error at most $\epsilon$ with gate complexity
\begin{equation}\label{eq:end-to-end-gates}
G=\widetilde{\Or}\!\left(T^2N^7\epsilon^{-1}\right),
\end{equation}
for $T\ge1$ and $0<\epsilon\le1$, including time, spatial, and implementation errors. Here the spatial approximation uses the stated regularity and localization assumptions; the continuum time-convergence theorem itself requires neither. 
Together with first-quantized circuit constructions~\cite{SuBerryWiebeRubinBabbush2021,LowChuang2019}, this avoids an additional particle-number factor in converting the step count into gates. 

When initial states are in $H^2(\mathbb{R}^{3N})$, the Hamiltonian domain, the time-error bound of our algorithm improves to $\Or(T^2N^7/L)$, so $L=\Or(T^2N^7/\epsilon)$ steps suffice, proved in~\cref{subsection:H^2data_eta7}.
Each step samples only one interaction, and kinetic factors on unaffected registers cancel, giving a gate count with the same leading scaling as the step count. The derivation are given in \cref{sec:end-to-end-app}. Under the same input assumptions, \cref{tab:complexity} compares this bound with spatial discretization followed by standard simulation algorithms. Our bound improves on those for direct qubitization and every fixed-order Trotter formula, while interaction-picture simulation retains a better asymptotic bound. 

We also note that stronger initial-state assumptions, such as states in $D(H^2)$ rather than merely the domain of the Hamiltonian $D(H)=H^2 (\mathbb{R}^{3N})$, can yield sharper spatial discretization bounds~\cite{BerryTikkuSimonBabbush2026}. Whether this additional regularity can also improve the convergence and particle dependence of supersmoothing qDRIFT remains an interesting future direction.

\section{Conclusion and Remarks}
Supersmoothing qDRIFT is uniformly first order in the diamond norm for continuum many-body Coulomb dynamics, with no regularity or energy assumption on the initial state. The mechanism is that exact kinetic evolution combined with a randomized interaction insertion time generates precisely the time integrals bounded by Kato supersmoothing. Randomization alone does not suffice: Trotter formulas and conventional qDRIFT have order barriers on the Hamiltonian domain $H^2(\mathbb{R}^{3N})$ and no algebraic rate on general quantum states $L^2(\mathbb{R}^{3N})$.

The continuum convergence theorem is assumption-free; regularity and localization, and parameter assumptions enter through spatial discretization, as in existing spatial-discretization analyses. Combined with circuit implementations of the underlying unitaries, it gives an end-to-end gate count of $\widetilde{\mathcal{O}}(T^2N^7\epsilon^{-1})$ with all spatial errors explicitly controlled. Although our method is only first order in time, this improves on prior dynamics results on the discretize-first bounds for qubitization and fixed-order Trotter formulas~\cite{FangZhang2026}, whose costs depend on norms of the discretized Hamiltonian that grow as the grid is refined. Thus for unbounded dynamics, higher-order bounded-operator algorithms need not yield lower end-to-end cost. Although interaction-picture with high-order Dyson series still retains a better asymptotic scaling, our algorithm inherits the merits of product formulas that do not need additional ancillary qubits for this time discretization and may also be more near-term friendly.

Quantum simulation of continuum systems conventionally discretizes space first and then applies algorithms designed for bounded operators. The physics of the continuum is thereby partially discarded, resurfacing only as operator norms that grow with grid resolution. Here we reversed the order: building a continuum mechanism directly into the algorithm and discretizing the space only at the end yields convergence for every initial state and improved end-to-end complexity. In this sense, the continuum acts as a resource rather than an obstacle. We expect this continuum-inspired design to tame what a discretized description sees only as large operator norms, extending to other continuum systems whose cumulative effect is controlled by the continuum dynamics itself.

\section*{Acknowledgments}
D.F. acknowledges the support by the NSF CAREER award DMS-2438074, and the U.S. Department of Energy, Office of Science, Accelerated Research in Quantum Computing Centers, Quantum Utility through Advanced Computational Quantum Algorithms, grant no. DE-SC0025572. W.Y. acknowledges the support of the National Research Foundation of Korea (NRF) grant funded by the Korea government (MSIT), grant no. RS-2026-25484464.

\appendix
\section{Proof of the Main Result 1}

We use the Hamiltonian and randomized product formula defined in \cref{eq:hamiltonian,eq:overview-probability,eq:overview-step}. Throughout this section, $L^2=L^2(\mathbb R^{3N})$ and $H^2=H^2(\mathbb R^{3N})$. For $g\in H^2$, the $H^2$-norm is defined as
\[
\|g\|_{H^2}:=\sqrt{\|(-\Delta)g\|^2+\|g\|^2}.
\]
We also write $U(s):=e^{-isH}$ for $s\in\mathbb R$, and
\begin{equation*}
v_{jk}(x):=\frac{1}{|x_j-x_k|},
\quad
w_{jk}:=|c_{jk}|,
\quad
W:=\sum_{j<k}w_{jk}\le \frac{c_0N(N-1)}{2},
\end{equation*}
where $c_0:=\max_{1\leq j<k\leq N}|c_{jk}|<\infty$, as defined in \cref{sec:algorithm-main-results}. Thus $B_{jk}=c_{jk}v_{jk}$ and the sampling probabilities in \cref{eq:overview-probability} satisfy $\pi_{jk}=w_{jk}/W$. 
In the algorithm, we further introduce one additional random parameter $\theta\sim \mathrm{Unif}[0,1]$ so that at each step, we apply
\[S_{jk,\theta}(t):=e^{-i\theta t A}e^{-itB_{jk}/\pi_{jk}}e^{-i(1-\theta)t A}.\]
We randomly choose $S_1,\ldots, S_L$ which are $L$ independent copies of $S_{jk,\theta}$. Then the full random propagator can be written as
\[S=S_L\cdots S_1.\]
Note that these $S_i$'s and $S$ depend on $t$. 
As given in \cref{sec:algorithm-main-results}, 
\begin{equation} \label{eq:Et_def}
    E_t:=\bbe S_1=\sum_{j<k}\pi_{jk}\int_0^1e^{-i\theta t A}e^{-itB_{jk}/\pi_{jk}}e^{-i(1-\theta)t A}d\theta, 
\end{equation}
then the independence of $S_i$'s gives 
\[\bbe S=E_t^L.\] 

Our first goal is to derive the error estimate between $E_t^L$ and the exact operator $U(T)=e^{-iHT}$. We state the main result in the following theorem. 
\begin{thm} \label{thm:main_result1}
    Let $H=A+B$ be the $N$-body Hamiltonian with Coulomb interactions, where $A=-\Delta$ denotes the kinetic part and $B=V(x)$ the Coulomb interaction potential. Then for a total evolution time $T>0$ using $L$ times steps, set $t=T/L$ and assume that 
    \begin{equation}\label{eq:main_result1_L2_cond}
        t\leq\frac{1}{4\pi W^2}.
    \end{equation}
    Then
    \begin{equation}\label{eq:main_result1_L2}
    \|E_t^L-e^{-iTH}\|_{L_x^2\to L_x^2}\leq 
    8\pi^2tW^2(1+TW^2).
\end{equation}
\end{thm}
Note that \cref{thm:main_result1} immediately implies our main result, \cref{thm:overview-channel}, through the following mixing lemma~\cite[Lemma 3.1]{ChenHuangKuengEtAl2020}.

\begin{lemma}\label{lem:mean-to-diamond}
For every normalized $\psi_0\in L^2(\mathbb{R}^{3N})$ and $\rho_0=|\psi_0\rangle\langle\psi_0|$,
\begin{equation}
\norm{\Phi(\rho_0)-\mathcal U_T(\rho_0)}_1
\le4\norm{(\mathbb E S-U(T))\psi_0}_{L_x^2}.
\end{equation}
The channels also satisfy
\begin{equation}
\norm{\Phi-\mathcal U_T}_\diamond
\le4\norm{\mathbb E S-U(T)}_{L^2\to L^2}.
\end{equation}
\end{lemma}
The proof of this lemma is independent of the Hilbert-space dimension, and its proof extension to the unitaries on $L^2$ considered here is straightforward, which we do not detail here. Combining \cref{thm:main_result1,lem:mean-to-diamond} and use the fact that $W\le c_0N(N-1)/2$ immediately yields
\begin{equation}\label{eq:diamond_channel_thm_to_thm_main}
\begin{aligned}
\norm{\Phi-\mathcal U_T}_\diamond
&\le 32\pi^2tW^2(1+TW^2)\\
&\le \frac{C}{L}\left(c_0^2N^4T+c_0^4N^8T^2\right).
\end{aligned}
\end{equation}
Note that \cref{eq:main_result1_L2_cond} should not be considered as an additional condition, because for the error of \cref{eq:diamond_channel_thm_to_thm_main} to be bounded by the precision $\epsilon \in (0, 1)$, \cref{eq:main_result1_L2_cond} naturally holds. Alternatively, if $t>(4\pi W^2)^{-1}$, the same conclusion follows from the trivial channel bound
\begin{equation}
\norm{\Phi-\mathcal U_T}_\diamond
\le2<8\pi tW^2
\le2\pi c_0^2N^4t.
\end{equation}
We also note that the channel $\Phi$ describes the output averaged over the classical randomness of the algorithm. Its diamond-norm distance from $\mathcal U_T$ controls the worst-case error in measurement statistics, including measurements on an auxiliary system entangled with the input. Specifically, for any joint input state $\rho$ and measurement effect $0\le M\le I$,
\begin{equation}
\left|
\operatorname{Tr}\!\left(
M[(\Phi-\mathcal U_T)\otimes\mathrm{id}](\rho)
\right)
\right|
\le
\frac12\norm{\Phi-\mathcal U_T}_\diamond
\le
2\norm{\mathbb ES-U}_{L^2\to L^2}.
\end{equation}

It therefore suffices to prove \cref{thm:main_result1}, and we will split its proof in two parts. We introduce an intermediate operator $M_t$ defined as follows:
\begin{equation} \label{eq:P_B_M_definition}
    P_t:=\sum_{j<k} \pi_{jk} e^{-itB_{jk}/\pi_{jk}},\quad B_t:=i\frac{P_t-I}{t},\quad  M_t(s):=e^{-is(A+B_t)}.
\end{equation}
Heuristically, this $B_t$ can be regarded as an approximation of the original $B$, since
\begin{align*}
    P_t&=\sum_{j<k}\pi_{jk} e^{-itB_{jk}/\pi_{jk}}\\
    &= \sum_{j<k} \pi_{jk}\left(I-it\frac{B_{jk}}{\pi_{jk}}+\cO(t^2)\right)\\
    &=1-itB+\cO(t^2), 
\end{align*}
where we used 
\[\sum_{j<k} \pi_{jk}=1,\quad \sum_{j<k}B_{jk}=B.\]
However, things are not that simple for unbounded operators, as the above Taylor expansion argument does not hold in the operator norm sense. Moreover, the potential $B_t$ is generally complex-valued, so $A+B_t$ need not be self-adjoint and $M_t(s)$ need not be unitary. Nevertheless, $B_t$ is bounded and $\operatorname{Im}B_t(x)\le0$, so $M_t(s)$ is a strongly continuous contraction semigroup for $s\ge0$.
The crude operator-norm bound $\norm{B_t}_{L^2\to L^2}\le2/t$ deteriorates as $t\to0$. Instead, as is presented below, we use the pointwise bound $|B_t(x)|\le\sum_{j<k}w_{jk}v_{jk}(x)$, which preserves the Coulomb structure uniformly in $t$.

Using $M_t$, we decompose the error into two parts
\[E_t^L-U(T)=\left(M_t(T)-U(T)\right)+\left(E_t^L-M_t(T)\right).\]
In the following two subsections, we prove the estimates
    \begin{equation*}
        \|U(T)-M_t(T)\|_{L_x^2\to L_x^2}\leq \pi tW^2(1+2\pi TW^2), 
    \end{equation*}
\begin{equation*}
    \|E_t^L-M_t(T)\|_{L_x^2\to L_x^2}
    \leq
    2\pi tW^2
    \sqrt{2\pi(1+TW^2)(1+4\pi TW^2)}.
\end{equation*}
Combining them yields \cref{thm:main_result1}.

\subsection{Estimate of $\|U(T)-M_t(T)\|_{L_x^2\to L_x^2}$}
We begin with the first part, $\|U(T)-M_t(T)\|_{L_x^2\to L_x^2}$.
\begin{prop} \label{prop:U_M_error}
    Let $M_t$ be defined by \cref{eq:P_B_M_definition} and $U(T)=e^{-iTH}$. Then
    \begin{equation} \label{eq:U_M_error}
        \|U(T)-M_t(T)\|_{L_x^2\to L_x^2}\leq \pi tW^2(1+2\pi TW^2).
    \end{equation}
\end{prop}
We first prove the following smoothing lemmas.
\begin{lemma} \label{lem:smoothing1}
    For every pair $(j,k)$ and $f\in L^2_x(\bbr^{3N})$, we have
    \[\int_\bbr \|v_{jk}e^{-isA}f\|_{L^2_x}^2 ds\leq \frac{\pi}{2}\|f\|_{L^2_x}^2.\]
\end{lemma}
\begin{proof}
    Recall Kato's smoothing estimate \cite{KatoYajima1989,FrohlichKnowlesSchwarz2009}: for any $g\in L^2_y(\bbr^3)$,
    \begin{equation} \label{eq:smoothing1}
        \int_\bbr\left\|\frac{1}{|y|}e^{is\Delta_y}g\right\|^2_{L^2_y}ds\leq \pi\|g\|_{L^2_y}^2.
    \end{equation}
    
    \noindent We use the orthogonal change of coordinates
    \[
        (x_j,x_k)\longmapsto
        \left(
        \frac{x_j-x_k}{\sqrt2},
        \frac{x_j+x_k}{\sqrt2}
        \right)=:(y,w).
    \]
    Collect all coordinates other than $y$ into
    $z\in\bbr^{3N-3}$ and write
    $\tilde f(y,z):=f(x(y,z))$.
    Then
    \[
        v_{jk}=\frac{1}{\sqrt2|y|},
        \quad
        \Delta_x=\Delta_y+\Delta_z,
        \quad
        \|\tilde f\|_{L^2_{y,z}}=\|f\|_{L_x^2}.
    \]
    Define $v_M(y):=\min\{M,|y|^{-1}\}$. Since $e^{is\Delta_z}$ is unitary and commutes with multiplication by functions of $y$, we have
    \begin{align*}
        \frac12
        \left\|v_Me^{is\Delta_y}e^{is\Delta_z}
        \tilde f\right\|_{L^2_{y,z}}^2
        =\frac12
        \left\|e^{is\Delta_z}
        \bigl(v_M e^{is\Delta_y}\tilde f\bigr)
        \right\|_{L^2_{y,z}}^2
        =\frac12
        \left\|v_M e^{is\Delta_y}\tilde f
        \right\|_{L^2_{y,z}}^2.
    \end{align*}
    Taking $M\to \infty$ gives
    \begin{align*}
        \|v_{jk}e^{-isA}f\|_{L_x^2}^2
        =\frac12
        \left\|\frac{1}{|y|}e^{is\Delta_y}e^{is\Delta_z}
        \tilde f\right\|_{L^2_{y,z}}^2
        =\frac12
        \left\|\frac{1}{|y|}e^{is\Delta_y}\tilde f
        \right\|_{L^2_{y,z}}^2.
    \end{align*}
    Since $\tilde f(\cdot,z)\in L_y^2$ for almost every $z$,
    we can use \cref{eq:smoothing1} to derive
    \begin{align*}
        \int_\bbr\|v_{jk}e^{-isA}f\|_{L_x^2}^2 ds
        &=\frac12\int_{\bbr^{3N-3}}\int_\bbr
        \left\||y|^{-1}e^{is\Delta_y}
        \tilde f(\cdot,z)\right\|_{L_y^2}^2 ds dz\\
        &\leq\frac{\pi}{2}
        \int_{\bbr^{3N-3}}
        \|\tilde f(\cdot,z)\|_{L_y^2}^2 dz
        =\frac{\pi}{2}\|f\|_{L_x^2}^2.
    \end{align*}
\end{proof}
We use \cref{lem:smoothing1} to prove the following smoothing property which will be applied for $U$ and $M_t$.
\begin{lemma} \label{lem:smoothing2}
    Let $Z:\bbr^{3N}\to\mathbb{C}$ be a measurable function satisfying
    \begin{equation}\label{eq:smoothing2_assumption}
        |Z(x)|\leq\sum_{j<k}w_{jk}v_{jk}(x),
        \quad \operatorname{Im}Z(x)\leq0,
        \quad\text{for a.e. }x\in\bbr^{3N}.
    \end{equation}
    Set
    \[
        H_Z:=A+Z,\quad D(H_Z)=H_x^2(\bbr^{3N}).
    \]
    Then $-iH_Z$ generates a strongly continuous contraction
    semigroup $\left\{e^{-isH_Z}\right\}_{s\geq 0}$ on
    $L_x^2(\bbr^{3N})$. Moreover, for every $T\geq0$, every
    $1\leq j<k\leq N$,
    \begin{equation}\label{eq:smoothing2}
        \int_0^T\|v_{jk}e^{-isH_Z}f\|_{L_x^2}^2 ds
        \leq2\pi(1+2\pi TW^2)\|f\|_{L_x^2}^2,\quad \forall\,f\in L_x^2(\mathbb{R}^{3N}).
    \end{equation}
    The same conclusions hold with $H_Z$ replaced by $-A+Z$,
    with the same domain and the same constant in
    \cref{eq:smoothing2}.
\end{lemma}
We will apply \cref{lem:smoothing2} with $Z=B_t$, so that
$M_t(s)=e^{-isH_Z}$ for each fixed $t>0$.
Since $B_t$ is generally complex-valued, $A+B_t$ need not
be self-adjoint.
The assertion that $-iH_Z$, with domain $H_x^2$, generates
a strongly continuous contraction semigroup provides
the domain invariance and differentiability needed
for the subsequent arguments.
In particular, for every $f\in H_x^2$,
\[
    M_t(s)f\in H_x^2,
    \quad
    \frac{d}{ds}M_t(s)f
    =-i(A+B_t)M_t(s)f\in L_x^2.
\]
These properties justify differentiating the evolution
operators and deriving the Duhamel formulas used below. We recall the Lumer-Phillips theorem \cite[Chapter~1, Theorem~4.3]{Pazy2012}, which will be used in the proof of \cref{lem:smoothing2}.
\begin{lemma}[Lumer-Phillips theorem]
    Let $A$ be a linear operator with dense domain $D(A)$ in a Banach space X. If $A$ is dissipative and the range of $\lambda_0I-A$ is $X$ for some $\lambda_0>0$, then $A$ is the infinitesimal generator of a contraction semigroup $\left\{e^{sA}\right\}_{s\geq 0}$.
\end{lemma}

\vspace{.3cm}
\noindent\emph{Proof of \cref{lem:smoothing2}. }
    The case $T=0$ is immediate. If $W=0$, then $Z=0$ almost
    everywhere, so \cref{eq:smoothing2} follows from \cref{lem:smoothing1}.
    Hence, assume $T>0$ and $W>0$. The three-dimensional Hardy inequality states that
    \[
        \left\||y|^{-1}g\right\|_{L_y^2}
        \leq2\|\nabla_y g\|_{L_y^2},
        \quad g\in H_y^1(\bbr^3).
    \]
    Introduce the orthogonal coordinates
\[
    y=\frac{x_j-x_k}{\sqrt2},
    \quad
    w=\frac{x_j+x_k}{\sqrt2},
\]
and let $z\in\bbr^{3N-3}$ collect all coordinates other than $y$. Writing $\widetilde h(y,z)=h(x(y,z))$, we have
\[
    dx=dydz,
    \quad
    v_{jk}(x)=\frac{1}{\sqrt2|y|}.
\]
For $h\in H_x^1$, applying the Hardy inequality in $y$
for almost every fixed $z$ and then integrating in $z$
gives
\begin{align*}
    \|v_{jk}h\|_{L_x^2}^2
    &=
    \frac12
    \int_{\bbr^{3N-3}}\int_{\bbr^3}
    \frac{|\widetilde h(y,z)|^2}{|y|^2}\,dy\,dz\\
    &\leq
    2\int_{\bbr^{3N-3}}\int_{\bbr^3}
    |\nabla_y\widetilde h(y,z)|^2\,dy\,dz
    =2\|\nabla_y\widetilde h\|_{L_{y,z}^2}^2.
\end{align*}
Moreover, the chain rule and the change of variables yield
\[
    \|\nabla_y\widetilde h\|_{L_{y,z}^2}^2
    =\frac12
    \|\nabla_{x_j}h-\nabla_{x_k}h\|_{L_x^2}^2
    \leq
    \|\nabla_{x_j}h\|_{L_x^2}^2
    +\|\nabla_{x_k}h\|_{L_x^2}^2
    \leq\|\nabla_xh\|_{L_x^2}^2.
\]
Taking square roots, we conclude that
\begin{equation}    \label{eq:hardy1}
    \|v_{jk}h\|_{L_x^2}
    \leq\sqrt2\|\nabla_xh\|_{L_x^2},
    \quad h\in H_x^1.
\end{equation}
    Thus, for every $h\in H_x^2$,
    \begin{equation}    \label{eq:hardy2}
        \|Zh\|_{L_x^2}\leq \sum_{j<k}w_{jk}\|v_{jk}h\|_{L_x^2}\leq\sqrt{2}W\|\nabla_x h\|_{L_x^2}.
    \end{equation}
    Since $\|\nabla_x h\|^2_{L_x^2}= \la h, Ah\ra_{L_x^2}$, we have
    \begin{equation} \label{eq:hardy}
        \|Zh\|_{L_x^2}\leq\sqrt{2}W
        \|Ah\|_{L_x^2}^{1/2}\|h\|_{L_x^2}^{1/2}\leq\frac{1}{2}\|Ah\|_{L_x^2}
        +W^2\|h\|_{L_x^2},
    \end{equation}
    Then, we use $H_Z=A+Z$ to get
    \[\|Ah\|_{L_x^2}\leq \|H_Zh\|_{L_x^2}+\frac{1}{2}\|Ah\|_{L_x^2}+W^2 \|h\|_{L_x^2}.\]
    Therefore, we have
    \begin{equation}    \label{eq:Ah_norm}
        \|Ah\|_{L_x^2}\leq 2\|H_Zh\|_{L_x^2}+2W^2 \|h\|_{L_x^2}.
    \end{equation}
    Now, we first justify the existence and contractivity of $U_Z$. For every $h\in H_x^2$, we have
    \begin{align*}
        \la h, -iH_Z h\ra &= -i\la h, Ah\ra -i\la h,Zh\ra\\
        &=-i\|\nabla_x h\|_{L_x^2}^2 -i \la h, Zh \ra,
    \end{align*}
    which leads to
    \[
        \operatorname{Re}\langle h,-iH_Zh\rangle_{L_x^2}
        =\int_{\bbr^{3N}}\operatorname{Im}Z(x)|h(x)|^2 dx
        \leq0.
    \]
    Therefore, $-iH_Z$ is dissipative: for any $\lambda>0$ and $h\in H^2$, we have
    \begin{equation} \label{eq:dissipative}
        \|(\lambda I-(-iH_Z))h\|_{L_x^2}^2=\lambda^2\|h\|_{L_x^2}^2-2 \lambda\operatorname{Re} \la h, -iH_Z h\ra_{L_x^2}+\|-iH_Z h\|_{L_x^2}^2\geq \lambda^2 \|h\|_{L_x^2}^2.
    \end{equation}
    For $\lambda>0$ and $f\in L_x^2$, define $g$ by
\[
    \widehat g(\xi)
    :=\frac{\widehat f(\xi)}{\lambda+i|\xi|^2}.
\]
Then $g\in L^2_x$ since
\[
    \|g\|_{L_x^2}^2
    =
    \int_{\bbr^{3N}}
    \frac{|\widehat f(\xi)|^2}{\lambda^2+|\xi|^4}\,d\xi
    \leq\lambda^{-2}\|f\|_{L_x^2}^2.
\]
Moreover,
\[
    \int_{\bbr^{3N}}|\xi|^4|\widehat g(\xi)|^2\,d\xi
    =
    \int_{\bbr^{3N}}
    \frac{|\xi|^4}{\lambda^2+|\xi|^4}
    |\widehat f(\xi)|^2\,d\xi
    \leq\|f\|_{L_x^2}^2,
\]
which implies $g\in H_x^2$ with $\|Ag\|_{L_x^2}\leq\|f\|_{L_x^2}$.
The Fourier representation shows that $g$ is the unique
solution of $(\lambda+iA)g=f$. Thus $g=(\lambda+iA)^{-1}f\in H_x^2$ with
\[
    \|(\lambda+iA)^{-1}\|_{L_x^2\to L_x^2}
    \leq\lambda^{-1},
    \quad
    \|A(\lambda+iA)^{-1}\|_{L_x^2\to L_x^2}
    \leq1.
\]
We may apply
\cref{eq:hardy} to obtain
\begin{align*}
    & \|Z(\lambda+iA)^{-1}f\|_{L_x^2}
    =\|Zg\|_{L_x^2}\\
    \leq & \ \frac12\|Ag\|_{L_x^2}
        +W^2\|g\|_{L_x^2}
    \leq
    \left(\frac12+\frac{W^2}{\lambda}\right)
    \|f\|_{L_x^2}.
\end{align*}
Since $f\in L_x^2$ was arbitrary, this gives
\[
    \|Z(\lambda+iA)^{-1}\|_{L_x^2\to L_x^2}
    \leq\frac12+\frac{W^2}{\lambda}<1,
    \quad \lambda>2W^2.
\]
    Hence $I+iZ(\lambda+iA)^{-1}$ is invertible,
    and the factorization
    \[
        \lambda+iH_Z=(I+iZ(\lambda+iA)^{-1})(\lambda+iA)
    \]
    shows that $\lambda+iH_Z:H_x^2\to L_x^2$ is surjective.
    Since $H_x^2$ is dense in $L_x^2$, the dissipativity established in \cref{eq:dissipative} and the surjectivity proved above imply, by the Lumer-Phillips
    theorem, $-iH_Z$ generates a strongly continuous contraction semigroup $\left\{e^{-isH_Z}\right\}_{s\geq 0}$ on $L_x^2$. Now, fix $f\in D(H_Z)=H_x^2$ and let $u(s):=e^{-isH_Z}f\in H_x^2$. Observe that
\cref{eq:hardy} yields
\[
    \|H_Zh\|_{L_x^2}
    \leq \frac32\|Ah\|_{L_x^2}
        +W^2\|h\|_{L_x^2},
    \qquad h\in H_x^2.
\]
Together with \cref{eq:Ah_norm}, this gives
\[
    \|h\|_{L_x^2}+\|H_Zh\|_{L_x^2}
    \asymp
    \|h\|_{L_x^2}+\|Ah\|_{L_x^2}
    \asymp
    \|h\|_{H_x^2}.
\]
In other words, the graph norm of $H_Z$ is equivalent to
the $H_x^2$ norm.
In particular, for every $r\geq0$,
\[
    \|u(s)-u(r)\|_{H_x^2}
    \leq C\left(
        \|e^{-isH_Z}f-e^{-irH_Z}f\|_{L_x^2}
        +
        \|e^{-isH_Z}H_Zf-e^{-irH_Z}H_Zf\|_{L_x^2}
    \right)
    \longrightarrow0
    \quad\text{as }s\to r.
\]
We have therefore established
\begin{equation}\label{eq:semigroup_regularity}
    u\in C([0,\infty);H_x^2)
    \cap C^1([0,\infty);L_x^2),
    \qquad
    u'(s)=-iH_Zu(s).
\end{equation}
Finally, applying \cref{eq:hardy1} and \cref{eq:hardy2} to $h=u(s)-u(r)$, we conclude that
\[
    Zu,\;v_{jk}u\in C([0,\infty);L_x^2).
\]For fixed $s\geq0$, differention gives
    \[
        \frac{d}{dr}\bigl(e^{-i(s-r)A}u(r)\bigr)
        =ie^{-i(s-r)A}Au(r)-ie^{-i(s-r)A}H_Z u(r)=-i e^{-i(s-r)A}Zu(r).
    \]
    Integrating over $0\leq r\leq s$, we obtain
    \begin{equation}\label{eq:duhamel}
        u(s)=e^{-isA}f
        -i\int_0^s e^{-i(s-r)A}Zu(r)dr.
    \end{equation}
    Now, set 
    \[\delta:=\frac{1}{2\pi W^2}.\] 
    We first establish an estimate on the interval $[0,\delta]$. Fix $M>0$ and set $v_{jk,M}:=\min\{M,v_{jk}\}$.
    Applying this bounded multiplier to \cref{eq:duhamel}, we get
    \begin{align*}
    \left(\int_0^\delta
    \|v_{jk,M}u(s)\|_{L_x^2}^2\,ds\right)^{1/2}
    &\leq
    \left(\int_0^\delta
    \|v_{jk,M}e^{-isA}f\|_{L_x^2}^2\,ds\right)^{1/2}+
    \left(\int_0^\delta
    \left\|\int_0^\delta
    \mathbf{1}_{\{r\leq s\}}
    v_{jk,M}e^{-i(s-r)A}Zu(r)\,dr
    \right\|_{L_x^2}^2\,ds\right)^{1/2}\\
    &\leq
    \left(\int_0^\delta
    \|v_{jk,M}e^{-isA}f\|_{L_x^2}^2\,ds\right)^{1/2}+
    \int_0^\delta
    \left(\int_0^\delta
    \mathbf{1}_{\{r\leq s\}}
    \|v_{jk,M}e^{-i(s-r)A}Zu(r)\|_{L_x^2}^2\,ds
    \right)^{1/2}dr\\
    &=
    \left(\int_0^\delta
    \|v_{jk,M}e^{-isA}f\|_{L_x^2}^2\,ds\right)^{1/2}+
    \int_0^\delta
    \left(\int_r^\delta
    \|v_{jk,M}e^{-i(s-r)A}Zu(r)\|_{L_x^2}^2\,ds
    \right)^{1/2}dr\\
    &\leq
    \sqrt{\frac{\pi}{2}}\|f\|_{L_x^2}
    +\sqrt{\frac{\pi}{2}}
    \int_0^\delta\|Zu(r)\|_{L_x^2}\,dr,
\end{align*}
where we used 
\[\left\|\int_0^\delta F(r,s,x)\,dr\right\|_{L_s^2((0,\delta);L_x^2)}
\leq
\int_0^\delta\|F(r,s,x)\|_{L_s^2((0,\delta);L_x^2)}dr\]
in the second inequality and \cref{lem:smoothing1} in the last inequality. By monotone convergence theorem, taking $M\to\infty$ gives
    \begin{equation}\label{A-8}
        \left(\int_0^\delta\|v_{jk}u(s)\|_{L_x^2}^2ds\right)^{1/2}
        \leq\sqrt{\frac{\pi}{2}}\|f\|_{L_x^2}
        +\sqrt{\frac{\pi}{2}}\int_0^\delta\|Zu(r)\|_{L_x^2}dr.
    \end{equation}
    Define
    \[
        X_\delta:=\max_{1\leq j<k\leq N}
        \left(\int_0^\delta\|v_{jk}u(s)\|_{L_x^2}^2ds\right)^{1/2},
    \]
    which is finite since $v_{jk}u\in C([0,\infty);L_x^2)$.
    By \cref{eq:smoothing2_assumption}, we have
    \begin{equation*}
        \|Zu(r)\|_{L_x^2}\leq \left\|\sum_{j<k}w_{jk}v_{jk}|u(r)|\right\|_{L_x^2}\leq \sum_{j<k}w_{jk}\|v_{jk}u(r)\|_{L_x^2},
    \end{equation*}
    and Cauchy-Schwartz inequality gives
    \begin{equation*}
        \int_0^\delta\|Zu(r)\|_{L_x^2}dr
        \leq\sum_{j<k}w_{jk}
        \int_0^\delta\|v_{jk}u(r)\|_{L_x^2}dr\leq W\sqrt{\delta} X_\delta.
    \end{equation*}
    Taking the maximum in \cref{A-8}, we therefore obtain
    \[
        X_\delta\leq\sqrt{\frac{\pi}{2}}\|f\|_{L_x^2}
        +\sqrt{\frac{\pi}{2}}W\sqrt{\delta}X_\delta=\sqrt{\frac{\pi}{2}}\|f\|_{L_x^2}
        +\frac{1}{2}X_\delta.
    \]
    Therefore, we have
    \begin{equation}\label{eq:smoothing2_subinterval}
        \int_0^\delta\|v_{jk}e^{-isH_Z}f\|_{L_x^2}^2ds= \int_0^\delta \|v_{jk} u(s)\|_{L_x^2}^2 ds
        \leq X_\delta^2\leq2\pi\|f\|_{L_x^2}^2.
    \end{equation}
    Partition $[0,T]$ into $m:=\lceil T/\delta\rceil$ consecutive
    intervals of length at most $\delta$. Applying \cref{eq:smoothing2_subinterval} to each summand, we obtain
    \begin{align*}
    \int_0^T\|v_{jk}e^{-isH_Z}f\|_{L_x^2}^2ds
    &\leq
    \sum_{\ell=0}^{m-1}
    \int_{\ell\delta}^{(\ell+1)\delta}
    \|v_{jk}e^{-isH_Z}f\|_{L_x^2}^2ds\\
    &=
    \sum_{\ell=0}^{m-1}
    \int_0^\delta
    \|v_{jk}e^{-isH_Z}e^{-i\ell \delta H_Z}f\|_{L_x^2}^2 ds\\
    &\leq
    2\pi\sum_{\ell=0}^{m-1}
    \|e^{-i\ell \delta H_Z}f\|_{L_x^2}^2.
\end{align*}
Since $\{e^{-isH_Z}\}_{s\geq 0}$ is contractive, we get
\begin{equation}    \label{eq:smoothing2_H2}
    \int_0^T\|v_{jk}e^{-isH_Z}f\|_{L_x^2}^2ds\leq
    2\pi m\|f\|_{L_x^2}^2\leq
    2\pi(1+2\pi TW^2)\|f\|_{L_x^2}^2,
\end{equation}
where we used $m-1<\frac{T}{\delta}$.
    Now let $f\in L_x^2$ be arbitrary and choose $f_n\in H_x^2$
    with $f_n\to f$ in $L_x^2$. For fixed $M>0$, contractivity of $e^{-isH_Z}$ gives
    \[
        \int_0^T
        \|v_{jk,M}e^{-isH_Z}(f_n-f)\|_{L_x^2}^2ds\leq M^2\int_0^T \|e^{-isH_Z}(f_n-f)\|_{L_x^2}^2ds
        \leq TM^2\|f_n-f\|_{L_x^2}^2\to 0.
    \]
    Since
    \begin{align*}
    &\left(\int_0^T\|v_{jk,M}e^{-isH_Z} f\|_{L_x^2}^2ds\right)^{1/2}\\
       \leq & \left(\int_0^T\|v_{jk,M}e^{-isH_Z}(f_n-f)\|_{L_x^2}^2ds\right)^{1/2}+\left(\int_0^T \|v_{jk,M}e^{-isH_Z}f_n\|_{L_x^2}^2ds\right)^{1/2},
    \end{align*}
    we apply \cref{eq:smoothing2_H2} on $f_n$ and take $n\to \infty$ to obtain
    \begin{equation*}
        \int_0^T\|v_{jk,M}e^{-isH_Z}f\|_{L_x^2}^2ds\leq 2\pi(1+2\pi TW^2)\|f\|_{L_x^2}^2
    \end{equation*}
    Letting $M\to\infty$ proves \cref{eq:smoothing2} for every $f\in L_x^2$. Finally, observe that $-\overline Z$ also satisfies
\cref{eq:smoothing2_assumption}, since
\[
    |-\overline{Z}(x)|=|Z(x)|
    \leq\sum_{j<k}w_{jk}v_{jk}(x),
    \quad 
    \operatorname{Im}(-\overline{Z}(x))
    =\operatorname{Im}Z(x)\leq0.
\]
Therefore, $-i(A-\overline Z)$ generates
a strongly continuous contraction semigroup on $L_x^2$.
Consequently,
\begin{align*}
    \int_0^T
    \|v_{jk}e^{-is(-A+Z)}f\|_{L_x^2}^2\,ds
    &=
    \int_0^T
    \|v_{jk}e^{-is(A-\overline Z)}\overline f\|_{L_x^2}^2\,ds\\
    &\leq
    2\pi(1+2\pi TW^2)\|\overline f\|_{L_x^2}^2\\
    &=
    2\pi(1+2\pi TW^2)\|f\|_{L_x^2}^2.
\end{align*}
\hfill\qed

\vspace{.3cm}
Using \cref{lem:smoothing1,lem:smoothing2}, we derive the following estimate of $\|U(T)-M_t(T)\|_{L_x^2\to L_x^2}$.

\vspace{3mm}
\noindent\emph{Proof of \cref{prop:U_M_error}.}
Recall \cref{eq:P_B_M_definition}:
\begin{equation*}
    P_t:=\sum_{j<k} \pi_{jk} e^{-itB_{jk}/\pi_{jk}},\quad B_t:=i\frac{P_t-I}{t},\quad  M_t(s):=e^{-is(A+B_t)}.
\end{equation*}
For every $a\in\bbr$, we have
\begin{equation}    \label{eq:exp_ineq}
    |e^{-ia}-1|\leq |a|,
    \quad
    |e^{-ia}-1+ia|\leq a^2/2.
\end{equation}
Since $\sum_{j<k}\pi_{jk}=1$, we can write $B_t$ as 
\[B_t=\frac{i}{t}\sum_{j<k} \pi_{jk}(e^{-itB_{jk}/\pi_{jk}}-1).\]
Using the first inequality of \cref{eq:exp_ineq}, we obtain
\begin{equation}\label{eq:Bt_bound}
\begin{aligned}
    |B_t(x)|
    &\leq\frac{1}{t}\sum_{j<k}\pi_{jk}
    \left|e^{-itB_{jk}(x)/\pi_{jk}}-1\right|\\
    &\leq \frac{1}{t}\sum_{j<k}\pi_{jk}\cdot t\frac{|B_{jk}|}{\pi_{jk}}
    =\sum_{j<k}w_{jk}v_{jk}(x).
\end{aligned}
\end{equation}
Moreover, since each $B_{jk}$ is real-valued,
\[
    \operatorname{Im}B_t(x)
    =\frac{1}{t}\sum_{j<k}\pi_{jk}
    \left(\cos\left(\frac{tB_{jk}(x)}{\pi_{jk}}\right)-1\right)
    \leq0.
\]
Thus $Z=B_t$ satisfies \cref{eq:smoothing2_assumption},
and \cref{lem:smoothing2} shows that $M_t(s):=e^{-is(A+B_t)}$ is a strongly
continuous contraction semigroup with generator domain $H_x^2$. Applying \cref{lem:smoothing2} with $Z=B_t$ and $Z=B$, respectively,
we obtain, for every $f\in L_x^2$,
\begin{align}\label{eq:smoothing_U_M_t}
\begin{aligned}
    \int_0^T\|v_{jk}M_t(s)f\|_{L_x^2}^2ds
    &\leq 2\pi(1+2\pi TW^2)\|f\|_{L_x^2}^2,\\
    \int_0^T\|v_{jk}U(s)f\|_{L_x^2}^2ds
    &\leq 2\pi(1+2\pi TW^2)\|f\|_{L_x^2}^2.
\end{aligned}
\end{align}
In particular, using $U(s-T)=U(s)U(-T)$ and unitarity, we have
\begin{align}\label{eq:smoothing_shifted}
\begin{aligned}
    \int_0^T\|v_{jk}U(s-T)f\|_{L_x^2}^2ds
    &=\int_0^T\|v_{jk}U(s)U(-T)f\|_{L_x^2}^2ds\\
    &\leq 2\pi(1+2\pi TW^2)\|U(-T)f\|_{L_x^2}^2\\
    &=2\pi(1+2\pi TW^2)\|f\|_{L_x^2}^2.
\end{aligned}
\end{align}

Now, let $\psi_0\in H_x^2$ and set $u(s):=M_t(s)\psi_0$.
The semigroup properties and the equivalence of the graph norm
of $A+B_t$ with the $H_x^2$ norm imply that
\[
    u\in C([0,T];H_x^2)\cap C^1([0,T];L_x^2),
    \quad u'(s)=-i(A+B_t)u(s).
\]
Moreover, \cref{eq:Bt_bound} gives
\[
    |B_t(x)-B(x)|\leq2\sum_{j<k}w_{jk}v_{jk}(x).
\]
Together with Hardy's inequality, this shows that
$(B_t-B)M_t(s)\psi_0$ is continuous in $L_x^2$.
Differentiating in $L_x^2$, we obtain
\begin{align*}
    \frac{d}{ds}\bigl(U(T-s)M_t(s)\psi_0\bigr)
    &=iU(T-s)(A+B)M_t(s)\psi_0-iU(T-s)(A+B_t)M_t(s)\psi_0\\
    &=-iU(T-s)(B_t-B)M_t(s)\psi_0,
\end{align*}
which gives
\begin{align*}
    \frac{d}{ds}\langle\phi_0, U(T-s)M_t(s)\psi_0\rangle = \langle\phi_0, -iU(T-s)(B_t-B)M_t(s)\psi_0\rangle.
\end{align*}
Integrating over $[0,T]$ yields
\begin{equation}\label{eq:duhamel_comparison}
    \langle \phi_0, (M_t(T)-U(T))\psi_0\rangle_{L_x^2}
    =-i\int_0^T\la \phi_0, U(T-s)(B_t-B)M_t(s)\psi_0\ra_{L_x^2} ds.
\end{equation}
Since $B=\sum_{j<k}B_{jk}$ and $\pi_{jk}=w_{jk}/W$, the second inequality in \cref{eq:exp_ineq} yields
\begin{equation}\label{eq:Bt_error}
\begin{aligned}
    |B_t(x)-B(x)|
    &\leq\frac{1}{t}\sum_{j<k}\pi_{jk}
    \left|e^{-itB_{jk}(x)/\pi_{jk}}-1
    +\frac{itB_{jk}}{\pi_{jk}}\right|\\
    &\leq\frac{t}{2}\sum_{j<k}
    \frac{|B_{jk}|^2}{\pi_{jk}}
    =\frac{tW}{2}\sum_{j<k}w_{jk}v_{jk}^2.
\end{aligned}
\end{equation}
Using $U(T-s)^*=U(s-T)$
and combining \cref{eq:duhamel_comparison} and \cref{eq:Bt_error}, we find
\begin{align*}
    &\left|\langle\phi_0,(M_t(T)-U(T))\psi_0\rangle_{L_x^2}\right|\\
   \leq & \int_0^T
    \left|\langle U(s-T)\phi_0,(B_t-B)M_t(s)\psi_0
    \rangle_{L_x^2}\right|ds\\
    \leq & \frac{tW}{2}\sum_{j<k}w_{jk}
    \int_0^T\int_{\bbr^{3N}}v_{jk}(x)^2
    |(U(s-T)\phi_0)(x)||(M_t(s)\psi_0)(x)|dxds.
\end{align*}
For each pair $(j,k)$, we use Cauchy-Schwarz inequality, \cref{eq:smoothing_U_M_t}, and
\cref{eq:smoothing_shifted} to get
\begin{align*}
    &\int_0^T\int_{\bbr^{3N}}v_{jk}(x)^2
    |(U(s-T)\phi_0)(x)||(M_t(s)\psi_0)(x)|dxds\\
    \leq & \int_0^T
    \|v_{jk}U(s-T)\phi_0\|_{L_x^2}
    \|v_{jk}M_t(s)\psi_0\|_{L_x^2}ds\\
    \leq &
    \left(\int_0^T\|v_{jk}U(s-T)\phi_0\|_{L_x^2}^2ds\right)^{1/2} \left(\int_0^T\|v_{jk}M_t(s)\psi_0\|_{L_x^2}^2ds\right)^{1/2}\\
    \leq & 2\pi(1+2\pi TW^2)\|\phi_0\|_{L_x^2}\|\psi_0\|_{L_x^2}.
\end{align*}
Summing over all pairs and using $\sum_{j<k}w_{jk}=W$, we obtain
\[
    \left|\langle\phi_0,(M_t(T)-U(T))\psi_0\rangle_{L_x^2}\right| \leq \pi tW^2(1+2\pi TW^2)\|\phi_0\|_{L_x^2}\|\psi_0\|_{L_x^2},\quad \phi_0, \psi_0\in H^2_x.
\]
We extend this estimate to arbitrary
$\phi_0,\psi_0\in L_x^2$.
Since $H_x^2$ is dense in $L_x^2$, there exist sequences
$\{\phi_n\},\{\psi_n\}\subset H_x^2$ such that
\[
    \phi_n\to \phi_0,
    \quad
    \psi_n\to \psi_0
    \quad\text{in }L_x^2.
\]
Moreover, contractivity of $M_t(T)$ and unitarity of $U(T)$
imply
\[
    \|M_t(T)-U(T)\|_{L_x^2\to L_x^2}
    \leq
    \|M_t(T)\|_{L_x^2\to L_x^2}
    +\|U(T)\|_{L_x^2\to L_x^2}
    \leq2.
\]
Consequently,
\begin{align*}
    &\left|
        \langle\phi_n,(M_t(T)-U(T))\psi_n\rangle_{L_x^2}
        -\langle\phi_0,(M_t(T)-U(T))\psi_0\rangle_{L_x^2}
    \right|\\
    \leq &
    \left|
        \langle\phi_n-\phi_0,
        (M_t(T)-U(T))\psi_n\rangle_{L_x^2}
    \right|+
    \left|
        \langle\phi_0,
        (M_t(T)-U(T))(\psi_n-\psi_0)\rangle_{L_x^2}
    \right|\\
   \leq &
    2\|\phi_n-\phi_0\|_{L_x^2}\|\psi_n\|_{L_x^2}
    +2\|\phi_0\|_{L_x^2}\|\psi_n-\psi_0\|_{L_x^2}
    \to 0.
\end{align*}
Here $\{\|\psi_n\|_{L_x^2}\}$ is bounded because
$\psi_n\to\psi_0$ in $L_x^2$. Applying the preceding estimate to $\phi_n,\psi_n$
and passing to the limit, we obtain
\begin{align*}
    \left|
        \langle\phi_0,(M_t(T)-U(T))\psi_0\rangle_{L_x^2}
    \right|&=
    \lim_{n\to\infty}
    \left|
        \langle\phi_n,(M_t(T)-U(T))\psi_n\rangle_{L_x^2}
    \right|\\
    &\leq
    \pi tW^2(1+2\pi TW^2)
    \lim_{n\to\infty}
    \|\phi_n\|_{L_x^2}\|\psi_n\|_{L_x^2}\\
    &=
    \pi tW^2(1+2\pi TW^2)
    \|\phi_0\|_{L_x^2}\|\psi_0\|_{L_x^2}.
\end{align*}
For every $\psi_0\in L_x^2$, we have
\begin{align*}
    \|(M_t(T)-U(T))\psi_0\|_{L_x^2}
    &=
    \sup_{\substack{\phi_0\in L_x^2\\
                    \|\phi_0\|_{L_x^2}=1}}
    \left|
        \langle\phi_0,(M_t(T)-U(T))\psi_0\rangle_{L_x^2}
    \right|\\
    &\leq
    \pi tW^2(1+2\pi TW^2)\|\psi_0\|_{L_x^2}.
\end{align*}
Taking the supremum over all unit vectors $\psi_0\in L_x^2$,
we conclude that
\[
    \|M_t(T)-U(T)\|_{L_x^2\to L_x^2}
    \leq \pi tW^2(1+2\pi TW^2).
\]
\hfill \qed

\subsection{Estimate of $\|E_t^L-M_t(T)\|_{L_x^2\to L_x^2}$}
Next, we present the estimate of the second error term, $\|E_t^L-M_t(T)\|_{L_x^2\to L_x^2}$.
\begin{prop} \label{prop:E_M_error}
Let $E_t$ be defined by \cref{eq:Et_def} and $M_t$ be defined by \cref{eq:P_B_M_definition}. Then
\begin{equation*}\label{eq:discrete_continuous_bound}
    \|E_t^L-M_t(T)\|_{L_x^2\to L_x^2}
    \leq
    2\pi tW^2
    \sqrt{2\pi(1+TW^2)(1+4\pi TW^2)}.
\end{equation*}
\end{prop}
We begin by recalling \cref{eq:Et_def}:
\begin{equation*}
    E_tf=\sum_{j<k} \pi_{jk}\int_0^1 e^{-i\theta t A}e^{-itB_{jk}/\pi_{jk}}e^{-i(1-\theta)tA}fd\theta,\quad f\in L_x^2.
\end{equation*}
By unitarity, we have
\begin{align*}
    \|E_t f\|_{L_x^2}&\leq \sum_{j<k} \pi_{jk}\int_0^1 \left\|e^{-i\theta tA}e^{-itB_{jk}/\pi_{jk}}e^{-i(1-\theta)tA}f\right\|_{L_x^2}d\theta\\
    &=\sum_{j<k}\pi_{jk}\|f\|_{L_x^2}=\|f\|_{L_x^2},
\end{align*}
which gives
\begin{equation} \label{eq:Et_norm}
    \|E_t\|_{L_x^2\to L_x^2}\leq 1.
\end{equation}
Using the definition of $P_t$ and $P_t=I-itB_t$, we can also write
\begin{align} \label{eq:Et_representation}
    \begin{aligned}
    E_tf&=\int_0^1 e^{-i\theta t A}\left(\sum_{j<k}\pi_{jk} e^{-itB_{jk}/\pi_{jk}}\right)e^{-i(1-\theta)tA}fd\theta\\
    &=\int_0^1 e^{-i\theta t A}P_te^{-i(1-\theta)tA}fd\theta\\
    &=e^{-itA}f-it\int_0^1 e^{-i\theta t A}B_te^{-i(1-\theta)tA}fd\theta.
    \end{aligned}
\end{align}
For $\psi_0\in L_x^2$, define
\[\psi_\ell:=E_t^\ell\psi_0,\quad \ell=0,\ldots,L,
\]
and a piecewise interpolation $\tilde{\psi}$ by
\begin{equation}    \label{eq:psitilde_def}
    \tilde{\psi}(s):= e^{-i(s-\ell t)A}\psi_\ell,
    \quad \ell t\leq s<(\ell+1)t, \quad \ell=0,\ldots,L-1,
\end{equation}
and set $\tilde{\psi}(T):=\psi_L$.
Note that $\tilde{\psi}(\ell t)=\psi_\ell$ at every grid point. Also, unitarity of $e^{-i(s-\ell t)A}$ and \cref{eq:Et_norm} gives
\[\|\tilde{\psi}(s)\|_{L_x^2}=\|\psi_\ell\|_{L_x^2}
    \leq\|\psi_0\|_{L_x^2},
    \quad 0\leq s\leq T.
\]
The following lemma provides an integral representation of the piecewise interpolation $\tilde{\psi}$.
\begin{lemma} \label{lem:tildepsi_representation}
Let $\tilde{\psi}$ be defined by \cref{eq:psitilde_def}. Then
    \[
    \tilde{\psi}(s)
    =
    e^{-isA}\psi_0
    -i\int_0^{t\left\lfloor s/t\right\rfloor}
    e^{-i(s-r)A}B_t\tilde{\psi}(r) dr,
    \quad 0\leq s\leq T.
\]
\end{lemma}
\begin{proof}
Since $\psi_{\ell+1}=E_t\psi_\ell$, we use \cref{eq:Et_representation} and change of variables $r=(\ell+1-\theta)t$ to obtain
\begin{align*}
    \psi_{\ell+1} &=e^{-itA}\psi_\ell
    -it\int_0^1 e^{-i\theta tA}B_t e^{-i(1-\theta)tA}\psi_\ell d\theta\\
    &=e^{-itA}\psi_\ell-i\int_{\ell t}^{(\ell+1)t} e^{-i((\ell+1)t-r)A}B_t e^{-i(r-\ell t)A}\psi_\ell dr\\
    &=e^{-itA}\psi_\ell-i\int_{\ell t}^{(\ell+1)t} e^{-i((\ell+1)t-r)A}B_t \tilde{\psi}(r) dr.
\end{align*}
Thus, iterating this recurrence, we obtain the integral representation at the grid points:
\begin{align*}
    \psi_\ell
    &=e^{-i\ell tA}\psi_0
    -i\sum_{j=0}^{\ell-1}e^{-i(\ell-j-1)tA}
    \int_{jt}^{(j+1)t}
    e^{-i((j+1)t-r)A}B_t\tilde{\psi}(r)\,dr\\
    &=e^{-i\ell tA}\psi_0
    -i\sum_{j=0}^{\ell-1}\int_{jt}^{(j+1)t}
    e^{-i(\ell t-r)A}B_t\tilde{\psi}(r)\,dr\\
    &=e^{-i\ell tA}\psi_0
    -i\int_0^{\ell t}
    e^{-i(\ell t-r)A}B_t\tilde{\psi}(r)\,dr,
    \quad \ell=0,\ldots,L.
\end{align*}
Now, for $\ell t\leq s<(\ell+1)t$, we have
\begin{align*}
    \tilde{\psi}(s)
    &=
    e^{-i(s-\ell t)A}\psi_\ell\\
    &=
    e^{-isA}\psi_0
    -i\int_0^{\ell t}
    e^{-i(s-r)A}B_t\tilde{\psi}(r) dr\\
    &=
    e^{-isA}\psi_0
    -i\int_0^{t\left\lfloor s/t\right\rfloor}
    e^{-i(s-r)A}B_t\tilde{\psi}(r) dr.
\end{align*}
The same identity holds at $s=T$, since
$t\left\lfloor\frac{T}{t}\right\rfloor=Lt=T$ and $\tilde{\psi}(T)=\psi_L$.
\end{proof}
To control the integral term in \cref{lem:tildepsi_representation}, we establish the following smoothing estimate for $\tilde{\psi}$.
\begin{lemma} \label{lem:psitilde_smoothing}
    Let $B_t$ be defined by \cref{eq:P_B_M_definition} and $\tilde{\psi}$ be defined by \cref{eq:psitilde_def}. Then for all $t>0$,
    \begin{equation*}
    \|B_t\tilde{\psi}\|_{L_s^2(0,T;L_x^2)}\leq
    W\sqrt{2\pi\left(1+4\pi TW^2\right)}
     \|\psi_0\|_{L_x^2}.
\end{equation*}
\end{lemma}
\noindent Indeed, by \cref{eq:Bt_bound}, we have
\begin{align*}
    \|B_t\tilde{\psi}\|_{L_s^2(0,T;L_x^2)}
    &\leq
    \sum_{j<k}
    w_{jk}
    \|v_{jk}\tilde{\psi}\|_{L_s^2(0,T;L_x^2)}.
\end{align*}
Then, \cref{lem:smoothing1} and \cref{eq:Et_norm} give
\begin{align}   \label{eq:vtildepsi_bound}
\begin{aligned}
    \int_0^T
    \|v_{jk}\tilde{\psi}(s)\|_{L_x^2}^2 ds
    &=
    \sum_{n=0}^{L-1}
    \int_0^t
    \|v_{jk}e^{-isA}\psi_n\|_{L_x^2}^2 ds\\
    &\leq
    \frac{\pi}{2}
    \sum_{n=0}^{L-1}\|\psi_n\|_{L_x^2}^2
    \leq
    \frac{\pi L}{2}\|\psi_0\|_{L_x^2}^2
    <\infty.
\end{aligned}
\end{align}
Although this yields a bound for $\|B_t\tilde{\psi}\|_{L_s^2(0,T;L_x^2)}$, the resulting bound depends on the number of time steps $L$. \cref{lem:psitilde_smoothing} provides a bound that is uniform in $L$.

\vspace{.3cm}
\noindent\emph{Proof of \cref{lem:psitilde_smoothing}.} If $t>1/(4\pi W^2)$, then
\cref{eq:Bt_bound} and \cref{eq:vtildepsi_bound} give
\begin{align*}
    \|B_t\tilde{\psi}\|_{L_s^2(0,T;L_x^2)}
    &\leq
    W\sqrt{\frac{\pi L}{2}}\|\psi_0\|_{L_x^2}
    =
    W\sqrt{\frac{\pi T}{2t}}\|\psi_0\|_{L_x^2}\\
    &\leq
    W\sqrt{2\pi(1+4\pi TW^2)}\|\psi_0\|_{L_x^2}.
\end{align*}
It therefore remains to consider
\[
    0<t\leq\frac{1}{4\pi W^2}.
\]
Now, fix integers $0\leq n<L$ and $1\leq\ell\leq L-n$.
We first establish a local estimate on $[nt,(n+\ell)t]$:
\[
    \int_{nt}^{(n+\ell)t}
    \|v_{jk}\tilde{\psi}(s)\|_{L_x^2}^2\,ds
    =
    \int_0^{\ell t}
    \|v_{jk}\tilde{\psi}(nt+s)\|_{L_x^2}^2\,ds.
\]
Note that the function $s\mapsto\tilde{\psi}(nt+s)$ is the piecewise interpolation starting from $\psi_n$, with the same time step $t$ and final time $(L-n)t$. Moreover, 
by \cref{eq:vtildepsi_bound},
\begin{equation} \label{eq:maxvtildepsi_finite}
\begin{aligned}
    \max_{j<k}
    \|v_{jk}\tilde{\psi}(nt+s)\|_{L_s^2(0,\ell t;L_x^2)}
    &=
    \max_{j<k}
    \left(
        \int_{nt}^{(n+\ell)t}
        \|v_{jk}\tilde{\psi}(r)\|_{L_x^2}^2\,dr
    \right)^{1/2}\\
    &\leq
    \max_{j<k}
    \left(
        \int_0^T
        \|v_{jk}\tilde{\psi}(r)\|_{L_x^2}^2\,dr
    \right)^{1/2}\\
    &\leq
    \sqrt{\frac{\pi L}{2}}\,
    \|\psi_0\|_{L_x^2}
    <\infty.
    \end{aligned}
\end{equation}
Applying \cref{lem:tildepsi_representation} to this interpolation, we obtain
\begin{equation}\label{eq:interpolation_integral}
\begin{aligned}
    \tilde{\psi}(nt+s)
    &=e^{-i(nt+s)A}\psi_0 - i\int_0^{nt+t\lfloor s/t\rfloor}e^{-i(nt+s-r)A}B_t\tilde{\psi} (r)dr\\
    &= e^{-isA}\underbrace{\left(e^{-intA}\psi_0 -i\int_0^{nt}e^{-i(nt-r)A}B_t\tilde{\psi}(r)dr\right)}_{=\psi_n}-i\int_{nt}^{nt+t\lfloor s/t\rfloor}e^{-i(nt+s-r)A}B_t \tilde{\psi}(r)dr\\
    &=e^{-isA}\psi_n
    -i\int_0^{t\left\lfloor s/t\right\rfloor}
    e^{-i(s-r)A}B_t\tilde{\psi}(nt+r)\,dr,
    \quad 0\leq s\leq\ell t,
    \end{aligned}
\end{equation}
which leads to 
\begin{equation*}
    \|v_{jk}\tilde{\psi}(nt+s)\|_{L_s^2(0,\ell t;L_x^2)}
    \leq 
    \|v_{jk}e^{-isA}\psi_n\|_{L_s^2(0,\ell t;L_x^2)}+\left\|v_{jk}\int_0^{t\left\lfloor s/t\right\rfloor}
    e^{-i(s-r)A}B_t\tilde{\psi}(nt+r) dr\right\|_{L_s^2(0,\ell t;L_x^2)}.
\end{equation*}
For the first term, \cref{lem:smoothing1} yields
\begin{align*}
    \|v_{jk}e^{-isA}\psi_n\|_{L_s^2(0,\ell t;L_x^2)}
    &\leq
    \left(
        \int_{\mathbb R}
        \|v_{jk}e^{-isA}\psi_n\|_{L_x^2}^2 ds
    \right)^{1/2}\\
    &\leq
    \sqrt{\frac{\pi}{2}}\|\psi_n\|_{L_x^2}.
\end{align*}
For the integral term, first define
\[
    v_{jk,M}(x):=\min\{v_{jk}(x),M\},
    \quad M>0.
\]
For $0\leq s\leq \ell t$, we have
\begin{align*}
    \left\|
        v_{jk,M}\int_0^{t\left\lfloor s/t\right\rfloor}
        e^{-i(s-r)A}B_t\tilde{\psi}(nt+r) dr
    \right\|_{L_x^2} &\leq
    \int_0^{t\left\lfloor s/t\right\rfloor}
    \|v_{jk,M}e^{-i(s-r)A}B_t\tilde{\psi}(nt+r)\|_{L_x^2} dr\\
    &\leq
    \int_0^{\ell t}
    \|v_{jk,M}e^{-i(s-r)A}B_t\tilde{\psi}(nt+r)\|_{L_x^2} dr.
\end{align*}
Using $0\leq v_{jk,M}\leq v_{jk}$,
the change of variables $u=s-r$, and \cref{lem:smoothing1},
we obtain
\begin{align*}
    \int_0^{\ell t}
    \|v_{jk,M}e^{-i(s-r)A}B_t\tilde{\psi}(nt+r)\|_{L_x^2}^2 ds
    &\leq
    \int_{\mathbb R}
    \|v_{jk}e^{-iuA}B_t\tilde{\psi}(nt+r)\|_{L_x^2}^2 du\\
    &\leq
    \frac{\pi}{2}
    \|B_t\tilde{\psi}(nt+r)\|_{L_x^2}^2.
\end{align*}
Therefore, taking the $L_s^2(0,\ell t)$ norm yields
\begin{align*}
    &\left\|
        v_{jk,M}\int_0^{t\left\lfloor s/t\right\rfloor}
        e^{-i(s-r)A}B_t\tilde{\psi}(nt+r) dr
    \right\|_{L_s^2(0,\ell t;L_x^2)}\\
    \leq &
    \int_0^{\ell t}
    \left(
        \int_0^{\ell t}
        \|v_{jk,M}e^{-i(s-r)A}B_t\tilde{\psi}(nt+r)\|_{L_x^2}^2 ds
    \right)^{1/2} dr\\
     \leq &
    \sqrt{\frac{\pi}{2}}
    \int_0^{\ell t}
    \|B_t\tilde{\psi}(nt+r)\|_{L_x^2} dr.
\end{align*}
Letting $M\to\infty$ and applying the monotone convergence
theorem gives the same estimate with $v_{jk,M}$ replaced
by $v_{jk}$. Combining the estimates for the two terms in
\cref{eq:interpolation_integral}, we obtain
\begin{equation*}
    \|v_{jk}\tilde{\psi}(nt+s)\|_{L_s^2(0,\ell t;L_x^2)}\leq
    \sqrt{\frac{\pi}{2}}\|\psi_n\|_{L_x^2}
    +
    \sqrt{\frac{\pi}{2}}
    \int_0^{\ell t}
    \|B_t\tilde{\psi}(nt+r)\|_{L_x^2} dr.
\end{equation*}
Moreover, using \cref{eq:Bt_bound} and applying the Cauchy-Schwarz inequality, we find
\begin{align*}
    \int_0^{\ell t}
    \|B_t\tilde{\psi}(nt+r)\|_{L_x^2} dr
    &\leq
    \sum_{j<k}
    w_{jk}\int_0^{\ell t}
    \|v_{jk}\tilde{\psi}(nt+r)\|_{L_x^2} dr\\
    &\leq
    \sqrt{\ell t}
    \sum_{j<k}
    w_{jk}
    \|v_{jk}\tilde{\psi}(nt+s)\|_{L_s^2(0,\ell t;L_x^2)}\\
    &\leq
    W\sqrt{\ell t}
    \max_{1\leq j<k\leq N}
    \|v_{jk}\tilde{\psi}(nt+s)\|_{L_s^2(0,\ell t;L_x^2)}.
\end{align*}
If $\ell t\leq 1/(2\pi W^2)$, then
\begin{equation*}
    \max_{1\leq j<k\leq N}
    \|v_{jk}\tilde{\psi}(nt+s)\|_{L_s^2(0,\ell t;L_x^2)}\leq
    \sqrt{\frac{\pi}{2}}\|\psi_n\|_{L_x^2}
    +
    \frac12
    \max_{1\leq j<k\leq N}
    \|v_{jk}\tilde{\psi}(nt+s)\|_{L_s^2(0,\ell t;L_x^2)}.
\end{equation*}
Absorbing the last term into the left-hand side, we obtain, for every $0\leq n<L$ and $1\leq\ell\leq L-n$ with $\ell t\leq1/(2\pi W^2)$,
\begin{equation}\label{eq:interpolation_local_smoothing}
    \begin{aligned}
        \int_{nt}^{(n+\ell)t}
        \|v_{jk}\tilde{\psi}(s)\|_{L_x^2}^2 ds
        =
        \int_0^{\ell t}
        \|v_{jk}\tilde{\psi}(nt+s)\|_{L_x^2}^2 ds
        \leq
        2\pi\|\psi_n\|_{L_x^2}^2,
        \quad 1\leq j<k\leq N.
    \end{aligned}
\end{equation}
We now extend \cref{eq:interpolation_local_smoothing}
to $[0,T]$. Choose
\[\ell:=\left\lfloor\frac{1}{2\pi tW^2}\right\rfloor.\]
Since $t\leq 1/(4\pi W^2)$, we have
\[
    \frac{1}{4\pi W^2}
    \leq\frac{1}{2\pi W^2}-t
    <\ell t
    \leq\frac{1}{2\pi W^2}.
\]
We partition $[0,T]$ as
\[
    [0,T]
    =
    \bigcup_{\substack{n\in\{0,\ell,2\ell,\ldots\}\\ n<L}}
    \left[nt,\min\{(n+\ell)t,T\}\right].
\]
For each such $n$,
\[
    0<
    \min\{(n+\ell)t,T\}-nt
    =
    t\min\{\ell,L-n\}
    \leq\ell t\leq\frac{1}{2\pi W^2}.
\]
Applying \cref{eq:interpolation_local_smoothing}
with $\ell$ replaced by $\min\{\ell,L-n\}$,
and using \cref{eq:Et_norm}, we obtain
\begin{equation*}
    \int_{nt}^{\min\{(n+\ell)t,T\}}
    \|v_{jk}\tilde{\psi}(s)\|_{L_x^2}^2 ds\leq
    2\pi\|\psi_n\|_{L_x^2}^2\leq
    2\pi\|\psi_0\|_{L_x^2}^2.
\end{equation*}
Summing over the $\lceil L/\ell\rceil$ intervals and using
$T=Lt$ and $\ell t\geq 1/(4\pi W^2)$, we have
\begin{align*}
    \int_0^T
    \|v_{jk}\tilde{\psi}(s)\|_{L_x^2}^2 ds
    &=
    \sum_{\substack{n\in\{0,\ell,2\ell,\ldots\}\\ n<L}}
    \int_{nt}^{\min\{(n+\ell)t,T\}}
    \|v_{jk}\tilde{\psi}(s)\|_{L_x^2}^2 ds\\
    &\leq
    2\pi\left\lceil\frac{L}{\ell}\right\rceil
    \|\psi_0\|_{L_x^2}^2\\
    &\leq
    2\pi\left(1+\frac{T}{\ell t}\right)
    \|\psi_0\|_{L_x^2}^2\\
    &\leq
    2\pi\left(1+4\pi TW^2\right)
    \|\psi_0\|_{L_x^2}^2.
\end{align*}
Finally, we apply \cref{eq:Bt_bound} to conclude
\begin{align*}
    \|B_t\tilde{\psi}\|_{L_s^2(0,T;L_x^2)}
    &\leq
    \sum_{j<k}
    w_{jk}
    \|v_{jk}\tilde{\psi}\|_{L_s^2(0,T;L_x^2)}\\
    &\leq
    W\sqrt{2\pi\left(1+4\pi TW^2\right)}
     \|\psi_0\|_{L_x^2}.
\end{align*}
\hfill \qed

\vspace{.3cm}
Replacing the upper integration limit $t\left\lfloor s/t\right\rfloor$ by $s$
in the integral representation of $\tilde{\psi}$,
we define $\psi_c$ as
\begin{equation}\label{eq:continuous_interpolation}
    \psi_c(s):=
    e^{-isA}\psi_0
    -i\int_0^s
    e^{-i(s-r)A}B_t\tilde{\psi}(r)\,dr,
    \quad
    d(s):=\psi_c(s)-\tilde{\psi}(s),
    \quad 0\leq s\leq T.
\end{equation}
Since $B_t\tilde{\psi}\in L_s^2(0,T;L_x^2)
\subset L_s^1(0,T;L_x^2)$,
this defines a continuous $L_x^2$-valued function. Then by \cref{eq:psitilde_def} and \cref{eq:continuous_interpolation}, we can write $d(s)$ as 
\begin{equation}\label{eq:interpolation_correction}
    d(s)=
    -i\int_{t\lfloor s/t\rfloor}^s
    e^{-i(s-r)A}B_t\tilde{\psi}(r)\,dr,
    \quad 0\leq s\leq T.
\end{equation}
and at every grid point,
\begin{equation}\label{eq:continuous_interpolation_grid}
    d(\ell t)=0,
    \quad
    \psi_c(\ell t)=\tilde{\psi}(\ell t)=\psi_\ell,
    \quad \ell=0,\ldots,L.
\end{equation}
We next derive an integral representation of $E_t^L-M_t(T)$
involving $d(s)$, which we estimate using \cref{lem:psitilde_smoothing}.

\begin{lemma}\label{lem:discrete_continuous_identity}
For $\psi_0\in L_x^2(\mathbb{R}^{3N})$, let $\tilde{\psi}$ be defined by
\cref{eq:psitilde_def} and $d(s)$ be defined by \cref{eq:continuous_interpolation}. Then
\begin{equation}\label{eq:interpolation_correction_bound}
        \|d\|_{L_s^2(0,T;L_x^2)}
        \leq tW\sqrt{\pi(1+4\pi TW^2)} \|\psi_0\|_{L_x^2},
\end{equation}
and
\begin{equation}\label{eq:discrete_continuous_identity}
    (E_t^L-M_t(T))\psi_0
    =
    i\int_0^T M_t(T-s)B_td(s)\,ds.
\end{equation}
\end{lemma}

\begin{proof}
For $\ell=0,\ldots,L-1$ and
$\ell t\leq s<(\ell+1)t$,
unitarity and the Cauchy-Schwarz inequality give
\begin{align*}
    \|d(s)\|_{L_x^2}^2
    &\leq
    \left(
        \int_{\ell t}^s
        \|B_t\tilde{\psi}(r)\|_{L_x^2}\,dr
    \right)^2\\
    &\leq
    (s-\ell t)
    \int_{\ell t}^s
    \|B_t\tilde{\psi}(r)\|_{L_x^2}^2\,dr.
\end{align*}
Integrating over $[\ell t, (\ell+1)t]$ and using Fubini's theorem, we obtain
\begin{align*}
    \int_{\ell t}^{(\ell+1)t}
    \|d(s)\|_{L_x^2}^2\,ds
    &\leq
    \int_{\ell t}^{(\ell+1)t}
    \|B_t\tilde{\psi}(r)\|_{L_x^2}^2
    \left(
        \int_r^{(\ell+1)t}(s-\ell t)\,ds
    \right)\,dr\\
    &\leq
    \frac{t^2}{2}
    \int_{\ell t}^{(\ell+1)t}
    \|B_t\tilde{\psi}(r)\|_{L_x^2}^2\,dr.
\end{align*}
Summing over $\ell$ and applying \cref{lem:psitilde_smoothing} gives \cref{eq:interpolation_correction_bound}:
\begin{equation*}
        \|d\|_{L_s^2(0,T;L_x^2)}\leq
        \frac{t}{\sqrt{2}}
        \|B_t\tilde{\psi}\|_{L_s^2(0,T;L_x^2)}\leq
        tW\sqrt{\pi(1+4\pi TW^2)}
        \|\psi_0\|_{L_x^2}.
\end{equation*}
Now, substituting $\tilde{\psi}=\psi_c-d$ into
\cref{eq:continuous_interpolation}, we obtain
\begin{equation}\label{eq:continuous_interpolation_equation}
        \psi_c(s)=
        e^{-isA}\psi_0
        -i\int_0^s
        e^{-i(s-r)A}B_t\psi_c(r)\,dr+
        i\int_0^s
        e^{-i(s-r)A}B_td(r)\,dr.
\end{equation}
Since 
\[|B_t(x)|\leq \frac{1}{t}\sum_{j<k}\pi_{jk}\left|e^{-itB_{jk}(x)/\pi_{jk}}-1\right|\leq \frac{2}{t},\]
we have
\[
    \int_0^T\|B_td(r)\|_{L_x^2}\,dr
    \leq
    \frac{2}{t}\sqrt{T}
    \|d\|_{L_s^2(0,T;L_x^2)}
    <\infty.
\]
We claim that
\begin{equation}\label{eq:continuous_interpolation_variation}
    \psi_c(s)
    =
    M_t(s)\psi_0
    +i\int_0^s M_t(s-r)B_td(r)\,dr.
\end{equation}
Recall \cref{eq:duhamel} with $Z=B_t$: for $f\in H_x^2$,
\begin{equation} \label{eq:duhamel1}
    M_t(u)f
    =
    e^{-iuA}f
    -i\int_0^u
    e^{-i(u-a)A}B_tM_t(a)f\,da,\quad u\geq0.
\end{equation}
For $f\in L_x^2$, choose $f_m\in H_x^2$ such that $f_m\to f$ in $L_x^2$. Since $M_t$ is contractive, for every fixed $u\geq0$, we have
\begin{align*}
    &\left\|
        \int_0^u e^{-i(u-a)A}B_tM_t(a)(f_m-f)\,da
    \right\|_{L_x^2}\\
    \leq &
    \int_0^u
    \|B_tM_t(a)(f_m-f)\|_{L_x^2}\,da
    \leq 
    u\frac{2}{t}\|f_m-f\|_{L_x^2}
    \to 0.
\end{align*}
Moreover,
\[
    M_t(u)f_m\to M_t(u)f,
    \quad 
    e^{-iuA}f_m\to e^{-iuA}f
    \quad\text{in }L_x^2.
\]
Thus, \cref{eq:duhamel1} also holds for every $f\in L_x^2$. Therefore, we have
\begin{equation} \label{eq:duhamel2}
M_t(s)\psi_0 = e^{-isA}\psi_0 -i \int_0^s e^{-i(s-r)A}B_tM_t(r)\psi_0 dr.
\end{equation}
Also, we can put $u=s-a$, $f=B_t d(a)$ in \cref{eq:duhamel1} to get
\[M_t(s-a)B_td(a)
    =
    e^{-i(s-a)A}B_td(a)
    -i\int_a^{s}
    e^{-i(s-r)A}B_tM_t(r-a)B_td(a)\,dr\]
which leads to 
\begin{align} \label{eq:duhamel3}
\begin{aligned}
    i\int_0^s M_t(s-a)B_td(a)\,da
    &=
    i\int_0^s e^{-i(s-a)A}B_td(a)\,da\\
    &\quad+
    \int_0^s e^{-i(s-r)A}B_t
    \left(
        \int_0^r M_t(r-a)B_td(a)\,da
    \right)\,dr.
    \end{aligned}
\end{align}
Here, we applied Fubini's theorem since
\begin{align*}
    \int_0^s\int_0^a
    \|e^{-i(s-a)A}B_tM_t(a-r)B_td(r)\|_{L_x^2}
    \,dr\,da\leq
    s\frac{4}{t^2}
    \int_0^s\|d(r)\|_{L_x^2}\,dr
    <\infty.
\end{align*}
Adding \cref{eq:duhamel2} and \cref{eq:duhamel3} gives
\begin{align*}
    &M_t(s)\psi_0+i\int_0^s M_t(s-a)B_td(a)da\\
     = & e^{-isA}\psi_0 -i\int_0^s e^{-i(s-r)A}B_t\left(M_t(r)\psi_0+i\int_0^r M_t(r-a)B_td(a)da\right)dr+i\int_0^se^{-i(s-r)A}B_td(r)dr.
\end{align*}
The integral equation \cref{eq:continuous_interpolation_equation} for $\psi_c$ has at most one continuous solution. Indeed, the difference $h$ of two such solutions satisfies
\[
    h(s)
    =
    -i\int_0^s e^{-i(s-r)A}B_th(r)\,dr,
\]
and hence
\[
    \|h(s)\|_{L_x^2}
    \leq
    \frac{2}{t}
    \int_0^s\|h(r)\|_{L_x^2}\,dr.
\]
Gr\"onwall's lemma gives $h=0$, proving \cref{eq:continuous_interpolation_variation}. Finally, evaluating \cref{eq:continuous_interpolation_variation} at $s=T$ and using $\psi_c(T)=\psi_L=E_t^L\psi_0$ yields \cref{eq:discrete_continuous_identity}.
\end{proof}
\noindent\emph{Proof of \cref{prop:E_M_error}.}
To estimate this integral, we use smoothing for the adjoint
semigroup. Since $A$ is self-adjoint and $B_t$ is a bounded
multiplication operator,
\[
    M_t(u)^*
    =
    e^{iu(A+\overline{B_t})}
    =
    e^{-iu(-A-\overline{B_t})},
    \quad u\geq0.
\]
Moreover,
\[
    |-\overline{B_t}(x)|
    =
    |B_t(x)|
    \leq
    \sum_{j<k}w_{jk}v_{jk}(x),
    \quad
    \operatorname{Im}(-\overline{B_t}(x))
    =
    \operatorname{Im}B_t(x)
    \leq0.
\]
Thus, \cref{lem:smoothing2}, with free generator $-A$
and potential $-\overline{B_t}$, gives
\begin{equation}\label{eq:adjoint_smoothing}
    \begin{aligned}
        \int_0^T
        \|v_{jk}M_t(T-s)^*\phi_0\|_{L_x^2}^2\,ds
        &=
        \int_0^T
        \|v_{jk}M_t(u)^*\phi_0\|_{L_x^2}^2\,du\\
        &\leq
        4\pi^2(1+TW^2)\|\phi_0\|_{L_x^2}^2,
        \quad \phi_0\in L_x^2.
    \end{aligned}
\end{equation}

For almost every $s$, the pointwise bound on $B_t$
and the Cauchy-Schwarz inequality in space imply
\begin{align*}
    &\left|
        \langle M_t(T-s)^*\phi_0,B_td(s)\rangle_{L_x^2}
    \right|\\
    \leq  &
    \sum_{j<k}w_{jk}
    \int_{\bbr^{3N}}
    v_{jk}(x)
    |(M_t(T-s)^*\phi_0)(x)|\,|d(s,x)|\,dx\\
   \leq  &
    \sum_{j<k}w_{jk}
    \|v_{jk}M_t(T-s)^*\phi_0\|_{L_x^2}
    \|d(s)\|_{L_x^2}.
\end{align*}
Pairing \cref{eq:discrete_continuous_identity}
with $\phi_0$, applying the Cauchy-Schwarz inequality
in time, and using \cref{eq:interpolation_correction_bound} and \cref{eq:adjoint_smoothing}, we obtain
\begin{align*}
    &\left|
        \langle\phi_0,(E_t^L-M_t(T))\psi_0\rangle_{L_x^2}
    \right|\\
    \leq &
    \sum_{j<k}w_{jk}
    \int_0^T
    \|v_{jk}M_t(T-s)^*\phi_0\|_{L_x^2}
    \|d(s)\|_{L_x^2}\,ds\\
    \leq &
    \sum_{j<k}w_{jk}
    \left(
        \int_0^T
        \|v_{jk}M_t(T-s)^*\phi_0\|_{L_x^2}^2\,ds
    \right)^{1/2}
    \|d\|_{L_s^2(0,T;L_x^2)}\\
    \leq &
    2\pi W\sqrt{1+TW^2}
    \|\phi_0\|_{L_x^2}
    \|d\|_{L_s^2(0,T;L_x^2)}\\
    \leq &
    2\pi tW^2
    \sqrt{2\pi(1+TW^2)(1+4\pi TW^2)}
    \|\phi_0\|_{L_x^2}\|\psi_0\|_{L_x^2}.
\end{align*}
Taking the supremum over all unit vectors
$\phi_0,\psi_0\in L_x^2$ gives
\begin{equation*}
    \|E_t^L-M_t(T)\|_{L_x^2\to L_x^2}
    \leq
    2\pi tW^2
    \sqrt{2\pi(1+TW^2)(1+4\pi TW^2)}.
\end{equation*}
\hfill\qed

\vspace{.3cm}
\noindent\emph{Proof of \cref{thm:main_result1}.} Combining \cref{prop:U_M_error} and \cref{prop:E_M_error}, we obtain
\begin{align*}
    \|E_t^L-U(T)\|_{L_x^2\to L_x^2}
    &\leq
    \|E_t^L-M_t(T)\|_{L_x^2\to L_x^2}
    +
    \|M_t(T)-U(T)\|_{L_x^2\to L_x^2}\\
    &\leq
    tW^2\left[
        2\pi\sqrt{2\pi(1+TW^2)(1+4\pi TW^2)}
        +\pi(1+2\pi TW^2)
    \right].
\end{align*}
Since 
\[
    1+4\pi TW^2\leq4\pi(1+TW^2),
    \quad
    1+2\pi TW^2\leq2\pi(1+TW^2),
\]
it follows that
\begin{align*}
    \|E_t^L-U(T)\|_{L_x^2\to L_x^2}
    &\leq
    (2+4\sqrt{2})\pi^2tW^2(1+TW^2)\\
    &\leq
    8\pi^2tW^2(1+TW^2).
\end{align*}
\hfill \qed

\subsection{Error estimate for $H^2$ initial states}
\label{subsection:H^2data_eta7}

If we consider the initial states $\psi_0$ in the domain of the Hamiltonian ($H^2$ Sobolev space), then we can obtain an improved system-size dependence. We estimate the exact solution using energy conservation and apply smoothing to the adjoint evolution. 

\begin{prop}\label{prop:E_U_state_error}
Under the assumptions of \cref{thm:main_result1}, let $\psi_0\in H^2(\mathbb{R}^{3N})$ be normalized. Then
\begin{equation}\label{eq:E_U_state_error}
 \norm{(E_t^L-U(T))\psi_0}_{L_x^2}
 \le CtW^{3/2}(1+TW^2)^{1/2}
 \left(\sum_{j<k}w_{jk}\int_0^T\norm{v_{jk}U(s)\psi_0}_{L_x^2}^2\dd s\right)^{1/2}.
\end{equation}
Here $C$ is independent of the particle number and the interaction coefficients.
\end{prop}

Before proceeding with the proof, we again note that this immediately implies the trace norm error
for $H^2$ initial states.
\begin{thm}[Convergence for $H^2$ initial states]
\label{thm:H2-channel}
Under the assumptions of \cref{thm:main_result1}, suppose $T\ge1$. For every normalized $\psi_0\in H^2(\mathbb{R}^{3N})$, set $\rho_0=|\psi_0\rangle\langle\psi_0|$. Then
\begin{equation}\label{eq:state_channel_error}
\norm{\Phi(\rho_0)-\mathcal U_T(\rho_0)}_1
\le C \left(c_0^2+c_0^4 \right) tTN^7\norm{\psi_0}_{H_x^2}
\end{equation}
where $C$ is an absolute constant.
\end{thm}

\begin{proof}[Proof of \cref{thm:H2-channel}]
For $\rho_0=|\psi_0\rangle\langle\psi_0|$, the averaged output satisfies
\begin{equation}
\begin{aligned}
& \norm{\Phi(\rho_0)-\mathcal U_T(\rho_0)}_1
 \le 4\norm{(E_t^L-U(T))\psi_0}_{L_x^2}
 \\
 \le &  C t (c_0 N^2)^{3/2}(1+  T c_0^2 N^4)^{1/2} \left( c_0 T N (1+ c_0^2 N^3) \norm{\psi_0}_{H_x^2}^2\right)^{1/2} 
 \\
 \le & 
    C t T N^7 c_0^2 \left(1 + c_0^2 \right) \norm{\psi_0}_{H_x^2},
 \end{aligned}
\end{equation}
for some constant $C>0$, where we first use the mixing lemma in~\cref{lem:mean-to-diamond}, then \cref{prop:E_U_state_error}, and finally the fact that for every $\psi_0\in H^2(\mathbb{R}^{3N})$ and $T>0$, 
\begin{equation*}
\sum_{j<k}w_{jk}\int_0^T
\norm{v_{jk}U(s)\psi_0}_{L_x^2}^2\,\dd s
\le Cc_0TN(1+c_0^2N^3)\norm{\psi_0}_{H_x^2}^2,
\end{equation*}
where $C$ is an absolute constant.
Indeed, the three-dimensional Hardy inequality in the relative coordinate gives, for every $f\in H^1(\mathbb{R}^{3N})$,
\begin{equation*}
\begin{aligned}
\sum_{j<k}w_{jk}\norm{v_{jk}f}_{L_x^2}^2
&\le c_0\sum_{j<k}
\norm{(\nabla_{x_j}-\nabla_{x_k})f}_{L_x^2}^2
\le c_0N\norm{\nabla f}_{L_x^2}^2.
\end{aligned}
\end{equation*}
The gradient propagation estimate proved in~\cite[Eq.~(41)]{FangWuSoffer2025} yields
\begin{equation*}
\sup_{s\in\mathbb R}\norm{\nabla U(s)\psi_0}_{L_x^2}
\le C(1+c_0N^{3/2})\norm{\psi_0}_{H_x^2}.
\end{equation*}
Consequently,
\begin{equation*}
\begin{aligned}
\sum_{j<k}w_{jk}\int_0^T
\norm{v_{jk}U(s)\psi_0}_{L_x^2}^2\,\dd s
&\le c_0NT\sup_{s\in[0,T]}
\norm{\nabla U(s)\psi_0}_{L_x^2}^2\\
&\le Cc_0TN(1+c_0^2N^3)\norm{\psi_0}_{H_x^2}^2,
\end{aligned}
\end{equation*}
which completes the proof.
\end{proof}

\begin{proof}[Proof of \cref{prop:E_U_state_error}]
By \cref{eq:Bt_bound,eq:Bt_error} and Cauchy-Schwarz inequality, we have
\begin{equation}\label{eq:Bt_weighted_bounds}
 \begin{aligned}
 \norm{Bf}_{L_x^2}^2,\ \norm{B_tf}_{L_x^2}^2,\ \norm{B_t^*f}_{L_x^2}^2
 &\le W\sum_{j<k}w_{jk}\norm{v_{jk}f}_{L_x^2}^2,\\
 |\langle g,(B_t-B)f\rangle_{L_x^2}|
 &\le\frac{tW}{2}\sum_{j<k}w_{jk}\norm{v_{jk}g}_{L_x^2}\norm{v_{jk}f}_{L_x^2}.
 \end{aligned}
\end{equation}
Recall that $\psi(s)=U(s)\psi_0$. For a test vector $\phi_0\in L_x^2$, define the adjoint piecewise interpolation, on $\ell t\le s<(\ell+1)t$, by
\begin{equation*}
 \tilde\phi(s)=e^{i((\ell+1)t-s)A}(E_t^*)^{L-1-\ell}\phi_0.
\end{equation*}
The identity $E_t^L-U(T)=\sum_{\ell=0}^{L-1}E_t^{L-1-\ell}(E_t-U(t))U(\ell t)$, the integral representation of $E_t$, and Duhamel's formula give
\begin{equation}\label{eq:E_U_state_identity}
 \begin{aligned}
 \langle\phi_0,(E_t^L-U(T))\psi_0\rangle_{L_x^2}
 &=-i\int_0^T\langle\tilde\phi(s),(B_t-B)\psi(s)\rangle_{L_x^2}\dd s\\
 &\quad+\sum_{\ell=0}^{L-1}\int_{\ell t}^{(\ell+1)t}\int_{\ell t}^s
 \langle\tilde\phi(s),B_te^{-i(s-r)A}B\psi(r)\rangle_{L_x^2}\dd r\dd s.
 \end{aligned}
\end{equation}
The function $\tilde\phi$ is the time reversal of the piecewise interpolation for $E_t^*$. Its step representation has $A,B_t$ replaced by $-A,-B_t^*$, with the same contraction and majorant bounds. Applying \cref{eq:interpolation_local_smoothing} and its block argument therefore gives
\begin{equation}\label{eq:adjoint_interpolation_smoothing}
 \int_0^T\norm{v_{jk}\tilde\phi(s)}_{L_x^2}^2\dd s
 \le C(1+TW^2)\norm{\phi_0}_{L_x^2}^2.
\end{equation}
Young's inequality on each time interval and \cref{eq:Bt_weighted_bounds} yield
\begin{equation}\label{eq:within_step_duhamel_bound}
 \begin{aligned}
 & \sum_{\ell=0}^{L-1}\int_{\ell t}^{(\ell+1)t}
 \left\|\int_{\ell t}^s e^{-i(s-r)A}B\psi(r)\dd r\right\|_{L_x^2}^2\dd s
 \\
 \le & t^2\norm{B\psi}_{L_s^2(0,T;L_x^2)}^2
 \le t^2W\sum_{j<k}w_{jk}\int_0^T\norm{v_{jk}\psi(s)}_{L_x^2}^2\dd s.
 \end{aligned}
\end{equation}
For the first term in \cref{eq:E_U_state_identity}, apply \cref{eq:Bt_weighted_bounds,eq:adjoint_interpolation_smoothing} and Cauchy-Schwarz in time and in the interaction indices. For the second, move $B_t$ to the test vector and use \cref{eq:within_step_duhamel_bound} together with 
\[
\norm{B_t^*\tilde\phi}_{L_s^2(0,T;L_x^2)}\le CW\sqrt{1+TW^2}\norm{\phi_0}_{L_x^2}.
\]
Both terms are bounded by the right-hand side of \cref{eq:E_U_state_error} times $\norm{\phi_0}_{L_x^2}$. Taking the supremum over normalized $\phi_0$ completes the proof. No regularity of the randomized trajectories is used.
\end{proof}

We now specialize to a fixed-nuclei Coulomb system, as typically appears in the setting of electronic structures. In this subsection, use the physical kinetic normalization and write
\begin{equation}\label{eq:eta7-electronic-model}
 \begin{aligned}
 H&=A+B,\quad A=-\frac12\sum_{j=1}^{\eta}\Delta_{x_j},\\
 B&=\sum_{j<k}\frac1{|x_j-x_k|}-\sum_{j,\kappa}\frac{Z_\kappa}{|x_j-R_\kappa|}
 =\sum_\alpha c_\alpha v_\alpha.
 \end{aligned}
\end{equation}
where $Z_\kappa>0$, $Z_{\mathrm{tot}}=\sum_\kappa Z_\kappa$, $w_\alpha=|c_\alpha|$, $W=\sum_\alpha w_\alpha=\eta(\eta-1)/2+\eta Z_{\mathrm{tot}}$, and $\pi_\alpha=w_\alpha/W$. Here $\alpha$ indexes the electron-electron and electron-nucleus interactions, extending the notation $c_{jk},v_{jk},w_{jk},\pi_{jk}$ above. Each step still samples one interaction and one uniform insertion time. The preceding proof applies with $(j,k)$ replaced by $\alpha$, for $t\le cW^{-2}$: translation and the change from $-\Delta$ to $-\Delta/2$ only change universal constants in free smoothing and the small-step condition.

\begin{cor}\label{cor:eta7-error}
Consider \cref{eq:eta7-electronic-model} with $Z_{\mathrm{tot}}=\eta$. Let $\psi_0\in H^2(\bbr^{3\eta})$ satisfy $\norm{\psi_0}_{L_x^2}=1$, and set $\rho_0=|\psi_0\rangle\langle\psi_0|$. For $T\ge1$ and $t=T/L\le cW^{-2}$,
\begin{equation}\label{eq:eta7-continuum-error}
 \begin{aligned}
 \norm{(E_t^L-e^{-iTH})\psi_0}_{L_x^2}
 &\le C\frac{T^2\eta^7}{L}\norm{\psi_0}_{H_x^2},\\
 \norm{\Phi(\rho_0)-\mathcal U_T(\rho_0)}_1
 &\le C\frac{T^2\eta^7}{L}\norm{\psi_0}_{H_x^2}.
 \end{aligned}
\end{equation}
Here $c,C>0$ are universal constants.
\end{cor}

\begin{proof}
Three-dimensional Hardy and $\sum_{j<k}\norm{(\nabla_{x_j}-\nabla_{x_k})f}_{L_x^2}^2\le\eta\norm{\nabla f}_{L_x^2}^2$ give
\begin{equation}\label{eq:eta7-weighted-hardy}
 \sum_\alpha w_\alpha\norm{v_\alpha f}_{L_x^2}^2
 \le C(\eta+Z_{\mathrm{tot}})\norm{\nabla f}_{L_x^2}^2.
\end{equation}
For normalized $f$, Cauchy-Schwarz bounds $|\langle f,Bf\rangle|$ by $C\sqrt{W(\eta+Z_{\mathrm{tot}})}\norm{\nabla f}_{L_x^2}$. Conservation of energy and completion of the square therefore imply
\begin{equation}\label{eq:eta7-exact-energy}
 \sup_s\norm{\nabla e^{-isH}\psi_0}_{L_x^2}^2
 \le C\left(\norm{\nabla\psi_0}_{L_x^2}^2+W(\eta+Z_{\mathrm{tot}})\right).
\end{equation}
Under neutrality, $W\asymp\eta^2$. Since $\norm{\psi_0}_{H_x^2}\ge\norm{\psi_0}_{L_x^2}=1$, the preceding estimates give
\begin{equation*}
 \sum_\alpha w_\alpha\int_0^T\norm{v_\alpha e^{-isH}\psi_0}_{L_x^2}^2\dd s
 \le CT\eta\left(\norm{\nabla\psi_0}_{L_x^2}^2+\eta^3\right)
 \le CT\eta^4\norm{\psi_0}_{H_x^2}^2.
\end{equation*}
Substituting into \cref{prop:E_U_state_error} and using $T\ge1$ gives $CtT\eta^7\norm{\psi_0}_{H_x^2}=CT^2\eta^7\norm{\psi_0}_{H_x^2}/L$ for both errors. This uses no fermionic assumption or bound on the individual nuclear charges beyond their total charge.
\end{proof}

\section{Proof of the Main Result 2 and 3}
\subsection{Proof of Trotterization part of \cref{thm:barriers-L2}}
For $c\in\mathbb{R}\setminus\{0\}$, consider the two-body Coulomb system
\begin{equation}\label{eq:operators}
 A=-\Delta_{x_1}-\Delta_{x_2},\quad
 B=\frac{c}{|x_1-x_2|},\quad H=A+B,
 \quad D(H)=H^2(\mathbb{R}^6).
\end{equation}
Under the orthogonal change of coordinates
\[
 y=\frac{x_1-x_2}{\sqrt2},\quad
 z=\frac{x_1+x_2}{\sqrt2},
\]
we have
\[
A=-\Delta_y-\Delta_z,\quad B=\frac{c}{\sqrt2 |y|}.
\]
Now, fix $T>0$ and normalized functions
\[
 \phi\in C_c^\infty(\{y\in\mathbb{R}^3:1<|y|<2\}),\quad
 \chi\in C_c^\infty(B(0,1)),\quad
 \left\|\phi\right\|_2=\left\|\chi\right\|_2=1.
\]
Define the initial state
\begin{equation*}
 \Psi_1(y,z):=\frac1{\sqrt{\zeta(3)}}
 \sum_{k=1}^{\infty}\frac1{k^{3/2}}
 \left[e^{-iT\Delta_y}\bigl(e^{3k/2}\phi(e^k\,\cdot)\bigr)\right](y)
 \chi(z-3ke_1),
\end{equation*}
where $e_1=(1,0,0)$ and $\zeta(3)=\sum_{k=1}^{\infty}k^{-3}$.
The translated functions $\chi(\cdot-3ke_1)$ have disjoint supports.
Since
\[\left\|e^{-iT\Delta_y}\bigl(e^{3k/2}\phi(e^k\,\cdot)\bigr)\right\|^2_{L_x^2}=\int_{\bbr^3}e^{3k}|\phi(e^ky)|^2dy=\|\phi\|_2^2=1,\]
the series converges in $L^2(\mathbb{R}^6)$ with $\left\|\Psi_1\right\|_2=1$.
Note that $\Psi_1$ depends on the fixed final time $T$, but not on the
number of time steps. Write
\begin{equation}\label{eq:summand}
 \psi_k(y,z)=
 \left[e^{-iT\Delta_y}\bigl(e^{3k/2}\phi(e^k\,\cdot)\bigr)\right](y)
 \chi(z-3ke_1),\quad k\geq1,
\end{equation}
then each $\psi_k$ is a Schwartz function of norm one. All constants denoted by
$C$ or $C_T$ below are independent of $k$ and $L$; they may depend on the
fixed coupling and test functions.

\smallskip\noindent\emph{Step 1. Concentration at the final time.}
Recall Hardy's inequality:
\[      \left\||y|^{-1}f\right\|_2\le2\left\|\nabla f\right\|_2,\quad f\in H^1(\bbr^3).
\]
For $g\in \mathcal{S}(\bbr^3),$ let $f(y)=e^{-\frac{i|y|^2}{4s}}e^{is\Delta_y}g(y)$. Then $f\in H^1(\bbr^3)$ with 
\[\nabla_yf=e^{-\frac{i|y|^2}{4s}}\left(\nabla_y -\frac{iy}{2s}\right)e^{is\Delta_y}g(y)=\frac{1}{2is}e^{\frac{-i|y|^2}{4s}}e^{is\Delta_y}(yg),\]
where we used
\[
 (y+2is\nabla)e^{is\Delta_y}g=e^{is\Delta_y}(yg).
\]
in the last equality. We apply Hardy's inequality on this $f$ to obtain
\begin{equation} \label{eq:Hardy_modified}
 \left\||y|^{-1}e^{is\Delta_y}g\right\|_2\leq \frac1{|s|}\left\|yg\right\|_2,\quad g\in \mathcal{S}(\bbr^3).
\end{equation}
Now, the free evolution of \cref{eq:summand} is
\begin{equation}\label{eq:free-summand}
 e^{-isA}\psi_k(y,z)=
 \left[e^{i(s-T)\Delta_y}\big(e^{3k/2}\phi(e^k\,\cdot)\big)\right](y)
 \left[e^{is\Delta_z}\chi(\cdot-3ke_1)\right](z).
\end{equation}
Scaling the preceding two estimates and using unitarity in $z$ yields
\begin{equation*}
 \left\|Be^{-isA}\psi_k\right\|_{L_{y,z}^2}=\frac{|c|}{\sqrt{2}}\left\||y|^{-1}e^{i(s-T)\Delta_y}\big(e^{3k/2}\phi(e^k\,\cdot)\big)\right\|_{L_y^2}.
\end{equation*}
First, directly applying Hardy's inequality and using $x=e^k y$ gives
\begin{align*}
    \left\||y|^{-1}e^{i(s-T)\Delta_y}\big(e^{3k/2}\phi(e^k\,\cdot)\big)\right\|_{L_y^2}&\leq 2\left\|\nabla_ye^{i(s-T)\Delta_y}\big(e^{3k/2}\phi(e^k\,\cdot)\big)\right\|_{L_y^2}\\
    &=2\left\|e^{i(s-T)\Delta_y}\nabla_y\big(e^{3k/2}\phi(e^k\,\cdot)\big)\right\|_{L_y^2}\\
 &=2e^{k}\|\nabla\phi\|_2
\end{align*}
Next, applying \cref{eq:Hardy_modified} leads to 
\begin{align*}
    \left\||y|^{-1}e^{i(s-T)\Delta_y}\big(e^{3k/2}\phi(e^k\,\cdot)\big)\right\|_{L_y^2}&\leq \frac{1}{|s-T|}\left\|y\,e^{3k/2}\phi(e^ky)\right\|_{L_y^2}\\
 &=\frac{e^{-k}}{|s-T|}\|y\phi\|_{L_y^2}
\end{align*}
By combining these two estimates, we obtain
\begin{align}\label{eq:weighted}
\begin{aligned}
 \left\|Be^{-isA}\psi_k\right\|_{L_{y,z}^2}
 &\leq
 \frac{|c|}{\sqrt{2}}
 \min\left\{
 2e^k\|\nabla\phi\|_2,\,
 \frac{e^{-k}}{|s-T|}\|y\phi\|_2
 \right\}\\
 &\leq
 C\min\left\{e^k,\frac{e^{-k}}{|s-T|}\right\},
 \quad s\ne T,
\end{aligned}
\end{align}
where $C>0$ depends only on $c$ and $\phi$. At $s=T$, the first bound still applies.

Since $\psi_k\in H^2$, Duhamel's formula is justified in $L^2$:
\[
 (e^{-iTH}-e^{-iTA})\psi_k
 =-i\int_0^T e^{-i(T-s)H}Be^{-isA}\psi_k\,ds.
\]
It follows from \cref{eq:weighted} that
\begin{equation}\label{eq:exact-free}
 \begin{aligned}
 \left\|(e^{-iTH}-e^{-iTA})\psi_k\right\|_2
 &\leq C\int_0^T\min\left\{e^k,\frac{e^{-k}}{T-s}\right\}ds\\
 &\leq C_T(1+k)e^{-k}.
 \end{aligned}
\end{equation}

\smallskip\noindent\emph{Step 2. The final potential factor.}
Using the telescoping sum
\[
\left(e^{-itB}e^{-itA}\right)^L-e^{-itB}e^{-iTA}=
\sum_{j=1}^{L-1}
\left(e^{-itB}e^{-itA}\right)^j
\left(e^{-itB}-I\right)e^{-i(T-jt)A},
\]
we obtain the following estimate:
\begin{align*}
 \left\|\left[(e^{-itB}e^{-itA})^L-e^{-itB}e^{-iTA}\right]\psi_k\right\|_2
 &\leq\sum_{j=1}^{L-1}
     \left\|(e^{-itB}-I)e^{-i(T-jt)A}\psi_k\right\|_2\\
 &\leq t\sum_{j=1}^{L-1}\left\|Be^{-i(T-jt)A}\psi_k\right\|_2,
\end{align*}
where we used $|e^{-itB}-1|\le t|B|$. Since \cref{eq:weighted} gives
\[
\left\|Be^{-i(T-jt)A}\psi_k\right\|_2\leq C\frac{e^{-k}}{jt},
\]
we get
\begin{equation}\label{eq:split-free}
   \left\|\left[(e^{-itB}e^{-itA})^L-e^{-itB}e^{-iTA}\right]\psi_k\right\|_2\leq \sum_{j=1}^{L-1}C\frac{e^{-k}}{j}\leq Ce^{-k}(1+\log L)
\end{equation}
At time $T$, we have
\begin{align*}
\left(e^{-iTA}\psi_k\right)(y,z)=e^{3k/2}\phi(e^ky)
\left[e^{iT\Delta_z}\chi(\cdot-3ke_1)\right](z),
\end{align*}
which leads to
\begin{align*}
\left\langle
e^{-iTA}\psi_k,e^{-itB}e^{-iTA}\psi_k
\right\rangle=
\iint
e^{-itc/(\sqrt{2}|y|)}
e^{3k}|\phi(e^ky)|^2\left|
\left[e^{iT\Delta_z}\chi(\cdot-3ke_1)\right](z)
\right|^2\,dz\,dy.
\end{align*}
Since $\|\chi\|_2=1$ gives
\begin{equation*}
\int_{\mathbb R^3}
\left|
\left[e^{iT\Delta_z}\chi(\cdot-3ke_1)\right](z)
\right|^2\,dz=1,
\end{equation*}
we obtain
\begin{align*}
\left\langle
e^{-iTA}\psi_k,e^{-itB}e^{-iTA}\psi_k
\right\rangle
&=
e^{3k}\int_{\mathbb R^3}
e^{-itc/(\sqrt{2}|y|)}|\phi(e^ky)|^2\,dy\\
&=\int_{\mathbb R^3}
e^{-ite^kc/(\sqrt{2}|x|)}|\phi(x)|^2\,dx.
\end{align*}
Note that since $\phi$ is supported in $1<|x|<2$ and $c\ne0$,
\begin{equation*}
 \int_{\mathbb R^3}
e^{-i\lambda c/(\sqrt{2}|x|)}|\phi(x)|^2\,dx
 \to 0\quad\text{as}\quad\lambda\to\infty.
\end{equation*}
Indeed, polar coordinates followed by the substitution $q=1/|x|$ reduce
this to the Riemann-Lebesgue lemma for an integrable function supported
in $[1/2,1]$.

Now restrict to $k\ge\lceil2\log L\rceil$. Then $te^k\ge TL\to\infty$,
so the integral tends to zero uniformly over these
indices. Meanwhile, \cref{eq:exact-free} and \cref{eq:split-free} imply
\begin{align*}
 &\left|
 \left\langle e^{-iTH}\psi_k,(e^{-itB}e^{-itA})^L\psi_k\right\rangle
 -\left\langle e^{-iTA}\psi_k,e^{-itB}e^{-iTA}\psi_k\right\rangle
 \right|\\
\leq &  \left|
\left\langle
(e^{-iTH}-e^{-iTA})\psi_k,\,
(e^{-itB}e^{-itA})^L\psi_k
\right\rangle\right|+
\left|\left\langle
e^{-iTA}\psi_k,\,
\left[(e^{-itB}e^{-itA})^L-e^{-itB}e^{-iTA}\right]\psi_k
\right\rangle\right|
\\
\leq & \left\|(e^{-iTH}-e^{-iTA})\psi_k\right\|_2+ \left\|\left[(e^{-itB}e^{-itA})^L-e^{-itB}e^{-iTA}\right]\psi_k\right\|_2 
\\
\leq &  C_Te^{-k}(1+k+\log L)
 \le C_T\frac{1+\log L}{L^2},
\end{align*}
where we used Cauchy-Schwarz inequality in the second inequality. Consequently, for every sufficiently large $L$ and all $k\geq\lceil2\log L\rceil$,
\begin{equation}\label{eq:overlap}
 \left|\left\langle e^{-iTH}\psi_k,(e^{-itB}e^{-itA})^L\psi_k\right\rangle\right|
 \le\frac12.
\end{equation}
Therefore, we have
\begin{equation}\label{eq:component-lower}
\left\|
\left[(e^{-itB}e^{-itA})^L-e^{-iTH}\right]\psi_k
\right\|_2^2\geq
2-2\left|
\left\langle
e^{-iTH}\psi_k,(e^{-itB}e^{-itA})^L\psi_k
\right\rangle
\right|\geq 1.
\end{equation}

\smallskip\noindent\emph{Step 3. The logarithmic lower bound.}
The operator $B$ is independent of $z$, so both evolutions act on the
center-of-mass factor of $\psi_k$ as $e^{iT\Delta_z}\chi(\cdot-3ke_1)$.
These factors remain orthonormal. Hence
\begin{equation}\label{eq:orthogonal-error}
\left\|\left[(e^{-itB}e^{-itA})^L-e^{-iTH}\right]\Psi_1\right\|_2^2
 =\frac1{\zeta(3)}\sum_{k=1}^{\infty}\frac1{k^3}
 \left\|\left[(e^{-itB}e^{-itA})^L-e^{-iTH}\right]\psi_k\right\|_2^2.
\end{equation}
The identity follows first for finite sums and then by boundedness of the
evolution operators and convergence in $L^2$. Using
\cref{eq:component-lower}, we obtain
\begin{equation}\label{eq:lower}
 \left\|\left[(e^{-itB}e^{-itA})^L-e^{-iTH}\right]\Psi_1\right\|_2^2
 \ge\frac1{\zeta(3)}\sum_{k\ge\lceil2\log L\rceil}\frac1{k^3}
 \ge\frac{a_T}{\log^2 L}.
\end{equation}
The last estimate is the elementary integral comparison for the series
$\sum k^{-3}$.

For $\rho_1=|\Psi_1\rangle\langle\Psi_1|$, the same orthogonality and \cref{eq:overlap} give
\begin{equation}\label{eq:trotter-full-overlap}
\begin{aligned}
&\left|\left\langle
e^{-iTH}\Psi_1,(e^{-itB}e^{-itA})^L\Psi_1
\right\rangle\right|\\
\le &
\frac1{\zeta(3)}
\left(
\sum_{1\le k<\lceil2\log L\rceil}\frac1{k^3}
+\frac12\sum_{k\ge\lceil2\log L\rceil}\frac1{k^3}
\right)\\
= &
1-\frac1{2\zeta(3)}
\sum_{k\ge\lceil2\log L\rceil}\frac1{k^3}.
\end{aligned}
\end{equation}
The trace-norm identity for two normalized pure states therefore yields
\begin{equation}\label{eq:trotter-trace-lower}
\begin{aligned}
\norm{\mathcal T_t^{\,L}(\rho_1)-\mathcal U_T(\rho_1)}_1
&=2\sqrt{1-\left|\left\langle e^{-iTH}\Psi_1,(e^{-itB}e^{-itA})^L\Psi_1\right\rangle\right|^2}\\
&\ge\left(\frac{2}{\zeta(3)}\sum_{k\ge\lceil2\log L\rceil}\frac1{k^3}\right)^{1/2}
\ge\frac{c_T}{\log L}.
\end{aligned}
\end{equation}
Taking $T=1$ proves the Trotter claim in \cref{thm:barriers-L2}.
\hfill \qed

\subsection{Proof of qDRIFT part of \cref{thm:barriers-L2}}

\noindent\emph{Step 1. Separation of the center-of-mass motion.}
Recall the coordinates
\[
    y=\frac{x_1-x_2}{\sqrt2},
    \quad
    z=\frac{x_1+x_2}{\sqrt2},
\]
and set
\begin{equation*}\label{eq:counterexample-hamiltonian}
    A_y:=-\Delta_y,
    \quad
    A_z:=-\Delta_z,
    \quad
    A=A_y+A_z,
    \quad
    B(y)=\frac{c}{\sqrt2|y|}.
\end{equation*}
Since $B$ depends only on $y$, under the identification
$L^2(\bbr^6)=L^2_z(\bbr^3)\otimes L^2_y(\bbr^3)$, we have
\begin{equation}\label{eq:counterexample-separation}
    e^{-isH}
    =e^{-isA_z}\otimes e^{-is(A_y+B)}.
\end{equation}
In particular, the center-of-mass motion is free and is
unaffected by the Coulomb interaction.

Choose normalized functions $f_k\in L^2_z(\bbr^3)$ whose Fourier
transforms are smooth and 
\[\mathrm{supp}\hat{f}_k\subset\{\xi:|\xi|^2\in I_k\},\] 
where
\[
    I_0=[0,1],
    \quad
    I_k=[4^k,4^k+1],
    \quad k\geq1.
\]
For a fixed normalized $\phi\in\Sc(\bbr^3)$, define
\begin{equation}\label{eq:counterexample-fixed-state}
    \psi(z):=\frac{f_0(z)}{\sqrt2}
    +\frac{\sqrt3}{\pi}\sum_{k=1}^{\infty}\frac{f_k(z)}{k},
    \quad
    \Psi_2(y,z):=\phi(y)\psi(z).
\end{equation}
Orthogonality gives
\[
    \|\Psi_2\|_2^2
    =\|\psi\|_2^2
    =\frac12+\frac3{\pi^2}\sum_{k=1}^{\infty}\frac1{k^2}
    =1.
\]

\smallskip\noindent\emph{Step 2. The random kinetic time.}
Fix $p\in(0,1)$.
Let $K_L$ be the number of kinetic steps among the $L$
independent choices. Then
\begin{equation}\label{eq:counterexample-random-time}
    K_L\sim\operatorname{Bin}(L,p),
    \quad
    \tau_L:=\frac{K_L}{Lp},
    \quad
    \bbe\tau_L=1,
    \quad
    \operatorname{Var}(\tau_L)=\frac{1-p}{Lp}.
\end{equation}
Since
\[
    e^{-itA/p}
    =e^{-itA_z/p}\otimes e^{-itA_y/p},
    \quad
    e^{-itB/(1-p)}
    =I\otimes e^{-itB/(1-p)},
\]
the full random propagator has the form
\[
    V_L(\omega)
    =\left(e^{-itA_z/p}\right)^{K_L}\otimes V_y(\omega)=e^{-i\tau_L(\omega)A_z}\otimes V_y(\omega),
\]
where $V_y(\omega)$ is unitary. Set $U:=e^{-iH}$ and define
\[
    \widetilde V_L:=U^*V_L.
\]
By \cref{eq:counterexample-separation},
\begin{equation}\label{eq:counterexample-relative-propagator}
    \widetilde V_L(\omega)
    =
    e^{-i(\tau_L(\omega)-1)A_z}
    \otimes
    \left(e^{i(A_y+B)}V_y(\omega)\right).
\end{equation}
Note that both factors are unitary.

\smallskip\noindent\emph{Step 3. The test operator.} 
Let $\Psi$ be normalized, set $\rho=|\Psi\rangle\langle\Psi|$, and let $J$ be a bounded self-adjoint operator satisfying $\|J\|_{\op}\leq1$. Then using $\mathcal Q_{1/L,p}^{\,L}(\rho)=\bbe[V_L\rho V_L^*]$, we obtain
\begin{equation}\label{eq:counterexample-quadratic-test}
\begin{aligned}
\left|\bbe\langle\widetilde V_L\Psi,J\widetilde V_L\Psi\rangle-\langle\Psi,J\Psi\rangle\right|
&=  \left|\operatorname{Tr}\left(J\left[U^*\mathcal Q_{1/L,p}^{\,L}(\rho)U-\rho\right]\right)\right|\\
&\leq\left\|U^*\mathcal Q_{1/L,p}^{\,L}(\rho)U-\rho\right\|_1\\
&=\norm{\mathcal Q_{1/L,p}^{\,L}(\rho)-\mathcal U_1(\rho)}_1.
\end{aligned}
\end{equation}
Write
\[
    a_0:=\frac1{\sqrt2},
    \quad
    a_j:=\frac{\sqrt3}{\pi j},
    \quad j\geq1,
\]
so that $\psi=\sum_{j=0}^{\infty}a_jf_j$.
For each $k\geq1$, define
\[
    j_k:=|f_0\rangle\langle f_k|
         +|f_k\rangle\langle f_0|
\]
on $L_z^2$, and set $J_k:=j_k\otimes I$
on $L_z^2\otimes L_y^2$.
By the orthonormality of $f_0$ and $f_k$,
\begin{align*}
    J_k^*
    &=
    \left(
        |f_k\rangle\langle f_0|
        +|f_0\rangle\langle f_k|
    \right)\otimes I
    =J_k,\\
    J_k^*J_k
    &=
    \left(
        |f_0\rangle\langle f_0|
        +|f_k\rangle\langle f_k|
    \right)\otimes I.
\end{align*}
Since $J_k^*J_k$ is a nonzero orthogonal projection,
\[
    \|J_k\|_{\op}^2
    =\|J_k^*J_k\|_{\op}
    =1.
\]
Moreover, the orthonormality of $\{f_j\}_{j\geq0}$
and the normalization of $\phi$ give
\begin{align*}
    \langle\Psi_2,J_k\Psi_2\rangle
    &=
    \left\langle
        \psi\otimes\phi,\,
        (a_kf_0+a_0f_k)\otimes\phi
    \right\rangle\\
    &=
    \left(
        a_k\langle\psi,f_0\rangle
        +a_0\langle\psi,f_k\rangle
    \right)\|\phi\|_2^2\\
    &=2a_0a_k.
\end{align*}
\smallskip\noindent
\emph{Step 4. Spectral representation of the quadratic form.}
By \cref{eq:counterexample-relative-propagator},
\[
    \widetilde V_L\Psi_2
    =
    \left(e^{-i(K_L/(Lp)-1)A_z}\psi\right)
    \otimes
    \left(e^{i(A_y+B)}V_y\phi\right).
\]
By unitarity, for every sequence of random choices,
\[
    \left\|e^{i(A_y+B)}V_y\phi\right\|_2^2
    =\|V_y\phi\|_2^2
    =\|\phi\|_2^2
    =1.
\]
Consequently,
\begin{align*}
    \left\langle
        \widetilde V_L\Psi_2,J_k\widetilde V_L\Psi_2
    \right\rangle
    &=
    \left\langle
        e^{-i(K_L/(Lp)-1)A_z}\psi,\,
        j_k e^{-i(K_L/(Lp)-1)A_z}\psi
    \right\rangle
    \left\|e^{i(A_y+B)}V_y\phi\right\|_2^2\\
    &=
    2\Re\!\left(
        \left\langle
            f_0,e^{-i(K_L/(Lp)-1)A_z}\psi
        \right\rangle
        \left\langle
            e^{-i(K_L/(Lp)-1)A_z}\psi,f_k
        \right\rangle
    \right).
\end{align*}
Let $\mu_j$ be the spectral probability measure of $A_z$
in $f_j$, defined by
\[
    \int F(E)\,d\mu_j(E)
    =
    \int_{\bbr^3}
        F(|\xi|^2)|\widehat f_j(\xi)|^2\,d\xi
\]
for every bounded Borel function $F$.
The measure $\mu_j$ is supported in $I_j$.
Since the free evolution preserves each spectral band
and the bands are disjoint,
\begin{align*}
    \left\langle
        f_j,e^{-i(K_L/(Lp)-1)A_z}\psi
    \right\rangle
    &=
    \sum_{\ell=0}^{\infty}
        a_\ell
        \left\langle
            f_j,e^{-i(K_L/(Lp)-1)A_z}f_\ell
        \right\rangle\\
    &=
    a_j\left\langle
        f_j,e^{-i(K_L/(Lp)-1)A_z}f_j
    \right\rangle\\
    &=
    a_j\int_{I_j}
        e^{-i(K_L/(Lp)-1)\lambda}\,d\mu_j(\lambda).
\end{align*}
The binomial law of $K_L$ gives
\begin{equation}\label{eq:random-clock-multiplier}
    m_{L,p}(s)
    :=\bbe e^{-i(K_L/(Lp)-1)s}
    =
    e^{is}\left(1-p+p e^{-is/(Lp)}\right)^L,
    \quad s\in\bbr.
\end{equation}
Taking expectations in the quadratic form therefore yields
\begin{equation}\label{eq:counterexample-quadratic-coherence}
    \bbe\left\langle
        \widetilde V_L\Psi_2,J_k\widetilde V_L\Psi_2
    \right\rangle=
    2a_0a_k\Re
    \int_{I_0}\int_{I_k}
        m_{L,p}(\lambda-\lambda')\,d\mu_k(\lambda')\,d\mu_0(\lambda).
\end{equation}

\smallskip\noindent
\emph{Step 5. The logarithmic lower bound.}
For each integer $L\geq16$, choose
\[
    k=k(L):=\left\lceil\frac{\log L}{\log16}\right\rceil.
\]
Then
\[
    \sqrt L\leq4^k<4\sqrt L,
    \quad
    k\leq1+\frac{\log L}{\log16}
    \leq\frac{2\log L}{\log16}.
\]
For $\lambda\in I_0$ and $\lambda'\in I_k$, the energy gap
$g:=|\lambda-\lambda'|$ satisfies
\[
    \sqrt L-1\leq g\leq4\sqrt L+1,
    \quad
    g\geq\frac{\sqrt L}{2}.
\]
Since $p\in(0,1)$ is fixed, for all sufficiently large $L$,
depending on $p$, we also have
\[
    \frac{g}{2Lp}
    \leq\frac{4\sqrt L+1}{2Lp}
    \leq\frac{\pi}{2}
\]
uniformly over $\lambda\in I_0$ and $\lambda'\in I_k$.
Using $1-x\leq e^{-x}$ for $x\geq0$ and
$|\sin x|\geq2|x|/\pi$ for $|x|\leq\pi/2$, we obtain
\begin{equation}\label{eq:counterexample-coherence-loss}
\begin{aligned}
    |m_{L,p}(\lambda-\lambda')|
    &=
    \left(
        1-4p(1-p)\sin^2\frac{g}{2Lp}
    \right)^{L/2}\\
    &\leq
    \exp\!\left(
        -2Lp(1-p)\sin^2\frac{g}{2Lp}
    \right)\\
    &\leq
    \exp\!\left(
        -\frac{2(1-p)g^2}{\pi^2Lp}
    \right)\\
    &\leq
    r_p:=\exp\!\left(
        -\frac{1-p}{2\pi^2p}
    \right)<1.
\end{aligned}
\end{equation}
Since $\mu_0$ and $\mu_k$ are probability measures,
\cref{eq:counterexample-quadratic-coherence,eq:counterexample-coherence-loss}
imply
\[
    \langle\Psi_2,J_k\Psi_2\rangle
    -
    \bbe\left\langle
        \widetilde V_L\Psi_2,J_k\widetilde V_L\Psi_2
    \right\rangle
    \geq2a_0a_k(1-r_p).
\]
Applying \cref{eq:counterexample-quadratic-test}
with $\rho_2=|\Psi_2\rangle\langle\Psi_2|$ and $J=J_k$,
we conclude that
\begin{align*}
    \norm{
        \mathcal Q_{1/L,p}^{\,L}(\rho_2)
        -\mathcal U_1(\rho_2)
    }_1
    &\geq2a_0a_k(1-r_p)\\
    &=\frac{\sqrt6(1-r_p)}{\pi k}\\
    &\geq
    \frac{\sqrt6(1-r_p)\log16}{2\pi\log L}.
\end{align*}
This holds for every fixed $p\in(0,1)$ and all sufficiently
large $L$, with the same initial state $\Psi_2$ independent
of both $p$ and $L$.
This proves the qDRIFT claim in \cref{thm:barriers-L2}.
Since $L^\alpha/\log L\to\infty$ for every $\alpha>0$,
no algebraic convergence rate holds on this fixed initial state.
\hfill\qed

\subsection{Proof of \cref{thm:barrier-H2-qdrift}}

\emph{Step 1. Construction of an $H^2$ initial state.}
Let $\phi\in\Sc(\bbr^3)$ be normalized and define
\[
    \Psi_3(y,z):=\phi(y)\psi(z),
    \quad
    \widehat\psi(\xi):=C_\psi(1+|\xi|^2)^{-9/4},
\]
where $C_\psi>0$ is chosen so that $\|\psi\|_2=1$.
Note that
\begin{align*}
    \|\psi\|_{H_z^2}^2
    &=\|\psi\|_2^2+\|-\Delta_z\psi\|_2^2\\
    &=C_\psi^2\int_{\bbr^3}
        (1+|\xi|^4)(1+|\xi|^2)^{-9/2}\,d\xi\\
    &\leq C_\psi^2\int_{\bbr^3}
        (1+|\xi|^2)^{-5/2}\,d\xi
    <\infty.
\end{align*}
Since $\phi$ is Schwartz,
$\Psi_3\in H^2(\bbr^6)=D(H)$.

\smallskip\noindent\emph{Step 2. The size of the high-energy component.}
Let $P_I:=\mathbf1_I(A_z)$ be the spectral projection onto
an energy interval $I$, and set
\[
    a_0:=\|P_{[0,1]}\psi\|_2>0,
    \quad
    a_L:=\|P_{[\sqrt L,2\sqrt L]}\psi\|_2.
\]
By Plancherel's theorem and polar coordinates,
\begin{align*}
    a_L^2
    &=C_\psi^2
    \int_{\sqrt L\leq|\xi|^2\leq2\sqrt L}
        (1+|\xi|^2)^{-9/2}\,d\xi\\
    &=4\pi C_\psi^2
    \int_{L^{1/4}}^{\sqrt2 L^{1/4}}
        r^2(1+r^2)^{-9/2}\,dr\\
    &\asymp
    \int_{L^{1/4}}^{\sqrt2 L^{1/4}}r^{-7}\,dr
    \asymp L^{-3/2}.
\end{align*}
Here $\asymp$ means that the ratio is bounded above and below by positive constants independent of $L$. In particular, $a_L\asymp L^{-3/4}$.

For sufficiently large $L$, define the normalized,
orthogonal vectors
\[
    f_0:=\frac{P_{[0,1]}\psi}{a_0},
    \quad
    f_L:=\frac{P_{[\sqrt L,2\sqrt L]}\psi}{a_L}.
\]
Only the test vector $f_L$ depends on $L$; the initial state
$\Psi_3$ remains fixed.

\smallskip\noindent\emph{Step 3. The trace-norm lower bound.}
Let $\widetilde V_L=e^{iH}V_L$ as in the preceding proof.
Define the bounded self-adjoint operator $J_L$ by
\[
    J_L(g\otimes h)
    :=
    \left(
        \langle f_L,g\rangle f_0
        +\langle f_0,g\rangle f_L
    \right)\otimes h.
\]
Since $f_0$ and $f_L$ are orthonormal, $\|J_L\|_{\op}=1$.
Moreover,
\[
    \langle\Psi_3,J_L\Psi_3\rangle=2a_0a_L.
\]
The energy gaps between the two selected bands satisfy
\[
    \sqrt L-1\leq|\lambda-\lambda'|\leq2\sqrt L.
\]
Thus the argument in \cref{eq:counterexample-coherence-loss}
gives
\[
    |m_{L,p}(\lambda-\lambda')|
    \leq r_p
    :=\exp\!\left(-\frac{1-p}{2\pi^2p}\right)<1
\]
for all sufficiently large $L$.
By \cref{eq:counterexample-relative-propagator},
\[
    \widetilde V_L\Psi_3
    =
    \left(e^{-i(K_L/(Lp)-1)A_z}\psi\right)
    \otimes
    \left(e^{i(A_y+B)}V_y\phi\right).
\]
Using the definition of $J_L$, we obtain
\begin{align*}
    &\langle\widetilde V_L\Psi_3,
             J_L\widetilde V_L\Psi_3\rangle\\
    &\quad=
    2\Re\!\left(
        \langle f_0,e^{-i(K_L/(Lp)-1)A_z}\psi\rangle
        \overline{
            \langle f_L,e^{-i(K_L/(Lp)-1)A_z}\psi\rangle
        }
    \right)
    \left\|e^{i(A_y+B)}V_y\phi\right\|_2^2\\
    &\quad=
    2a_0a_L\Re\!\left(
        \langle f_0,e^{-i(K_L/(Lp)-1)A_z}f_0\rangle
        \overline{
            \langle f_L,e^{-i(K_L/(Lp)-1)A_z}f_L\rangle
        }
    \right).
\end{align*}
Taking expectations and using the spectral representation
as in \cref{eq:counterexample-quadratic-coherence},
together with the bound $|m_{L,p}|\leq r_p$ on the
selected energy gaps, we obtain
\[
    \bbe\langle\widetilde V_L\Psi_3,
                 J_L\widetilde V_L\Psi_3\rangle
    \leq2r_pa_0a_L.
\]
Using \cref{eq:counterexample-quadratic-test}, we obtain, for $\rho_3=|\Psi_3\rangle\langle\Psi_3|$,
\begin{equation*}
\norm{\mathcal Q_{1/L,p}^{\,L}(\rho_3)-\mathcal U_1(\rho_3)}_1
\geq\langle\Psi_3,J_L\Psi_3\rangle
-\bbe\langle\widetilde V_L\Psi_3,J_L\widetilde V_L\Psi_3\rangle
\geq2(1-r_p)a_0a_L
\geq c_pL^{-3/4}.
\end{equation*}
Therefore, on this initial state, the error cannot be $O(L^{-\alpha})$ for any $\alpha>3/4$.
\hfill \qed

\subsection{Trotter formulas of arbitrary fixed order}\label{sec:trotter-all-orders}

In this section, we extend the $L^2$ convergence barrier in \cref{thm:barriers-L2} to Trotter formulas of arbitrary fixed order $k\geq 1$. We also show that their one-step trace-norm error on the hydrogen ground state scales as $t^{5/4}$, much more slowly than the expected $t^{k+1}$, generalizing the local error result for Lie and Strang splitting in~\cite{FangWu2026}.

Fix a Trotter formula of order $k\geq 1$ with finitely many factors and fixed real coefficients.
By first-order consistency, the kinetic coefficients and the potential coefficients each sum to one.
Since the kinetic coefficients sum to one and $B$ depends only on the relative coordinate $y$, the center of mass evolves exactly at every step. We write $P(t)$ for one step on the relative coordinate, where $A_y=-\Delta_y$ and $B=c/(\sqrt{2}|y|)$ as above.

\smallskip\noindent\emph{No algebraic rate for a fixed initial state.}
Fix $L=2^n$ and $t=1/L$. In the $L$ steps, let $s_i$ be the total kinetic time preceding the $i$th potential factor $e^{-itb_iB}$. Since $\sum_i b_i=L$, there is a time $u$ for which $\beta:=\sum_{s_i=u}b_i\ne0$. We use the same functions $\phi,\chi$ as in \cref{eq:summand}, but choose the relative initial state so that the free evolution equals $e^{3k/2}\phi(e^k y)$ at time $u$. Thus the relative initial state is $e^{iuA_y}[e^{3k/2}\phi(e^k\,\cdot)]$. As in \cref{eq:weighted,eq:exact-free}, the difference between its exact and free evolutions at time $1$ has $L^2$ norm bounded by
\begin{equation*}
 C\int_0^1\min\{e^k,e^{-k}/|s-u|\}\,\dd s\le C(1+k)e^{-k}.
\end{equation*}

First keep only the potential factors at $u$. Together they apply the phase $e^{-it\beta B}$ to the concentrated wave function. Before each omitted factor at $s_i\ne u$, the wave function has evolved freely for time $s_i-u$ from $e^{3k/2}\phi(e^k y)$ multiplied by the phases already applied at $u$. These phases preserve its absolute value. By \cref{eq:Hardy_modified} and $|e^{-ia}-1|\le |a|$, the contribution of that factor to the telescoping sum is at most
\begin{equation*}
 \frac{C t|b_i|}{|s_i-u|}\norm{|y|e^{3k/2}\phi(e^k y)}_{L_y^2}
 =\frac{C t|b_i|e^{-k}}{|s_i-u|}\norm{|y|\phi}_{L_y^2}.
\end{equation*}
For fixed $L$, there are only finitely many such terms, so their sum tends to zero as $k\to\infty$. The overlap with the free evolution when only the factors at $u$ are kept is $\int_{\mathbb R^3}e^{-it\beta e^k c/(\sqrt{2}|y|)}|\phi(y)|^2\,\dd y$. It tends to zero by the radial integration used in the proof of \cref{eq:overlap}, since $t\beta c\ne0$. Thus for each $L=2^n$ we can choose $k$ large enough that the corresponding normalized relative initial state, denoted by $f_n$, gives absolute overlap at most $1/2$ between the exact and Trotter outputs.

Now use the single initial state
\[
\Psi(y,z)=(\sqrt{6}/\pi)\sum_{n\ge1}n^{-1} f_n(y)\chi(z-3ne_1).
\]
The translates of $\chi$ are orthonormal and undergo the same free evolution in both outputs. Hence $\Psi$ is normalized, and at $L=2^n$ the absolute overlap is at most $1-3/(\pi^2 n^2)$, by the bound on the $n$th term. The pure-state trace-norm formula in \cref{eq:trotter-trace-lower} gives an error of at least $C/n=C\log 2/\log L$. For any prescribed countable family of formulas, the same construction gives a common initial state by using disjoint translates of $\chi$ for each pair $(j,n)$ and coefficients proportional to $2^{-j}/n$ for the $j$th formula.

For completeness, each fixed formula still converges strongly. On $D(H)$, telescoping its finitely many factors gives the same derivative at $t=0$ as the exact evolution, since the coefficients of both $A$ and $B$ sum to one. The difference quotients are uniformly bounded from $D(H)$ with its graph norm into $L^2$. Their difference therefore tends to zero uniformly on the compact set $\{e^{-isH}\psi_0:0\le s\le1\}$ for each $\psi_0\in D(H)$. Telescoping the $L$ steps gives convergence on $D(H)$, and unitarity and density extend it to $L^2(\mathbb R^6)$.

\smallskip\noindent\emph{The one-step error on the hydrogen ground state.}
For $c<0$, the normalized relative ground state is $f(y)=C e^{c|y|/(2\sqrt{2})}$, with $(A_y+B)f=Ef$. In particular, $\widehat{Bf}(\xi)=C'/(|\xi|^2+c^2/8)$ for a nonzero constant $C'$. Within one step, write $u_jt$ for the total kinetic time preceding the factor $e^{-itb_jB}$. We will show that
\begin{equation*}
 \norm{(P(t)-e^{-itE})f+it e^{-itA_y}D(tA_y)Bf}_{L_y^2}\le Ct^{3/2},
 \quad D(\lambda)=\sum_j b_j e^{iu_j\lambda}-\int_0^1e^{iv\lambda}\,\dd v.
\end{equation*}
Here $D$ is the difference between sampling at the finitely many interaction times and integrating over the step.

To verify the estimate, split $\widehat{Bf}$ at $|\xi|=|s|^{-1}$, where $0<|s|\le1$. From the explicit Fourier transform, the part below this cutoff has gradient norm at most $C|s|^{-1/2}$ and the part above it has $L^2$ norm at most $C|s|^{1/2}$. Free evolution preserves both norms. Using Hardy's inequality on the first part and unitarity on the second gives
\begin{equation*}
 \norm{(e^{-isB}-I)e^{-iaA_y}Bf}_{L_y^2}\le C|s|^{1/2},\quad a\in\mathbb R.
\end{equation*}
Integrating at $a=0$ gives 
\[
\norm{(e^{-isB}-I)f+isBf}_{L_y^2}\le C|s|^{3/2}.
\]
Also, Hardy's inequality and the free Fourier multiplier give 
\[
\norm{B(e^{-iaA_y}-I)f}_{L_y^2}\le C|a|^{1/2}\norm{f}_{H_y^2}.
\]
Thus $(e^{-itb_jB}-I)e^{-itu_jA_y}f$ differs from $-itb_jBf$ by at most $Ct^{3/2}$ in $L^2$. Telescoping $P(t)-e^{-itA_y}$ gives one such term for each $j$. On its leading term $-itb_jBf$, omitting the subsequent potential factors costs at most $Ct^{3/2}$ by the preceding uniform bound. The remaining free evolution is $e^{-it(1-u_j)A_y}$. In the exact Duhamel integral, replacing $e^{-isE}f$ by $f$ costs at most $Ct^2$ after integration. Subtracting the integral from the finite sum gives the asserted estimate.

Since $\sum_j b_j=1$, we have $|D(\lambda)|\le C\min\{|\lambda|,1\}$ for real $\lambda$. Moreover, $D$ cannot vanish on all positive $\lambda$. Otherwise, taking derivatives at zero would give $\sum_j b_j q(u_j)=\int_0^1q(v)\,\dd v$ for every polynomial $q$. Taking $q(v)=\prod_j(v-u_j)^2$ makes the sum zero and the integral strictly positive. By Plancherel's theorem, the change of variables $\xi\mapsto\sqrt{t}\,\xi$, and dominated convergence,
\begin{equation*}
 \lim_{t\downarrow0}t^{-5/2}\norm{tD(tA_y)Bf}_{L_y^2}^2
 =|C'|^2\int_{\mathbb R^3}\frac{|D(|\xi|^2)|^2}{|\xi|^4}\,\dd\xi\ \in(0,\infty).
\end{equation*}
Finally, $|D(\lambda)|\le C|\lambda|$ gives $|\langle f,e^{-itA_y}D(tA_y)Bf\rangle|\le Ct\norm{A_y f}_{L_y^2}\norm{Bf}_{L_y^2}$. Thus the component of $(P(t)-e^{-itE})f$ parallel to $f$ is $O(t^{3/2})$, while its $L^2$ norm is of order $t^{5/4}$. By the pure-state trace-norm formula, the one-step trace-norm error is twice the norm of the component orthogonal to $f$, and hence is of order $t^{5/4}$ as well. The same conclusion holds for the two-body state $f(y)\chi(z)$, since the center-of-mass evolution is exact.

\section{Spatial Error and End-to-end Complexity}\label{sec:end-to-end-app}

We derive the gate bound in \cref{eq:end-to-end-gates}. The spatial approximation is taken from the wave-function estimates of Fang and Zhang~\cite{FangZhang2026}. We prove the time-error estimate needed for the finite implementation and count its gates. Related basis-set estimates were obtained by Berry et al.~\cite{BerryTikkuSimonBabbush2026}, where sharper dynamical estimates are obtained with stronger initial-state assumptions. The end-to-end gate count analysis under such stronger regularities remains an interesting future direction.

\subsection{Problem setup and spatial parameters}

For this section, we specialize to a neutral molecule with $N$ electrons and fixed nuclei at $R_\kappa$, with charges $Z_\kappa>0$ and $\sum_\kappa Z_\kappa=N$. We retain the kinetic normalization of the continuum analysis and use
\begin{equation}\label{eq:complexity-molecular-model}
A=-\sum_{j=1}^N\Delta_{x_j},\quad B=\sum_{1\le j<k\le N}\frac1{|x_j-x_k|}-\sum_{j=1}^N\sum_\kappa\frac{Z_\kappa}{|x_j-R_\kappa|},\quad H=A+B.
\end{equation}
The standard kinetic coefficient $1/2$ changes only numerical constants in the estimates below. We continue to use the interaction indices $(j,k)$, placing the nuclear indices after the electron indices and interpreting $x_{N+\kappa}=R_\kappa$ as fixed parameters. All interaction sums in this section include only pairs containing an electron. Thus $B_{jk}=c_{jk}v_{jk}$, with $c_{jk}=1$ for electron pairs and $c_{j,N+\kappa}=-Z_\kappa$ for electron-nucleus pairs. The nucleus-nucleus energy is a scalar and does not affect the output density operator. In particular,
\begin{equation}\label{eq:complexity-total-charge}
\sum_{j<k}|c_{jk}|=\frac{N(N-1)}2+N^2\asymp N^2,\quad \pi_{jk}=\frac{|c_{jk}|}{\sum_{a<b}|c_{ab}|}.
\end{equation}

We assume $T\ge1$, $0<\epsilon\le1$, and
\begin{equation}\label{eq:complexity-input-assumptions}
\norm{\psi_0}_2=1,\quad \norm{\psi_0}_{H_x^2}\le C,\quad \norm{|x|\psi_0}_2\le CN^{3/2},\quad \max_\kappa|R_\kappa|\le C,
\end{equation}
where these constants are independent of $N,T,\epsilon$. No antisymmetry assumption is used. For the circuit comparison we adopt the molecular input model of~\cite{FangZhang2026,SuBerryWiebeRubinBabbush2021}: nuclear charges are positive integers, so there are at most $N$ nuclei, and nuclear positions are available as classical data to the required precision. The sampled interaction labels and insertion times are classical. Gates may act between arbitrary registers. State preparation, final measurement, and classical preprocessing are excluded from the quantum gate count.

Let $G_\ell$ be the mean-zero periodic Coulomb Green function on the cubic torus of side length $\ell$, and let $v_{jk,\ell}$ be the corresponding periodic interaction kernels. Let $p_m$ project onto the one-particle plane waves with momentum indices in $\{-m,\ldots,m\}^3$, and let $P_m=p_m^{\otimes N}$. The one-particle basis size and the first omitted physical momentum are
\begin{equation}\label{eq:complexity-basis}
M=(2m+1)^3,\quad \Lambda_m=\frac{2\pi(m+1)}{\ell}.
\end{equation}
The finite operators act on the range of $P_m$:
\begin{equation}\label{eq:complexity-finite-operators}
A_m=P_mAP_m,\quad v_{jk,m}=P_mv_{jk,\ell}P_m,\quad B_m=\sum_{j<k}c_{jk}v_{jk,m},\quad H_m=A_m+B_m.
\end{equation}
In finite space, we write 
$\|f\|_{2,\mathbb T_\ell}:=\|f\|_{L^2(\mathbb T_\ell^{3N})}$, as in the notation of~\cite{FangZhang2026}.

By~\cite[Theorem 16 and Corollary 17]{FangZhang2026}, sufficiently large constants in the choices
\begin{equation}\label{eq:complexity-spatial-parameters}
\rho\asymp\ell\asymp TN^{3/2}\epsilon^{-1},\quad \Lambda_m\asymp T^{1/2}N^3\epsilon^{-3/4},\quad M=\Or(T^{9/2}N^{27/2}\epsilon^{-21/4})
\end{equation}
control the finite-domain and finite-basis errors. Here $\rho$ is the localization radius. The constants in $\rho\asymp\ell$ are chosen so that the localized state and nuclei lie in the interior region required by the spatial theorem. If $F_\rho$ is its smooth cutoff and $\operatorname{Per}_\ell$ denotes periodization, the finite input is
\begin{equation}\label{eq:complexity-finite-input}
\psi_m=\frac{P_m\operatorname{Per}_\ell(F_\rho\psi_0)}{\norm{P_m\operatorname{Per}_\ell(F_\rho\psi_0)}_{2,\mathbb T_\ell}}.
\end{equation}
After the known scalar phase correction $c_\ell$ and zero extension from the centered cell (denoted $\operatorname{Ext}_\ell$), the exact finite evolution satisfies
\begin{equation}\label{eq:complexity-spatial-error}
\norm{\operatorname{Ext}_\ell e^{-iT(H_m-c_\ell I)}\psi_m-e^{-iTH}\psi_0}_{L^2_x}\le\frac{\epsilon}{12}.
\end{equation}
Normalization changes the unnormalized spatial error by at most a factor of two, by the reverse triangle inequality. The constants in \cref{eq:complexity-spatial-parameters} absorb this factor and the chosen error budget. Moreover, $\norm{\nabla\psi_m}_{2,\mathbb T_\ell}=\Or(1)$: localization costs $\Or(\rho^{-1})$ in the gradient norm, the Fourier projection contracts that norm, and the normalization denominator is bounded below. The scalar phase cancels in the output density operator. The spatial estimate is used for the exact finite evolution; it is not applied separately to the randomized interaction factors.

\subsection{Time error for the finite implementation}

The finite algorithm uses independent copies of the step
\begin{equation}\label{eq:complexity-finite-step}
S_{jk,\theta,m}(t)=e^{-i\theta tA_m}e^{-itc_{jk}v_{jk,m}/\pi_{jk}}e^{-i(1-\theta)tA_m},\quad t=T/L,
\end{equation}
with the same probabilities and uniform insertion parameter as in the continuum algorithm. We first verify that the smoothing argument remains available after projection.

\begin{lemma}\label{lem:complexity-finite-free-smoothing}
For every interaction $(j,k)$ and every $f$ in the finite space,
\begin{equation}\label{eq:complexity-finite-free-smoothing}
\int_0^T\norm{v_{jk,m}e^{-isA_m}f}_{2,\mathbb T_\ell}^2\dd s\le C\left[1+T\left(\frac{\Lambda_m}{\ell}+\ell^{-2}\right)\right]\norm{f}_{2,\mathbb T_\ell}^2,
\end{equation}
where constant $C$ is independent of $N,\ell,m$ and the nuclear positions. The same estimate holds with the opposite sign of the free evolution. 
\end{lemma}

\begin{proof}
In a coordinate chart containing the Coulomb singularity, write $G_\ell(y)=\chi_\ell(y)/|y|+r_\ell(y)$, where $\chi_\ell$ is supported inside the chart, equals one near the singularity, and $\norm{r_\ell}_\infty\le C/\ell$. Take a second cutoff equal to one on the support of $\chi_\ell$ and lift the localized free solution to $\mathbb R^3$. Its equation has the free Laplacian and a forcing term given by the cutoff commutator. Since the active momenta are bounded by $C\Lambda_m$, this forcing has norm at most $C(\Lambda_m/\ell+\ell^{-2})\norm f_{2,\mathbb T_\ell}$. Applying whole-space free smoothing to the initial term and to Duhamel's formula, followed by Minkowski's inequality, bounds the squared weighted norm on an interval of length $\tau$ by
\begin{equation}
C\left[\left(1+\tau\left(\frac{\Lambda_m}{\ell}+\ell^{-2}\right)\right)^2+\tau\ell^{-2}\right]\norm f_{2,\mathbb T_\ell}^2.
\end{equation}
The last term controls the bounded remainder $r_\ell$.

For an electron pair, use the torus coordinates $y=x_j-x_k$ and $z=x_k$. Fourier transformation in $z$ reduces the active kinetic operator to $2(-i\nabla_y-\xi/2)^2+|\xi|^2/2$. A gauge transformation on the localized whole-space lift removes the constant momentum shift; it need not be periodic. The product momentum cutoff gives $|\xi|\le C\Lambda_m$, so the same cutoff-commutator bound applies. The other particle coordinates contribute only unitary factors. Finally, $P_m$ is a contraction and commutes with free evolution. Partitioning the time interval into pieces of length at most $c(\Lambda_m/\ell+\ell^{-2})^{-1}$ proves the estimate.
\end{proof}

Let $E_{t,m}=\mathbb E S_{1,m}$ denote the mean step. To estimate its error, introduce the bounded operator
\begin{equation}\label{eq:complexity-effective-potential}
B_{t,m}=\frac{i}{t}\left(\sum_{j<k}\pi_{jk}e^{-itc_{jk}v_{jk,m}/\pi_{jk}}-I\right).
\end{equation}
This operator is used only in the proof. Averaging the insertion time gives
\begin{equation}\label{eq:complexity-mean-step}
E_{t,m}=e^{-itA_m}-i\int_0^t e^{-i(t-s)A_m}B_{t,m}e^{-isA_m}\dd s,\quad \norm{E_{t,m}}\le1.
\end{equation}

\begin{lemma}\label{lem:complexity-finite-time-error}
For $t\le c(N^4+\Lambda_m/\ell+\ell^{-2})^{-1}$ and normalized $\psi_m$, the finite mean error satisfies
\begin{equation}\label{eq:complexity-finite-time-error}
\begin{aligned}
\norm{(E_{t,m}^L-e^{-iTH_m})\psi_m}_{2,\mathbb T_\ell}
&\le CtN^3\left[1+T\left(N^4+\frac{\Lambda_m}{\ell}+\ell^{-2}\right)\right]^{1/2}\\
&\quad\times\left(\sum_{j<k}|c_{jk}|\int_0^T\norm{v_{jk,m}e^{-isH_m}\psi_m}_{2,\mathbb T_\ell}^2\dd s\right)^{1/2}.
\end{aligned}
\end{equation}
\end{lemma}

\begin{proof}
All operators in this proof act on the finite space, and we suppress its norm subscript. Functional calculus for each self-adjoint $v_{jk,m}$, the scalar exponential remainder, and \cref{eq:complexity-total-charge} give
\begin{equation}\label{eq:complexity-phase-bounds}
\begin{aligned}
\norm{B_mf}^2,\ \norm{B_{t,m}f}^2,\ \norm{B_{t,m}^*f}^2&\le CN^2\sum_{j<k}|c_{jk}|\norm{v_{jk,m}f}^2,\\
|\langle g,(B_{t,m}-B_m)f\rangle|&\le CtN^2\sum_{j<k}|c_{jk}|\norm{v_{jk,m}g}\norm{v_{jk,m}f}.
\end{aligned}
\end{equation}
The second estimate is a form bound, obtained term by term. It does not assume that the compressed interactions commute.

For a test vector $\phi$, define the adjoint free interpolation by $\widetilde\phi(s)=e^{i((q+1)t-s)A_m}(E_{t,m}^*)^{L-1-q}\phi$ on $qt\le s<(q+1)t$, $q=0,\ldots,L-1$. The time reversal of this interpolation satisfies the integral equation obtained by iterating \cref{eq:complexity-mean-step} with $A_m,B_{t,m}$ replaced by $-A_m,-B_{t,m}^*$. On a block of length $\tau$ with $\tau(\Lambda_m/\ell+\ell^{-2})\le c$, \cref{lem:complexity-finite-free-smoothing}, Minkowski, and \cref{eq:complexity-phase-bounds} bound its largest interaction-weighted spacetime norm by the norm of its starting vector plus $CN^2\sqrt\tau$ times that same spacetime norm. Choose $\tau\asymp(N^4+\Lambda_m/\ell+\ell^{-2})^{-1}$ so this term can be absorbed. Blocks can be chosen to contain whole steps by the assumed restriction on $t$. Contractivity of $E_{t,m}^*$ bounds every starting vector by $\norm\phi$. Summing the squared estimates over the blocks gives
\begin{equation}\label{eq:complexity-adjoint-smoothing}
\int_0^T\norm{v_{jk,m}\widetilde\phi(s)}^2\dd s\le C\left[1+T\left(N^4+\frac{\Lambda_m}{\ell}+\ell^{-2}\right)\right]\norm\phi^2.
\end{equation}

Write $\psi_m(s)=e^{-isH_m}\psi_m$. Telescoping the averaged steps and using Duhamel's formula yields the exact identity
\begin{equation}\label{eq:complexity-telescoping}
\begin{aligned}
\langle\phi,(E_{t,m}^L-e^{-iTH_m})\psi_m\rangle
= &-i\int_0^T\langle\widetilde\phi(s),(B_{t,m}-B_m)\psi_m(s)\rangle\dd s\\
&+\sum_{q=0}^{L-1}\int_{qt}^{(q+1)t}\int_{qt}^s\langle\widetilde\phi(s),B_{t,m}e^{-i(s-r)A_m}B_m\psi_m(r)\rangle\dd r\dd s.
\end{aligned}
\end{equation}
For the first term, apply the second estimate in \cref{eq:complexity-phase-bounds}, Cauchy-Schwarz in time and in the interaction indices, and \cref{eq:complexity-adjoint-smoothing}. The coefficient sum contributes at most $CN^2$.

For the second term, move $B_{t,m}$ onto the test vector. The integration operator within each step has $L^2$-in-time norm at most $t$, and therefore
\begin{equation}\label{eq:complexity-within-step}
\sum_{q=0}^{L-1}\int_{qt}^{(q+1)t}\left\|\int_{qt}^s e^{-i(s-r)A_m}B_m\psi_m(r)\dd r\right\|^2\dd s\le Ct^2N^2\sum_{j<k}|c_{jk}|\int_0^T\norm{v_{jk,m}\psi_m(s)}^2\dd s.
\end{equation}
Also, \cref{eq:complexity-phase-bounds,eq:complexity-adjoint-smoothing} imply
\begin{equation}
\norm{B_{t,m}^*\widetilde\phi}_{L^2(0,T;L^2)}\le CN^2\left[1+T\left(N^4+\frac{\Lambda_m}{\ell}+\ell^{-2}\right)\right]^{1/2}\norm\phi.
\end{equation}
Thus both terms in \cref{eq:complexity-telescoping} satisfy the claimed bound times $\norm\phi$. Taking the supremum over unit test vectors proves the lemma.
\end{proof}

We now use the initial-state assumption. Periodic Hardy's inequality, projection contraction, and charge neutrality give
\begin{equation}\label{eq:complexity-hardy}
\sum_{j<k}|c_{jk}|\norm{v_{jk,m}f}_{2,\mathbb T_\ell}^2\le CN\norm{\nabla f}_{2,\mathbb T_\ell}^2+CN^2\ell^{-2}\norm f_{2,\mathbb T_\ell}^2.
\end{equation}
For electron pairs, the derivative contribution follows by summing the relative-coordinate Hardy estimates and using $\sum_{j<k}\norm{(\nabla_{x_j}-\nabla_{x_k})f}^2\le N\norm{\nabla f}^2$. For electron-nucleus terms, summing the charges gives the same factor $N$. Cauchy-Schwarz then bounds the potential form on a normalized $f$ by $CN^{3/2}\norm{\nabla f}+CN^2/\ell$. Conservation of the exact finite energy and completion of the square imply
\begin{equation}\label{eq:complexity-energy}
\sup_s\norm{\nabla e^{-isH_m}\psi_m}_{2,\mathbb T_\ell}^2\le C\left(\norm{\nabla\psi_m}_{2,\mathbb T_\ell}^2+N^3+\frac{N^2}{\ell}\right)=\Or(N^3).
\end{equation}
Consequently,
\begin{equation}\label{eq:complexity-weighted-orbit}
\sum_{j<k}|c_{jk}|\int_0^T\norm{v_{jk,m}e^{-isH_m}\psi_m}_{2,\mathbb T_\ell}^2\dd s\le CTN^4.
\end{equation}
No regularity of the individual randomized intermediate states is used here.

The spatial choices in \cref{eq:complexity-spatial-parameters} satisfy
\begin{equation}\label{eq:complexity-cutoff-correction}
\frac{\Lambda_m}{\ell}+\ell^{-2}=\Or(T^{-1/2}N^{3/2}\epsilon^{1/4}+T^{-2}N^{-3}\epsilon^2).
\le CN^4.
\end{equation}
Substituting \cref{eq:complexity-weighted-orbit,eq:complexity-cutoff-correction} into \cref{lem:complexity-finite-time-error}, with $T\ge1$, gives
\begin{equation}\label{eq:complexity-N7-mean}
\norm{(E_{t,m}^L-e^{-iTH_m})\psi_m}_{2,\mathbb T_\ell}\le CtN^3(\sqrt{T}N^2)(\sqrt{T}N^2)=\frac{CT^2N^7}{L}.
\end{equation}
The same argument on whole space, using the free smoothing estimate without the cutoff correction, gives the continuum state-dependent bound $CT^2N^7\norm{\psi_0}_{H_x^2}/L$ for normalized $H^2$ initial data. This refinement uses initial kinetic-energy control; it does not replace the unrestricted continuum diamond-norm bound.

For $S_m=S_{L,m}\cdots S_{1,m}$, independence gives $\mathbb E S_m=E_{t,m}^L$. The mean-to-channel estimate is
\begin{equation}\label{eq:complexity-mean-to-channel}
\left\|\mathbb E|S_m\psi_m\rangle\langle S_m\psi_m|-|e^{-iTH_m}\psi_m\rangle\langle e^{-iTH_m}\psi_m|\right\|_1\le4\norm{(E_{t,m}^L-e^{-iTH_m})\psi_m}_{2,\mathbb T_\ell}.
\end{equation}
Indeed, subtract and add the rank-one operator formed from the mean vector. The covariance term is positive with trace $1-\norm{E_{t,m}^L\psi_m}^2$, at most twice the vector error; the remaining rank-one difference has trace norm at most twice that error.
Hence
\begin{equation}\label{eq:complexity-steps}
L=\left\lceil CT^2N^7\epsilon^{-1}\right\rceil
\end{equation}
satisfies the small-step condition and allocates any fixed fraction of the tolerance to time error, after increasing $C$. The spatial vector error in \cref{eq:complexity-spatial-error} contributes at most twice its size in trace norm.

\subsection{Circuit implementation and gate count}

The direct product contains $L$ sampled interaction exponentials and, after combining adjacent free segments, at most $L+1$ full kinetic exponentials. Its gate cost is reduced by an exact rearrangement. For the classical insertion times $s_j=(j-\theta_j)t$,
\begin{equation}\label{eq:complexity-kinetic-cancellation}
S_m=e^{-iTA_m}\prod_{j=L}^{1}\left(e^{is_jA_m}e^{-itc_{\alpha_j}v_{\alpha_j,m}/\pi_{\alpha_j}}e^{-is_jA_m}\right),
\end{equation}
where $\alpha_j$ denotes the interaction pair sampled at step $j$. Each interaction acts on one or two electron registers. Inside each conjugation, kinetic factors on all other registers commute through the interaction and cancel. The resulting circuit therefore uses $L$ selected interaction exponentials, $2L$ kinetic exponentials on at most two electron registers, and one full kinetic exponential. The kinetic exponential is diagonal in the momentum basis, with a phase given by a sum of $N$ quadratic one-particle phases. Reversible arithmetic and phase synthesis implement a full $e^{-isA_m}$ with $\widetilde{\Or}(N)$ gates, whereas a kinetic exponential acting on one or two particle registers costs $\widetilde{\Or}(1)$ gates. The suppressed factors are polynomial in $\log(2m+1)$ and the number of precision bits; the magnitude of the kinetic energies affects the required phase precision only logarithmically. After cancellation, the circuit contains $2L$ kinetic factors acting on at most two particle registers and one full kinetic exponential $e^{-iTA_m}$. Thus the total kinetic gate cost is $\widetilde{\Or}(L+N)$.

For a fixed classical interaction label, specialize the momentum-transfer construction of Su et al.~\cite[Sections II C-II D]{SuBerryWiebeRubinBabbush2021}. The corresponding block encoding of $v_{jk,m}$ has normalization
\begin{equation}\label{eq:complexity-pair-normalization}
\lambda_{\mathrm{pair}}=\Or\!\left(\frac1\ell\sum_{0<|\nu|_\infty\le2m}\frac1{|\nu|^2}\right)=\Or(\Lambda_m).
\end{equation}
The momentum-transfer state is prepared using the nested-box construction, and controlled momentum additions act only on the selected registers. An electron-nucleus term additionally uses the classical position $R_\kappa$ in its phase. Boundary checks reject transfers outside the basis: they implement the Galerkin compression rather than a cyclic wraparound. Since the interaction label is classical, no coherent selection or routing over all electrons and nuclei is required. The gate cost of this specialized encoding is polylogarithmic in the basis size and precision parameters.

One implementation of the sampled interaction exponential uses qubitization of its momentum-transfer block encoding~\cite{LowChuang2019}. Since $|c_{jk}|/\pi_{jk}=\Or(N^2)$, this implementation costs $\widetilde{\Or}(1+tN^2\Lambda_m)$ gates per sampled interaction, which reduces to $\widetilde{\Or}(1)$ with the parameters in \cref{eq:complexity-spatial-parameters,eq:complexity-steps}.

Choose error $\Or(\epsilon/L)$ for each compiled factor on zero-initialized auxiliary registers, including their full output. If the implementation controls only the unitary's signal block, require block error $\Or((\epsilon/L)^2)$, which also controls leakage at order $\epsilon/L$ and changes only logarithmic factors. Rounding each insertion parameter to $b$ bits causes at most $CT\Lambda_m^2 2^{-b}$ accumulated vector error, because only one or two registers occur in each kinetic conjugation. Thus $b=\Or(\log(1+T\Lambda_m^2/\epsilon))$ suffices. Sampling each discrete interaction label to total variation error $\Or(\epsilon/L)$ controls the joint label law to $\Or(\epsilon)$ and hence its contribution to averaged trace-norm error. Nuclear-data precision, arithmetic, and gate synthesis require only logarithmically many precision bits. Allocating suitably small fixed fractions of $\epsilon$ to these errors, time error, and spatial error proves the overall trace-norm guarantee.

Summing the kinetic and interaction costs gives
\begin{equation}\label{eq:complexity-gates-before-substitution}
G=\widetilde{\Or}(N+L+TN^2\Lambda_m).
\end{equation}
Plugging in the spatial parameter choices as in \cref{eq:complexity-spatial-parameters,eq:complexity-steps},
\begin{equation}\label{eq:complexity-final-substitution}
G=\widetilde{\Or}(N+T^2N^7\epsilon^{-1}+T^{3/2}N^5\epsilon^{-3/4})=\widetilde{\Or}(T^2N^7\epsilon^{-1}).
\end{equation}
The last step uses $T,N,\epsilon^{-1}\ge1$. The system register contains $3N\lceil\log_2(2m+1)\rceil$ qubits; the reusable arithmetic and simulation workspace is polylogarithmic. In particular, the total qubit count is $\widetilde{\Or}(N)$. The basis size enters the per-gate precision logarithmically, while its polynomial contribution through $\Lambda_m$ is displayed explicitly in \cref{eq:complexity-gates-before-substitution}.

\subsection{Comparison under the same spatial assumptions}

We use the same spatial parameters and normalized input class for every row of \cref{tab:complexity}. A deterministic vector error at most $\epsilon/2$ implies trace-norm error at most $\epsilon$, so using output trace norm throughout changes only constants in the bounds of~\cite{FangZhang2026}.

The first-quantized full-Hamiltonian block encoding has kinetic normalization $\Or(N\Lambda_m^2)$ and Coulomb normalization $\Or(N^2\Lambda_m)$, with $\widetilde{\Or}(N)$ gates per query~\cite{SuBerryWiebeRubinBabbush2021}. Direct qubitization therefore gives
\begin{equation}\label{eq:complexity-qubitization-comparison}
G_{\mathrm{qub}}=\widetilde{\Or}(N+TN^2\Lambda_m^2+TN^3\Lambda_m)=\widetilde{\Or}(T^2N^8\epsilon^{-3/2}),
\end{equation}
in agreement with~\cite[Corollary 21]{FangZhang2026}. The interaction-picture algorithm treats the kinetic evolution separately~\cite{LowWiebe2019,SuBerryWiebeRubinBabbush2021}, giving
\begin{equation}\label{eq:complexity-IP-comparison}
G_{\mathrm{IP}}=\widetilde{\Or}(N+TN^3\Lambda_m)=\widetilde{\Or}(T^{3/2}N^6\epsilon^{-3/4}),
\end{equation}
as in~\cite[Corollary 23]{FangZhang2026}.

For a fixed formula order $p$, the Galerkin nested-commutator estimate and product-formula analysis in~\cite[Section 6.2.1]{FangZhang2026}, using~\cite{ChildsSuTranEtAl2020}, give a sufficient step count
\begin{equation}\label{eq:complexity-trotter-steps}
L_p=\Or_p\!\left(1+T^{1+1/p}\epsilon^{-1/p}\left\{N^2\Lambda_m^3[\Lambda_m^2+N\Lambda_m]^{p-1}\right\}^{1/p}\right).
\end{equation}
Compiling the full-potential and kinetic factors costs $\widetilde{\Or}_p(NL_p+TN^3\Lambda_m)$ gates. Since the sufficient cutoff in \cref{eq:complexity-spatial-parameters} satisfies $\Lambda_m\gtrsim N$, its leading step contribution is $T^{1+1/p}\epsilon^{-1/p}N^{1+2/p}\Lambda_m^{2+1/p}$. Substitution gives
\begin{equation}\label{eq:complexity-trotter-comparison}
G_{\mathrm{Trot},p}=\widetilde{\Or}_p(T^{2+3/(2p)}N^{7+5/p}\epsilon^{-3/2-7/(4p)}),
\end{equation}
which is~\cite[Theorem 27]{FangZhang2026}. In particular, $p=1$ gives $\widetilde{\Or}(T^{7/2}N^{12}\epsilon^{-13/4})$. The order is fixed; the hidden constants need not be uniform as $p$ increases.

Thus \cref{eq:complexity-final-substitution} improves the stated upper bounds for direct qubitization and fixed-order Galerkin Trotter formulas. It does not improve the interaction-picture bound in \cref{eq:complexity-IP-comparison}. The comparison concerns the explicitly stated input class and implementation model, and does not assert optimality among all algorithms.

\subsection{Dynamical nuclei}\label{subsec:complexity-dynamical-nuclei}

We now allow the nuclei to move and derive the gate count in \cref{tab:complexity} for the joint electron-nuclear dynamics. Let $N=\eta+N_{\mathrm{nuc}}$, where $\eta$ is the number of electrons. Write $x=(x_1,\ldots,x_N)$ for the joint coordinates, with $x_{\eta+\kappa}=R_\kappa$. In atomic units, each electron has charge $q_j=-1$ and mass $m_j=1$, and nucleus $\kappa$ has charge $q_{\eta+\kappa}=Z_\kappa$ and mass $m_{\eta+\kappa}=M_\kappa$. The Hamiltonian is $H=A+B$, with
\[
 A=-\sum_{j=1}^{N}a_j\Delta_{x_j},\quad
 a_j=\frac{1}{2m_j},\quad
 B=\sum_{1\le j<k\le N}\frac{q_jq_k}{|x_j-x_k|}.
\]
Thus the sampled pairs include the nucleus-nucleus interactions $Z_\kappa Z_\lambda/|R_\kappa-R_\lambda|$. Under neutrality, $\sum_\kappa Z_\kappa=\eta$, the total absolute interaction weight and sampling probabilities are
\[
 W=\binom{\eta}{2}+\eta^2+\sum_{\kappa<\lambda} Z_\kappa Z_\lambda
 \asymp N^2,\quad \pi_{jk}=\frac{|q_jq_k|}{W},
\]
since the positive integer charges give $\eta\le N\le2\eta$.

For the full molecular Hamiltonian we assume $Z_\kappa\le Z_*$ and $0<M_-\le M_\kappa\le M_+$, with fixed $Z_*,M_-,M_+$. Let the joint initial state be normalized, with $\norm{\psi_0}_{H^2_x}=\Or(1)$ and $\norm{|x|\psi_0}_{L^2_x}=\Or(N^{3/2})$.
By~\cite[Theorem~38 and Corollary~39]{FangZhang2026}, the choices in \cref{eq:complexity-spatial-parameters} give \cref{eq:complexity-spatial-error} with the cutoff on the joint coordinates and $P_m=p_m^{\otimes N}$. Here the periodic scalar correction is $c_\ell=(c_{\rm cub}/\ell)\sum_{j<k}q_jq_k$, where $c_{\rm cub}$ is the regular part of $G_1$ at the origin.

To apply \cref{lem:complexity-finite-free-smoothing} to particles of different masses, use the same coordinates $y=x_j-x_k$, $z=x_k$. At momentum $\xi$ dual to $z$, their kinetic operator is
\[
 (a_j+a_k)\left(-\I\nabla_y-\frac{a_k}{a_j+a_k}\xi\right)^2
 +\frac{a_ja_k}{a_j+a_k}|\xi|^2 .
\]
On the localized lift the same gauge transformation removes the shift. Rescaling time by $a_j+a_k$ gives the free smoothing estimate with a factor $(a_j+a_k)^{-1}$ in the squared bound. The mass bounds and $|\xi|\le C\Lambda_m$ give the same cutoff commutator estimate as in the proof of \cref{lem:complexity-finite-free-smoothing}. Hence \cref{lem:complexity-finite-time-error} applies with constants independent of $N$.

For each $j$, $\sum_{k\ne j}|q_jq_k|\le 2\eta|q_j|\le CN$. Applying periodic Hardy in $x_j$ with $x_k$ fixed and summing with these weights gives \cref{eq:complexity-hardy} on the joint space. Since $\langle f,A_m f\rangle\ge(\min_j a_j)\norm{\nabla f}_2^2$, energy conservation gives \cref{eq:complexity-energy,eq:complexity-weighted-orbit} as before. Substitution in \cref{lem:complexity-finite-time-error} and the mean-to-channel estimate give the same $CT^2N^7/L$ joint trace-norm error under the step-size condition of \cref{lem:complexity-finite-time-error}. On the whole space, free smoothing and the weighted Hardy inequality give the same bound for these initial states.

For every sampled pair, the block encoding transfers $+\nu$ and $-\nu$ to the two quantum momentum registers and checks both basis boundaries. The nuclear masses and charges are classical data. Each kinetic phase uses the coefficient $a_j$, and the cancellation in \cref{eq:complexity-kinetic-cancellation} again leaves at most two active registers per conjugation. With $|q_jq_k|/\pi_{jk}=W$ and $\lambda_{\rm pair}=\Or(\Lambda_m)$, the gate count in \cref{eq:complexity-gates-before-substitution} is $\widetilde{\Or}(N+L+TN^2\Lambda_m)=\widetilde{\Or}(T^2N^7/\epsilon)$ with $L$ as in \cref{eq:complexity-steps}. With the same spatial and compilation error budgets, the averaged joint output has trace-norm error at most $\epsilon$. The qubitization, interaction-picture, and compiled Trotter bounds for the same model follow from~\cite[Theorems~40, 41, and~44]{FangZhang2026}, respectively.

\section{Numerical Details}\label{sec:numerical-details}

\subsection{Comparison with Trotter formulas}

We consider the two-body Hamiltonian $H=A+B$, with $A=-\Delta_y-\Delta_z$ and $B=-2/|y|$, where $y$ and $z$ are the relative and center-of-mass coordinates in $\mathbb R^3$. We fix $T=1$ and use $L=64,256,1024,4096,16384,65536$ steps of size $t=T/L$. We compare Lie-Trotter splitting, Strang splitting, and the fourth- and eighth-order recursive Suzuki formulas with our algorithm, which draws $u$ uniformly from $[0,t]$ at each step and applies $e^{-i(t-u)A}e^{-itB}e^{-iuA}$. The draws are independent across steps.

All five methods use the same normalized initial state
\begin{equation}\label{eq:numerical-initial-superposition}
\Psi_0=\tfrac12(\Psi_1+\Psi_2+\Psi_4+\Psi_8).
\end{equation}
Each $\Psi_p$ is normalized separately. The first family consists of $20$ shell packets focused at time $1$, and each of the other three consists of $8$ Gaussian packets. The four families are given by
\begin{equation}\label{eq:numerical-packet-families}
\begin{aligned}
\Psi_1(y,z)&\propto\sum_{k=1}^{20}k^{-3/2}\bigl(e^{-i\Delta_y}[e^{3k/2}\phi(e^k\,\cdot)]\bigr)(y)\,\chi_{1,k}(z),\\
\Psi_p(y,z)&\propto\sum_{k=1}^{8}k^{-1}\bigl(e^{-i\tau_{p,k}\Delta_y}g_{\sigma_{p,k}}\bigr)(y)\,\chi_{p,k}(z),\quad p=2,4,8.
\end{aligned}
\end{equation}
Here $\phi$ is the normalized projection onto the first $512$ radial harmonic-oscillator modes of the shell profile $\exp\{-1/[(r-1)(2-r)]\}$ on $1<r<2$, extended by zero elsewhere, and $g_\sigma(y)=(\pi\sigma^2)^{-3/4}e^{-|y|^2/(2\sigma^2)}$. The Gaussian widths and focusing times are $\sigma_{p,k}=0.05\,b_p4^{-k}$ and $\tau_{p,k}=1-b_p4^{-k}/2$, where $b_2=1$, $b_4=a_4$, $b_8=a_4a_6a_8$, and $a_q=(4-4^{1/(q-1)})^{-1}$. The $\chi_{p,k}$ are smooth, orthonormal center-of-mass states. Their free evolution is common to all methods, so only the relative coordinate requires a spatial basis. The initial state and the number of basis functions for each packet are fixed across all methods and step counts.

For the relative coordinate, we use moving radial harmonic-oscillator bases with $512$ modes for shell packets and $256$ for Gaussian packets. Free propagation between these moving bases is exact, while the Coulomb factors are evaluated using the corresponding finite Galerkin matrices. For the Trotter formulas, the final Coulomb multiplication is evaluated by oscillatory radial quadrature without projecting its output back into the basis. The reference state $\Psi_{\mathrm{ref}}$ is obtained by integrating the transformed Galerkin Schr\"odinger equation with an adaptive eighth-order Runge-Kutta solver (DOP853), with relative tolerance $2\times10^{-11}$ and absolute tolerance $2\times10^{-13}$. The reference integrator uses the variable $\operatorname{arsinh}(2(s-\tau)/\sigma^2)$, where $s$ is physical time and $(\sigma,\tau)$ are the packet width and focusing time, with maximum transformed step size $0.12$. For large $L$, the classical evaluation groups nearly commuting factors with an error budget per packet of $10^{-8}$ for our algorithm and $10^{-7}$ for the Trotter formulas.

For each method and each $L$, we plot the full trace-norm error $\|\rho_L-|\Psi_{\mathrm{ref}}\rangle\langle\Psi_{\mathrm{ref}}|\|_1$ on $\Psi_0$. For a Trotter formula with propagator $S$, the output is $\rho_L=|S\Psi_0\rangle\langle S\Psi_0|$. For our algorithm, we estimate $\rho_L=\mathbb E[|S\Psi_0\rangle\langle S\Psi_0|]$ using the numerically computed mean $m_L=\mathbb E[S\Psi_0]$ and the sample covariance of $128$ independent trajectories, rescaled to trace $1-\|m_L\|_2^2$.

For \cref{fig:unitary-comparison}, we count each exponential $e^{-isA}$ or $e^{-isB}$ as one unitary and combine adjacent exponentials of $A$. The counts are per circuit realization, excluding state preparation and measurement.

\begin{table*}[h]
\centering
\begin{tabular}{r|rrrrr}
\hline
$L$ & Our algorithm & Trotter 1 & Trotter 2 & Trotter 4 & Trotter 8 \\
\hline
 & $2L+1$ & $2L$ & $2L+1$ & $10L+1$ & $250L+1$ \\
$64$ & $129$ & $128$ & $129$ & $641$ & $16\,001$ \\
$256$ & $513$ & $512$ & $513$ & $2\,561$ & $64\,001$ \\
$1\,024$ & $2\,049$ & $2\,048$ & $2\,049$ & $10\,241$ & $256\,001$ \\
$4\,096$ & $8\,193$ & $8\,192$ & $8\,193$ & $40\,961$ & $1\,024\,001$ \\
$16\,384$ & $32\,769$ & $32\,768$ & $32\,769$ & $163\,841$ & $4\,096\,001$ \\
$65\,536$ & $131\,073$ & $131\,072$ & $131\,073$ & $655\,361$ & $16\,384\,001$ \\
\hline
\end{tabular}
\caption{Number of elementary unitary exponentials per circuit realization, after merging adjacent kinetic factors, at every step count shown in \cref{fig:unitary-comparison}.}
\label{tab:numerical-unitary-counts}
\end{table*}

\subsection{Comparison with conventional qDRIFT}

Here $A=-\tfrac12\Delta_y-\tfrac12\Delta_z$ and $B=-1/(\sqrt{2}|y|)$. We fix $T=1$, take $L=8,16,\ldots,1024$, and set $t=T/L$. Conventional qDRIFT independently applies $e^{-2itA}$ or $e^{-2itB}$ with equal probability at each step. Our algorithm draws a fresh uniform kick time at each step as above. Its total kinetic evolution time is exactly $T$, while that of conventional qDRIFT fluctuates between trajectories.

Both algorithms use the same normalized initial state
\begin{equation}\label{eq:numerics-clock-input}
\Psi_0(y,z)\propto e^{-|y|^2/2}\left(\frac{f_0(z)}{\sqrt{2}}+\frac{\sqrt{3}}{\pi}\sum_{k=1}^{J}\frac{f_k(z)}{k}\right).
\end{equation}
with $J=1024$. The center-of-mass packets occupy disjoint kinetic-energy bands $[E_k,E_k+1]$, where $E_0=0$ and $E_k=4^k$ for $k\ge1$. With the unitary Fourier transform, we take
\begin{equation}\label{eq:numerics-clock-bands}
\widehat f_k(\xi)=\left[\frac{\nu(|\xi|^2/2-E_k)}{4\pi|\xi|}\right]^{1/2},
\end{equation}
where $\nu$ is the probability density proportional to $\exp[-1/(s(1-s))]$ on $(0,1)$ and zero elsewhere. This finite truncation of the $L^2$ example is fixed throughout the experiment.

For the relative coordinate, we expand $\sqrt{4\pi}\,r\psi(r)$ in the first $2048$ Dirichlet sine modes on $(0,24)$. We evaluate the Coulomb matrix analytically and diagonalize the radial Galerkin Hamiltonian for the reference evolution. For conventional qDRIFT, we approximate the energy integrals in the center-of-mass overlaps using $16$ Gauss-Legendre nodes per band. For our algorithm, the center-of-mass evolution is identical for every trajectory and the reference state, and therefore cancels from the trace-norm error. For each $L$, we estimate the full trace-norm error on the joint state,
\begin{equation}\label{eq:numerics-clock-error}
\bigl\|\mathbb E[|S\Psi_0\rangle\langle S\Psi_0|]-|\Psi_{\mathrm{ref}}\rangle\langle\Psi_{\mathrm{ref}}|\bigr\|_1,
\end{equation}
where $S$ is the product of the $L$ sampled steps. We use $512$ independent trajectories per algorithm. For conventional qDRIFT we average the output projectors. For our algorithm we use the finite-basis mean output and the trace-rescaled sample covariance as above. The dashed lines are proportional to $L^{-1}$ and $1/\log L$.

\subsection{Two-body hydrogen in its internal ground state}

This unexpected order reduction of the Trotter formulas was first discovered in~\cite{BurgarthFacchiHahnJohnssonYuasa2024} for the hydrogen ground state with a fixed nucleus. The work also presents careful one-body numerical experiments for a variety of order downgrade behaviors. However, for the one-body hydrogen ground state, the dynamics is essentially stationary. Inspired by physically motivated wavepacket preparations in nonadiabatic dynamics, here we consider the two-body hydrogen atom with a moving nucleus. We prepare the atom in its internal ground state with a Gaussian center-of-mass wavepacket, so that the full two-body density evolves in time.
In atomic units, the Hamiltonian is $H=A+B$, with
\begin{equation}
A=-\frac12\Delta_{x_e}-\frac{1}{2M}\Delta_{x_n},
\quad B=-\frac{1}{|x_e-x_n|},
\quad M=1836.
\end{equation}
Writing $r=x_e-x_n$ for the relative coordinate, $X=(x_e+Mx_n)/(M+1)$ for the center of mass, and $\mu=M/(M+1)$ for the reduced mass, we take the internal hydrogen ground state times a Gaussian center-of-mass wavepacket as the initial state,
\begin{equation}
\Psi_0(x_e,x_n)=\left(\frac{\mu^3}{\pi}\right)^{1/2}e^{-\mu|r|}(\pi\sigma^2)^{-3/4}e^{-|X|^2/(2\sigma^2)},\quad \sigma=(M+1)^{-1/2}.
\end{equation}
The internal ground state evolves only by a global phase, while the center-of-mass wavepacket spreads freely. Together these give the exact two-body reference state $\Psi(T)$. This choice is motivated by wavepacket studies of nonadiabatic molecular dynamics, in which a nuclear wavepacket is prepared on a single electronic potential energy surface, such as the ground-state surface, and its subsequent evolution can transfer population to other electronic states. In our two-body analogue, the atom is prepared in its internal ground state and the Gaussian wavepacket describes its center-of-mass motion.

We fix $T=1$ and use $L=16,32,\ldots,1024$ steps of size $t=T/L$. We compare Lie-Trotter splitting, Strang splitting, and the fourth-order recursive Suzuki formula with conventional qDRIFT and supersmoothing qDRIFT. At each step, conventional qDRIFT applies $e^{-2itA}$ or $e^{-2itB}$ with equal probability. Our algorithm draws $u$ uniformly from $[0,t]$ and applies $e^{-i(t-u)A}e^{-itB}e^{-iuA}$. All five methods use the same radial sine grid for the relative coordinate, with $65{,}535$ interior points on $(0,128)$ and spacing $1/512$. This basis number is comparable to Cartesian $40$ grids per dimension ($40^3 = 64{,}000$).

\begin{figure}[t]
    \centering
    \includegraphics[width=.55\textwidth]{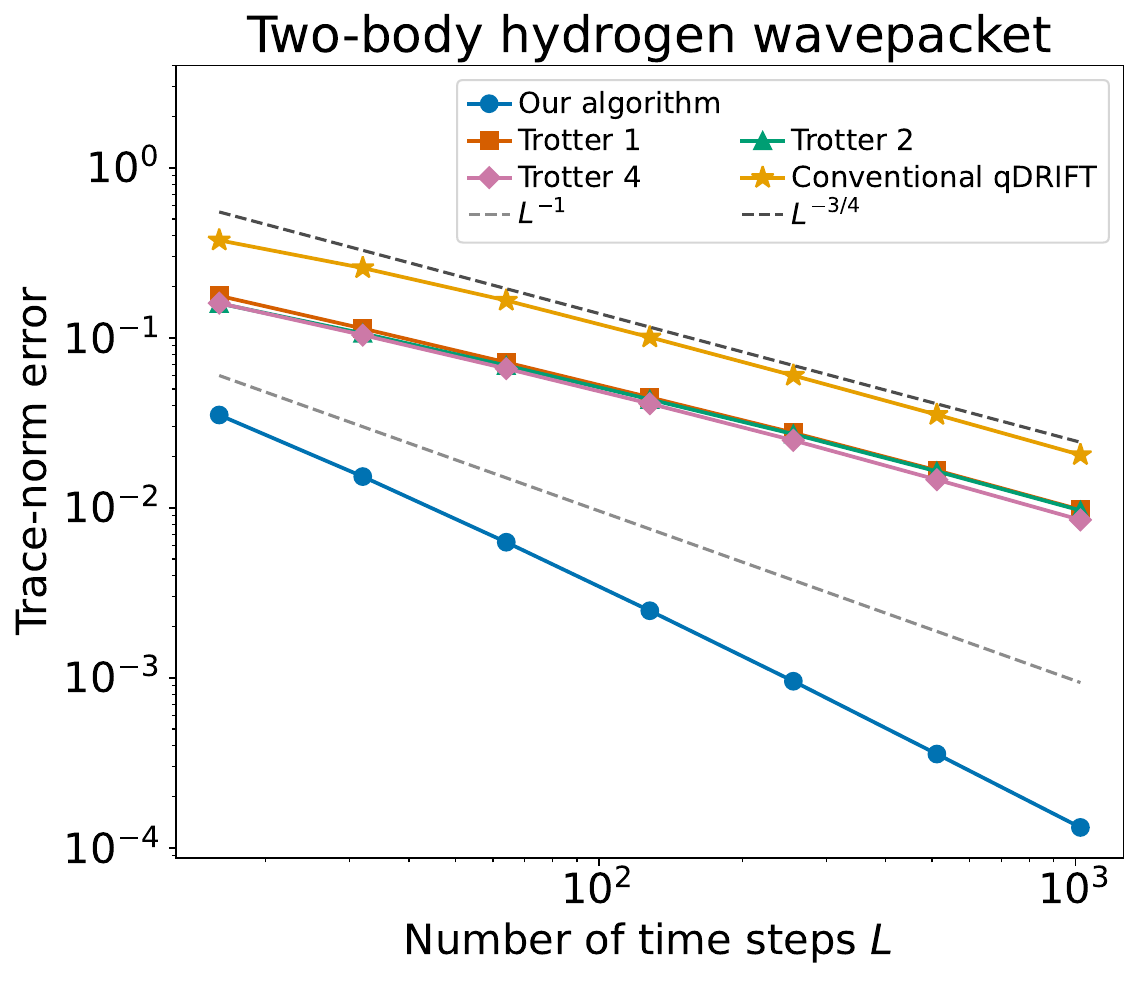}
    \caption{Trace-norm error at $T=1$ for the two-body hydrogen atom initialized in its internal ground state with a Gaussian center-of-mass wavepacket. The dashed lines show $L^{-1}$ and $L^{-3/4}$ for reference.}
    \label{fig:hydrogen-convergence}
\end{figure}

In \cref{fig:hydrogen-convergence}, we plot the full two-body trace-norm error $\|\rho_L-|\Psi(T)\rangle\langle\Psi(T)|\|_1$. For the Trotter formulas, $\rho_L$ is the propagated pure-state projector. For the randomized methods, we estimate the averaged output state using the finite-grid mean and the sample covariance, rescaled to the known trace of the covariance. We use $512$ independent trajectories per $L$ for conventional qDRIFT and $128$ for our algorithm. The Trotter formulas and our algorithm evolve the center of mass for a total time $T$, so the center-of-mass wavepacket agrees with the reference and cancels from the trace-norm error. For conventional qDRIFT, a trajectory with $n_A$ kinetic steps evolves the center of mass for time $2tn_A$. We use this time for each trajectory and evaluate the Gaussian center-of-mass energy integral by $32$-point generalized Gauss-Laguerre quadrature.

Fitting the errors over $L=128,256,512,1024$ gives observed orders of $0.733$, $0.724$, and $0.758$ for Lie-Trotter, Strang, and fourth-order Suzuki splitting, respectively. Conventional qDRIFT gives $0.766$, while supersmoothing qDRIFT gives $1.413$ over the same range.

\bibliographystyle{unsrtnat}
\bibliography{ref}
\end{document}